%% file: pgfa.tex
\documentclass[english,twoside,11pt]{article}

\usepackage{jmlr2e}

\usepackage{algorithm, algorithmicx}
\usepackage{algpseudocode}
\usepackage{amsmath}
\usepackage{amssymb}
\usepackage{booktabs}
\usepackage[margin=1in]{geometry}
\usepackage{hyperref}
\usepackage{longtable}
\usepackage{mathtools}
\usepackage{multirow}
\usepackage[section]{placeins}

\include{macros}

\begin{document}

\title{Particle-Gibbs Sampling For Bayesian Feature Allocation Models}

\author{Alexandre Bouchard-C\^{o}t\'{e}
\email bouchard@stat.ubc.ca\\
\addr Department of Statistics, University of British Columbia
\AND
Andrew Roth
\email aroth@cs.ubc.ca\\
\addr
Department of Computer Science, University of British Columbia \\
Department of Pathology and Laboratory Medicine, University of British Columbia \\
Department of Molecular Oncology, BC Cancer Agency \\
Corresponding address: 2366 Main Mall, Vancouver, BC, Canada V6T 1Z4
}

\editor{}

\maketitle

\begin{abstract}

Bayesian feature allocation models are a popular tool for modelling data with a combinatorial latent structure.
Exact inference in these models is generally intractable and so practitioners typically apply Markov Chain Monte Carlo (MCMC) methods for posterior inference.
The most widely used MCMC strategies rely on an element wise Gibbs update of the feature allocation matrix.
These element wise updates can be inefficient as features are typically strongly correlated.
To overcome this problem we have developed a Gibbs sampler that can update an entire row of the feature allocation matrix in a single move.
However, this sampler is impractical for models with a large number of features as the computational complexity scales exponentially in the number of features.
We develop a Particle Gibbs sampler that targets the same distribution as the row wise Gibbs updates, but has computational complexity that only grows linearly in the number of features.
We compare the performance of our proposed methods to the standard Gibbs sampler using synthetic data from a range of feature allocation models.
Our results suggest that row wise updates using the PG methodology can significantly improve the performance of samplers for feature allocation models.

\end{abstract}

\input{introduction}

\input{methods}

\input{results}

\input{discussion}

\clearpage

\bibliography{references}

\clearpage

\include{appendix}

\end{document}

%% file: macros.tex
\newcommand{\featMat}{Z}
\newcommand{\condCounts}{m_{k}^{(-n)}}
\newcommand{\condFeatMat}{\featMat^{(-n)}}
\newcommand{\perm}{\boldsymbol{\sigma}}
\newcommand{\set}[1]{\{ #1 \}}

\newcommand{\indicator}[1]{\mathbb{I}\left(#1\right)}
\newcommand{\featProb}{\rho}
\newcommand{\featOneProb}{\featProb_{n, k}}
\newcommand{\featZeroProb}{(1 - \featOneProb)}

\newcommand{\featProbVec}{\boldsymbol{\featProb}_{n}}

\newcommand{\datum}{x_n}
\newcommand{\data}{X}
\newcommand{\featVec}{\boldsymbol{z}}
\newcommand{\featParamVec}{\boldsymbol{\theta}}
\newcommand{\powerSet}{\{0, 1\}^{K}}
\newcommand{\rowCondDist}{p(\featVec_{n} = \featVec|\data, \condFeatMat, \featParamVec)}
\newcommand{\likelihood}[1]{p(\datum |\featVec_{#1}, \featParamVec)}
\newcommand{\featVecSMC}[1]{\boldsymbol{\xi}_{#1}}

\newcommand{\featScalarSMC}[1]{\xi_{#1}}
\newcommand{\featVecSMCConcat}{(\featVecSMC{t-1}, \featScalarSMC{t})}
\newcommand{\testPath}{\bar{\featVec}}
\newcommand{\testPathScalar}[1]{\bar{z}_{#1}}
\newcommand{\featVecFunc}[1]{\featVec(#1, \perm, \featVecSMC{#1}, \testPath) }
\newcommand{\targetDistTwoArg}[2]{\gamma_{#1}(#2|\featProbVec, \perm, \testPath)}
\newcommand{\targetDist}[1]{\targetDistTwoArg{#1}{\featVecSMC{#1}}}
\newcommand{\targetDistAnneal}[1]{\targetDistTwoArg{\beta,t}{\featVecSMC{#1}}}
\newcommand{\proposalDist}[1]{q_{#1}(\featScalarSMC{#1} |\featVecSMC{#1-1})}
\newcommand{\initProposalDist}{q_{1}(\featScalarSMC{1})}

\newcommand{\rowCondDistSMC}[1]{p(\datum|\featVecFunc{#1}, \featParamVec)}
\newcommand{\annealFactor}[1]{{\left(\frac{#1}{T} \right)^{\beta}}}
\newcommand{\rowCondDistSMCAnneal}[1]{p(\datum|\featVecSMC{#1}, \featParamVec)^{\annealFactor{#1}}}
\newcommand{\rawWeight}[2]{\widetilde{w}_{#1}^{#2}}

\newcommand{\normWeight}[2]{w_{#1}^{#2}}
\newcommand{\normWeightVec}[1]{\boldsymbol{w}_{#1}}
\newcommand{\ancestorVec}{\mathbf{a}}
\newcommand{\resampleDist}{r(\cdot|\mathbf{w}_{t-1})}
\newcommand{\algRef}[1]{Algorithm~\ref{#1}}
\newcommand{\eqnRef}[1]{Equation~\ref{#1}}
\newcommand{\figRef}[1]{Figure~\ref{#1}}
\newcommand{\figsRef}[2]{Figures~\ref{#1} and \ref{#2}}
\newcommand{\figRangeRef}[2]{Figures~\ref{#1}-\ref{#2}}

\newcommand{\tabsRef}[2]{Tables~\ref{#1} and \ref{#2}}
\newcommand{\tabRangeRef}[2]{Tables~\ref{#1}-\ref{#2}}
\newcommand{\statTabsRef}[1]{\tabsRef{tab:friedman_#1}{tab:nemenyi_#1}}
\newcommand{\includeTable}[1]{
	\begin{longtable}{lllrrrr}
		\csname @@input\endcsname #1
	\end{longtable}	
	\newpage
}
\newcommand{\traceComparisonCaption}[2]{
	Trace plots of sampling algorithms with simulated data from the #1 model with #2.
	See main text for other model parameters.	
	Rows are datasets and columns are initial parameter settings.
	Error bars are averaged over five restarts of the sampler with different random seeds.
}
\newcommand{\traceTuningCaption}[2]{
	Trace plots of #1 sampler using different #2.
	Rows are datasets and columns are initial parameter settings.
	Error bars are averaged over five restarts of the sampler with different random seeds.
}


%% file: introduction.tex
\section{Introduction}

\newcommand\abc[1]{{\bf *** #1 ***}}
\newcommand\ar[1]{{\it --- #1 ---}}

Bayesian feature allocation models posit that observed data is generated by a collection of latent features with the aim of obtaining an interpretable and sparse representation of the data.
A concrete way to represent a feature allocation is using a binary matrix, where the rows of this matrix represent data points or observations and the columns represent features.
Common prior distributions for these binary matrices include the finite dimensional Beta-Bernoulli (FBB) model and the non-parametric Indian Buffet process (IBP) \citep{griffiths2011indian}.
Exact inference for models which use these prior distributions is generally intractable, so practitioners often appeal to Monte Carlo Markov Chain (MCMC) approaches.
A straightforward Gibbs sampler can be derived for such models which proceeds by updating a single entry of the binary matrix conditioned on the values of the remaining entries.
While relatively simple to implement, this sampler can be extremely slow to mix due to the correlation among the feature allocation variables.
In this work we show that it is possible to derive a simple Gibbs sampler which updates the entire feature usage vector of a data point (row of the binary matrix) jointly.
When the number features (columns of the matrix), $K$, is small this sampler is practical and can significantly improve the mixing of the MCMC chain.
However, this sampler is computationally expensive, requiring $2^K$ evaluations of the likelihood.
We show that it is possible to sample efficiently from the distribution targeted by this row Gibbs update using the Particle Gibbs (PG) methodology \citep{andrieu2010particle}.
Our PG sampling approach has computational complexity that scales linearly with the number of features.

In sequel we will first review Bayesian feature allocation models and the relevant prior distributions on binary matrices.
Next we will describe the new row wise Gibbs update and explain how to use the PG methodology to efficiently sample from the target distribution.
We then compare these new samplers to existing approaches on a range of synthetic datasets using several previously published models.
Finally, we conclude with a discussion and some thoughts on future directions.

\subsection{Related work}

The widely used Gibbs sampler which updates the feature allocation of each data point sequentially was first introduced by \cite{ghahramani2006infinite}. 
Later \cite{meeds2007modeling} described the use of Metropolis-Hastings (MH) moves to update multiple components of the feature allocation vector for a data point.
They observed that larger moves in the space of feature allocations improved mixing, though this was never formally benchmarked.
While the MH move partially addresses the issue of highly correlated features, it becomes impractical as the number of features grows, as large moves proposed at random will increasingly be rejected.
An alternative approach to speeding up sampling for feature allocation models was proposed by \cite{doshi2009accelerated}.
The main idea of that work was to partially marginalize elements of the model to improve mixing.
This is not a general strategy however, as it requires conjugacy.
\cite{wood2007particle} proposed the use of particle filters to fit matrix factorization models using IBP priors.
In contrast to our approach, they used a single pass particle filter sampling the entire feature allocation matrix.
They showed that this approach could significantly outperform single entry Gibbs sampling.
However, the single pass particle filter approach does not scale well to models with large numbers of data points or features due to the degeneracy of standard particle filter methods.
In contrast PG algorithms are not subject to this degeneracy.
\cite{broderick2013cluster} pointed out the predictive distribution of the feature allocation models could be written as a product of Bernoulli distributions.
However, they did not appear to pursue the obvious row wise Gibbs sampler that this implies.
\cite{fox2014joint} proposed the use of split-merge moves to improve the mixing of features.
While the sequential nature of these proposals bear some similarity to our method, they differ in that this previous work updates the columns of the feature allocation matrix as opposed to the rows.
As a result they need to be interleaved with element wise Gibbs updates to obtain adequate mixing.
The methods we propose in this work can be used in place of the element wise Gibbs update with the moves proposed by \cite{fox2014joint} to further improve performance.

%% file: methods.tex
\section{Methods}

Here we review the basic background about feature allocation priors and introduce the standard Gibbs updating procedure.
We then explain how to implement an exact Gibbs sampler for updating an entire row of the feature allocation matrix.
Next we show how to construct a PG sampler to target the conditional distribution the row wise Gibbs samples from.
We then discuss two strategies for improving the efficiency of the basic PG algorithm.

\subsection{Notation}

We use bold letters for (random) vectors, capital letters for matrices and normal fonts for (random) scalars and sets.
For quantities such as an individual observation $\datum$, or a parameter $\theta$, which can be either scalars or vectors without affecting our methodology, we consider them as scalars without loss of generality. 
Given a vector $\boldsymbol{z}= (z_1, \ldots, z_K)$, and $i \leq j$, we use $\boldsymbol{z}_{i : j}$ to denote the sub-vector $\boldsymbol{z}_{i : j} = (z_i, z_{i + 1}, \ldots, z_j)$.
For a permutation, $\perm$, we let $\boldsymbol{y}[\perm] = (y_{\perm(1)}, \ldots, y_{\perm(K)})$ denote vector obtained by permuting the entries of $\boldsymbol{y}$ by $\perm$.
For a permutation $\perm$ we define the inverse permutation $\perm^{-1}$ to be the permutation such that $(\boldsymbol{y}[\perm])[\perm^{-1}] = \boldsymbol{y}$.
To simplify notation, we do not distinguish random variables from their realization.
We define discrete probability distributions with their probability mass functions, and continuous probability distributions with their density functions with respect to the Lebesgue measure. 

\subsection{Feature allocation}

Intuitively a feature allocation model ascribes a set of features that are exhibited by a set of data points $\datum$.
At the core of these models is the combinatorial stochastic feature allocation object.
To formally define a feature allocation we follow the description in \cite{broderick2013cluster}.
Let $[N] = \{ 1, \ldots, N \}$, then a feature allocation $f_N$ of $[N]$ is defined to be a multi-set of non-empty sets of $[N]$.
Let $f_N = \set{A_1^N, \ldots, A_K^N}$ where we refer to the elements $A_k^N$ as blocks. 
Each block represents the assignment of data points to a feature.
For example consider the feature allocation $f_3 = \{ \{ 1 \}, \{ 1, 2 \}, \{ 2, 3 \} \}$.
In this feature allocation the first feature is exhibited by data point 1, the second feature by data points 1 and 2, and the third feature by data points 2 and 3. 
In contrast to partitions which are frequently used in clustering models, feature allocations do not require data points to be in mutually exclusive blocks or in fact to be in any block.
If we let $z_{n, k} =\mathbb{I} (n \in A_k^N)$ then we can map the feature allocation $f_{N}$ to a binary matrix  $Z \in \{ 0, 1 \}^{N \times K}$.
The rows of $Z$ represent data points and the columns represent features.
We use the notation $\boldsymbol{z}_{n} = (z_{n, 1} \ldots z_{n, K})$ to denote the $n^{\text{th}}$ row of $Z$, that is the vector indicating which features data point $n$ uses.
Note that the ordering of features is arbitrary so that the matrix $Z$ is only defined up to a permutation of the columns and strictly speaking the feature allocation prior distribution is defined on the equivalence class of matrices that are identical up to a permutation of their columns.
An alternative way to define this equivalence class is as the set of matrices which are equivalent when put into left ordered form \citep{griffiths2011indian}.
In sequel we will abuse notation and not make the distinction between a feature allocation $f_{N}$ and its binary matrix representation $Z$.

\subsection{Feature allocation prior distributions}

To specify a Bayesian feature allocation model we need to define a prior distribution for the feature allocation.
In this work we consider the two most widely used prior distributions for feature allocations, the Finite Beta-Bernoulli (FBB) distribution and Indian Buffet Process (IBP).
Below we give: the probability mass function of these distributions; the probability that a data point $n$ exhibits feature $k$, $\featOneProb$; and the  predictive distribution when adding a new data point.
The predictive distribution is defined as $p(f_{N + 1} |f_N) = \frac{p (f_{N + 1})}{p (f_N)}$.
Let $K_N=|f_N|$ and $m_{k}=|A_{k}^{N}|=\sum_{n=1}^{N} z_{n, k}$, then these quantities are as follows: 
%
\begin{itemize}
  \item FBB with $K$ features 
  %
  \begin{eqnarray*}
  	p(f_{N}) & = & \mathbb{I}(K_{N} = K) \prod_{k=1}^{K} \frac{\Gamma (m_{k} + a) \Gamma (N - m_{k} + b)}{\Gamma (N + a + b)} \\
	\rho_{N + 1, k} & = & \frac{m_{k} + a}{N + a + b} \\
    p(f_{N + 1} |f_N) & = & \prod_{k = 1}^{K} \text{Bernoulli}\left( z_{N + 1, k} \middle| \rho_{N + 1, k} \right)
  \end{eqnarray*}
  \item Indian Buffet Process
  \begin{eqnarray*}
  	p(f_{N}) & = & \frac{\alpha^{K_{N}}}{K_{N}!} \prod_{k = 1}^{K_{N}} \frac{\Gamma(m_{k}) \Gamma(N - m_{k} + 1)}{\Gamma(N + 1)} \\
    \rho_{N + 1, k} & = & \frac{m_{k}}{N + 1} \\
    p (f_{N + 1} |f_N) & = & \text{Poisson}\left( K_{N + 1}^{+} \left|\frac{\alpha}{N + 1} \right. \right) \prod_{k = 1}^{K_N} \text{Bernoulli}\left( z_{N + 1, k} \middle| \rho_{N + 1, k} \right)
  \end{eqnarray*}
\end{itemize}
where $K_{N + 1}^{+}$ is the number of \textit{singletons} (unique) features exhibited by data point $N+1$.
We note that this definition of the IBP prior differs slightly from the original one defined in \cite{ghahramani2006infinite}.
This construction is due to \cite{broderick2013cluster} and results in an exchangeable prior as the probability mass functions only depend on the number of features and size of blocks. 
As we will see later this is useful for defining a Gibbs sampler for updating the feature allocation variable.

\subsection{Bayesian feature allocation models}

To fully specify a Bayesian feature allocation model we need two additional elements.
First, a set of parameters $\featParamVec = (\theta_{1}, \ldots, \theta_{K})$ associated with the features.
We will assume that $\theta_{k}$ are drawn i.i.d. from a common distribution so that the features are exchangeable.
Second, we need to define a likelihood for our data $\data = (x_{1}, \ldots, x_{N})^T$ which depends on our feature allocation matrix $Z$ and the feature parameters $\boldsymbol{\theta}$.
We also assume that the data points are exchangeable so the likelihood takes the form $p(\data|\featParamVec, \featMat) = \prod_{n = 1}^N \likelihood{n}$. 
In order for the model to remain exchangeable we require that for any permutation, $\perm$, that $p(\datum |\featParamVec, \featVec_n) = p (\boldsymbol{x}_n|\featParamVec[\perm], \featVec_n [\perm])$.
With these assumptions the full joint distribution is given by \eqnRef{eq:jointDist}.
\begin{equation}
\label{eq:jointDist}
  p(\data, \featMat, \featParamVec) = p(\featMat)  \left\{ \prod_{k = 1}^K p(\theta_k) \right\} \left\{ \prod_{n = 1}^N \likelihood{n} \right\}
\end{equation}
In general the component distributions will also depend on additional hyper-parameters which may also have prior distributions.
For notational clarity we have suppressed these terms and any dependencies on these hyper-parameters.

As a concrete example consider the linear Gaussian feature allocation model.
\begin{eqnarray*}
  Z & \sim & \text{IBP}(\alpha) \\
  \theta_k & \sim & \mathcal{N} (0, \mathbf{I}) \\
  \boldsymbol{x}_{n} | \sigma, \theta_k, \boldsymbol{z}_n & \sim & \mathcal{N} \left( \sum_k z_{n, k} \theta_k, \sigma^2 \mathbf{I} \right)
\end{eqnarray*}
This model assumes the feature parameters follow a multivariate normal distribution, and the data follow a multivariate normal distribution with a mean which is the sum of the features that a data point exhibits.
Note that the likelihood is invariant to permutations of the feature indexes due to the linear sum construction.
This model is thus exchangeable in both data points and features.

\subsection{Gibbs updates}

Since data points are exchangeable we can use $p(f_{N + 1} |f_N)$ to derive a simple Gibbs sampler to update the entries of $\featMat$ by assuming we are observing the last data point to be assigned.
Let $\condFeatMat = \{ \featVec_{i} \}_{i \neq n}$ indicate the entries of $\featMat$ minus the $n^{\text{th}}$ row.
Let $m_{k}^{(-n)} = \sum_{i \ne n} z_{i, k}$ and $\rho_{n, k}$ be defined by replacing $m_{k}$ in the definition of $\rho_{N+1, k}$ from the previous section with $m_{k}^{(-n)}$.
The element wise Gibbs update takes the form given by \eqnRef{eq:gibbsUpdate} for the FBB model leading to \algRef{alg:gibbs} for updating a row.
%
\begin{equation}\label{eq:gibbsUpdate}
 	p (z_{n, k} = 1| \datum, \condFeatMat, \featParamVec) \propto \featOneProb \times \likelihood{n}
\end{equation}
The update for the IBP prior is slightly more complex and is performed in two parts.
We update columns for features which are also exhibited by other data points using the Gibbs update in \algRef{alg:gibbs} as for the FBB.
The columns for features exhibited only by the current data point, singletons, are then updated with another move which leaves the target distribution invariant. 
The simplest of these is to use a Metropolis-Hastings update where the number of singletons is proposed from the Poisson with parameter $\frac{\alpha}{N}$, and the corresponding feature values from their prior distributions.
The methods we describe in this work only applies to the non-singleton updates, and can be used with any update for the singletons.

\begin{algorithm}
	\caption{Sample a row of the feature allocation using the element wise Gibbs update.}
	\begin{algorithmic}[1]
		\Function{ElementWiseGibbsUpdate}{$x_{n}$, $\featProbVec$, $\featVec_{n}$, $\perm$}
		
		\For{$k \in \perm$} \Comment{Iterate over columns in random order.}
			\State $z_{n, k} \gets 0$
			\State $p_0 \gets \featZeroProb \times \likelihood{n}$
			\State $z_{n k} \gets 1$
			\State $p_1 \gets \featOneProb \times \likelihood{n}$
			\State $p_1 \gets \frac{p_1}{p_0 + p_1}$
			\State $z_{n k} \sim \text{Bernoulli}(\cdot|p_1)$
		\EndFor
		
		\State \textbf{return} $\featVec_{n}$
		
		\EndFunction
	\end{algorithmic}
	\label{alg:gibbs}
\end{algorithm}

\subsection{Row wise Gibbs updates}

The element wise Gibbs update has been widely used.
It only requires $\mathcal{O}(K)$ evaluations of the likelihood function to update a row.
However, the resulting sampler can be extremely slow to mix due to correlations between the features.
The form of the predictive distributions for the FBB and IBP priors suggests an alternative Gibbs update that could potentially lead to better mixing.
Rather than sample a single entry at a time, instead update an entire row, $\featVec_{n}$, of the feature allocation matrix. This can be done by using the update defined by \eqnRef{eq:rowGibbsUpdate} leading to \algRef{alg:rowGibbs} for updating a row.
Again, this update only applies to the non-singleton entries when using the IBP prior.

\begin{eqnarray}\label{eq:rowGibbsUpdate}
	\rowCondDist & \propto & \likelihood{n} \prod_{k = 1}^{K} \text{Bernoulli}\left( z_{k} \middle| \featOneProb \right)
\end{eqnarray}

In order to sample from distribution defined by \eqnRef{eq:rowGibbsUpdate} we need to enumerate all possible binary vectors of length $K$ and evaluate the likelihood function.
This approach leads to a sampler with computational complexity $\mathcal{O}(2^K)$.
For moderate values of $K$, particularly if we are using the parametric FBB prior, this is a practical sampler.
However, the exponential scaling in $K$ will render this approach infeasible for larger numbers of features. 
This is especially problematic when using the IBP prior, as $K$ varies between iterations.

\begin{algorithm}
	\caption{Sample a row of the feature allocation using the row wise Gibbs update.}
	\begin{algorithmic}[1]
		\Function{RowWiseGibbsUpdate}{$x_{n}$, $\boldsymbol{\rho_{n}}$, $K$}
		
		\State $j$ $\gets$ $0$ \Comment{Counter for number of vectors}
		
		\State $S$ $\gets$ $()$ \Comment{List to store vectors}
				
		\For{$\featVec \in \powerSet$} \Comment{Iterate over all possible feature allocation vectors.}
			\State $j$ $\gets$ $j + 1$
			\State $S$ $\gets$ $(S, \featVec)$ \Comment{Add $\featVec$ to list of visited vectors}
			\State $p_{j}$ $\gets$ $\likelihood{} \prod_{k} \featZeroProb^{(1 - z_{k})} \featOneProb^{z_{k}}$
		\EndFor
		
		\For{$i \in \set{1, \dots, j}$}
			\State $p_{i}$ $\gets$ $\frac{p_{i}}{\sum_{l=1}^{j} p_{l}}$ \Comment{Normalize probabilities}
		\EndFor
		
		\State $i$ $\sim$ $\text{Categorical}(\cdot \mid \boldsymbol{p})$ \Comment{Sample vector index $i$ with probability $p_{i}$}
		
		\State $\featVec$ $\gets$ $S_{i}$
		
		\State \textbf{return} $\featVec$
		
		\EndFunction
	\end{algorithmic}
	\label{alg:rowGibbs}
\end{algorithm}

\subsection{Particle Gibbs updates}

\begin{figure}[h]
	\includegraphics[scale=1.0]{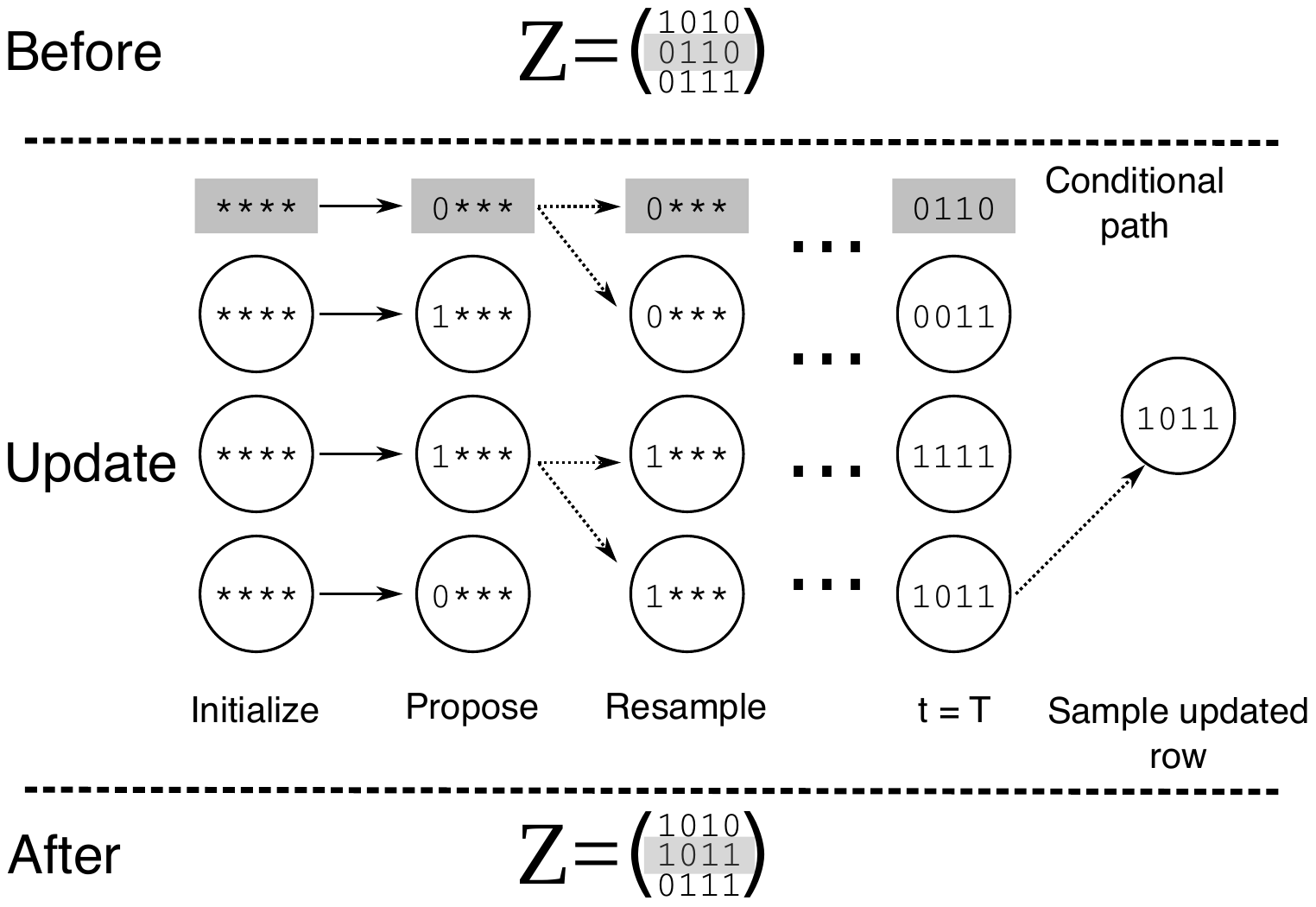}
	\center
	\caption{
		Illustration of the PG update procedure.
		Note we suppress the random ordering of features defined by $\perm$ for clarity.
		(top) Select a row for the update, shown in grey.
		(middle) Run a conditional particle filter and sample new row.
		(bottom) Update row with sample for particle filter, shown in grey.
		The stars (*) indicate values of the \textit{test path}.
		We discuss how these values can be selected in Section~\ref{sec:testPath}.
	}
	\label{fig:schematic}
\end{figure}

We now describe how to sample from $\rowCondDist$ with computational complexity $\mathcal{O}(K)$ by using the Particle Gibbs (PG) methodology \citep{andrieu2010particle}.
PG sampling is a form of Sequential Monte Carlo (SMC) sampling \citep{doucet2009tutorial}. 
Like all SMC algorithms the PG approach proceeds by approximating a sequence of distribution using a set of interacting particles.
Resampling is periodically used to prune particles which, informally, are exploring low probability regions. 
There are three key quantities that need to be defined when constructing an SMC sampler:
\begin{enumerate}
	\item The sequence of target distributions $\{ \gamma_{t} \}_{t=1}^T$ used to weigh the particles at each time step.
	\item The sequence of proposal distributions $\{ q_{t} \}_{t=1}^{T}$ used to extend particles between time steps.
	\item The resampling distribution $\resampleDist$.
\end{enumerate}
The key difference between PG and other SMC approaches is that we are updating a set of variables which have already been instantiated. 
We would like to do this in way that leads to a valid kernel targeting the conditional distribution $\rowCondDist$.
To accomplish this we need to include a conditional path, that is a particle trajectory which follows the sequence of choices required to generate the initial value before the update.
This trajectory will always be included after the resampling steps.
Thus the resampling step is conditional on including the particle representing this trajectory.
Intuitively this forces the sampler to explore regions of space around the existing value.
To simplify the bookkeeping and algorithm implementation we always assume the first particle is the conditional path.
The PG sampler is still valid when this is done as shown by \cite{chopin2015particle}.

SMC algorithms are commonly used for models with a natural sequential structure, such as state space models.
This in turn identifies a natural sequence of target distributions defined on an expanding state space.
Our setup is non-standard in that no natural sequential structure is defined.
To sample from $\rowCondDist$ we will define a sequence of distribution which updates one entry of the feature allocation vector $\featVec_{n}$ at each time step.
Thus if we have $K$ features we will define a sequence of $T = K$ target distributions.
For the FBB we take $T$ to be the fixed value of $K$ and update all elements.
For the IBP $T$ is taken to be the number of elements such $\condCounts > 0$ and we only update the corresponding feature assignments.

At time step $t$ of the algorithm the particles will take values $\featVecSMC{t} \in \{ 0, 1 \}^t$, that is we record the sequence of binary decisions up to point $t$.
In order to evaluate the likelihood term we need to set the values of the feature vector which have not been updated at time $t$.
To do this we introduce an auxiliary variable which we call the the \textit{test path} denoted $\testPath$.
We discuss and compare possible strategies for selecting $\testPath$ later.
An illustration of the method is given in \figRef{fig:schematic}.

We randomly order the features before each update by a permutation $\perm$ so that at time $t$ we sample component $\perm(t)$ of the feature allocation vector.
To obtain a complete feature vector to evaluate the likelihood we define the function given by \eqnRef{eq:featVecFunc} which returns a binary vector where the entries $\perm(1:t)$ have been set to the sampled values and the remaining entries are set to the test path.
The entries are then reordered by the inverse permutation $\perm^{-1}$.

\begin{eqnarray} \label{eq:featVecFunc}
	\featVecFunc{t} & = & (\featScalarSMC{1}, \ldots, \featScalarSMC{t},  \testPathScalar{\perm(t+1)}, \ldots, \testPathScalar{\perm(T)})[\perm^{-1}]
\end{eqnarray}
For the IBP the singleton entries are fixed to one and deterministically inserted when evaluating the likelihood.

We use the sequence of target distributions defined in \eqnRef{eq:targetDist}.
We have $\targetDist{T} \propto \rowCondDist$, that it the target density at the final iteration is proportional to the density of the distribution of interest.
This is the key constraint required to define a valid sequence of target distributions.
\begin{eqnarray} \label{eq:targetDist}
	\targetDist{t} & = & \rowCondDistSMC{t} \prod_{s=1}^{t} \rho_{n \perm(s)}^{(1 - \featScalarSMC{s})} \rho_{n \perm(s)}^{\featScalarSMC{s}}
\end{eqnarray}

The second component we need for our algorithm is a sequence of proposal distributions.
Here we exploit the fact that our proposal space is $\{0, 1\}$ and use the fully adapted proposal kernel defined in Equations~\ref{eq:initProposalDist} and \ref{eq:proposalDist}.
We use $\featVecSMCConcat$ to denote the concatenation $\featScalarSMC{t}$ to $\featVecSMC{t-1}$ and $\featVecSMC{t} = \featVecSMCConcat$.
\begin{eqnarray}
	\initProposalDist & = & \frac{\targetDist{1}}{\sum_{\featScalarSMC{1} \in {0, 1}} \targetDistTwoArg{1}{(\featScalarSMC{1})}} \label{eq:initProposalDist} \\
	\proposalDist{t} & = & \frac{\targetDist{t}}{\sum_{\featScalarSMC{t} \in {0, 1}} \targetDistTwoArg{t}{\featVecSMCConcat}} \label{eq:proposalDist}
\end{eqnarray}

Given our choice of proposal and target distributions the incremental weight functions are defined by Equations~\ref{eq:initWeight} and \ref{eq:weight}.
To reduce computational overhead $\rowCondDistSMC{t}$ can be cached to avoid re-evaluation of the likelihood term in the denominator of \eqnRef{eq:weight}.
\begin{eqnarray}
	w_{1}(\featVecSMC{1}) & = & \frac{\targetDist{1}}{\initProposalDist} \nonumber \\
	& = & \sum_{\featScalarSMC{1} \in \{ 0, 1 \}} \targetDistTwoArg{1}{(\featScalarSMC{1})} \label{eq:initWeight} \\
	w_{t}(\featVecSMC{t}|\featVecSMC{t-1}) & = & \frac{\targetDist{t}}{\targetDist{t-1} \proposalDist{t}} \nonumber \\
	& = & \sum_{\featScalarSMC{t} \in \{ 0, 1 \}} \frac{\targetDistTwoArg{t}{\featVecSMCConcat}}{\targetDist{t-1}} \label{eq:weight}
\end{eqnarray}

The final component we need to define our PG algorithm is a resampling distribution.
For simplicity we use multinomial resampling, however more sophisticated approaches such as stratified sampling could also be use.
Our resampling distribution deterministically includes the conditional path, which we arbitrarily assign to particle index 1.
The conditional multinomial resampling distribution is given by \eqnRef{eq:resamplingDist} where $\ancestorVec \in \set{1, \ldots, P}^{P}$ is the vector of ancestor indices, $\boldsymbol{w}$ the vector of normalized particle weights and $P$ is the number of particles.
\begin{eqnarray} \label{eq:resamplingDist}
	r(\ancestorVec|\boldsymbol{w}) & = & \indicator{a_1=1} \prod_{i=2}^{P} \prod_{j=1}^{P} w^{\indicator{a_i = j}}_{i}
\end{eqnarray}

\begin{algorithm}[h]
	\caption{Sample a row of the feature allocation using the particle Gibbs update.}
	\begin{algorithmic}[1]
		\Function{ParticleGibbsUpdate}{$x_{n}$, $\featVec_{n}$, $\perm$, $\boldsymbol{\rho^{n}}, \testPath$}
		
		\State $\featVecSMC{T}^{1} \gets \featVec_{n}[\perm]$ \Comment{Set conditional path}

		\For{$t \in \{1, \ldots, T-1\}$}
			\State $\featVecSMC{t}^{1} \gets (\featVecSMC{T}^{1})_{1:t}$ \Comment{First particle of each generation matches conditional path}
		\EndFor
		
		\For{$i \in \{2, \ldots, P\}$} \Comment{Initialize unconditional particles}
			\State $\featScalarSMC{1}^{i} \sim q_{1}(\cdot)$
			\State $\featVecSMC{1}^{i} \gets (\featScalarSMC{1}^{i})$
		\EndFor
		
		\For{$i \in \{1, \ldots, P\}$} \Comment{Initialize incremental importance weights}
			\State $\rawWeight{1}{i} \gets w_{1}(\featVecSMC{1}^{i})$
		\EndFor
		
		\For{$i \in \{1, \ldots, P\}$} \Comment{Compute normalized weights}
			\State $\normWeight{1}{i} \gets \frac{\rawWeight{1}{i}}{\sum_{j} \rawWeight{1}{j}}$
		\EndFor
		
		\For{$t \in \{2, \ldots, T\}$}
			\If{$(P \sum_{i} (\normWeight{t-1}{i})^2)^{-1} < \tau$} \Comment{Resample only if the relative ESS below threshold $\tau$}
				\State $\ancestorVec \sim \resampleDist$ \Comment{Conditional resampling}
				\State $\mathbf{w}_{t-1} \gets (1, \ldots, 1)$ \Comment{Reset incremental weights to one}
			\Else
				\State $\ancestorVec \gets (1, 2, \ldots, P)$ \Comment{Resampling skipped set $\ancestorVec$ to identity map}
			\EndIf
			
			\For{$i \in \{2, \ldots, P\}$} \Comment{Propose new feature usage for feature $\perm(t)$}
				\State $\featScalarSMC{t}^{i} \sim q_{t}(\cdot|\featVecSMC{t-1}^{a_i})$ 
				\State $\featVecSMC{t}^{i} \gets (\featVecSMC{t-1}^{a_i}, \featScalarSMC{t}^{i})$
			\EndFor
			
			\For{$i \in \{1, \ldots, P\}$}
				\State $\rawWeight{t}{i} \gets \normWeight{t-1}{a_i} w_{t}(\featVecSMC{t}^{i}|\featVecSMC{t-1}^{a_i})$ \Comment{Update incremental importance weights}
			\EndFor
			
			\For{$i \in \{1, \ldots, P\}$}
				\State $\normWeight{t}{i} \gets \frac{\rawWeight{t}{i}}{\sum_{j} \rawWeight{t}{j}}$ \Comment{Compute normalized weights}
			\EndFor
		\EndFor
		
		\State $\featVec \sim \sum_{i=1}^{P} \normWeight{T}{i} \delta_{\featVecSMC{T}^{i}}(\cdot)$ \Comment{Sample updated feature allocation}
		
		\State $\featVec \gets \featVec[\perm^{-1}]$ \Comment{Reorder sampled feature allocation vector by inverse of $\perm$}
		
		\State \textbf{return} $\featVec$
		
		\EndFunction
	\end{algorithmic}
	\label{alg:PG}
\end{algorithm}

\clearpage

\subsection{Annealed target distributions}

One potential pitfall of the target distribution is that due to the correlation among features, it is difficult to change a feature from its current values.
This will be particularly acute if there is a need to move through a low probability configuration.
A simple strategy to mitigate this is to consider an different family of target distributions which anneals the likelihood defined in \eqnRef{eq:annealedTargetDist}.
\begin{eqnarray} \label{eq:annealedTargetDist}
	\targetDistAnneal{t} = & \rowCondDistSMCAnneal{t} \prod_{s=1}^{t} \rho_{n \perm(s)}^{(1 - \featScalarSMC{s})} \rho_{n \perm(s)}^{\featScalarSMC{s}}
\end{eqnarray}
It can easily be checked that $\gamma_{\beta, T}(\featVec) \propto \rowCondDist$ so this sequence of target densities does indeed target the correct distribution.
Also note, the original sequence of densities is recovered if $\beta=0$.

\subsection{Discrete particle filtering}

SMC algorithms are known to be inefficient in cases where the target distribution is discrete.
This is due to the computation and storage of redundant particles. 
\cite{fearnhead2003line} addressed this problem by designing an SMC approach tailored to discrete state spaces. 
The key difference is that their approach deterministically expands each particle to test all available extensions, which is possible due to the discrete nature of the space.
In order to avoid storing an exponentially expanding system of particles, they introduce an approach to deterministically keep particles with high weights while resampling from those with low weights.
Their approach guarantees that no more that $| \mathcal{X} | M$ particles will be created, where $\mathcal{X}$ is the discrete state space and $M$ is a user specified value.
\cite{whiteley2010efficient} later showed that this approach could be adapted to the Particle Gibbs framework.
We refer to this approach as the discrete particle filter (DPF).
In practice we use a slightly different version which was proposed by \cite{barembruch2009approximate}.
This version sets the expected number of particles to $M$ instead of fixing it at exactly $M$.
We have found this implementation to be more stable numerically.
The resampling procedure is outlined in \algRef{alg:ResampleDPF}.

There is no proposal densities in the DPF algorithm so we obtain a different set of weight functions from the PG algorithm which are given by Equations~\ref{eq:initDPFWeight} and \ref{eq:DPFWeight}.
As for the PG algorithm it is useful to cache $\rowCondDistSMC{t}$ to avoid re-evaluation of the likelihood term in the denominator.
When using annealing the corresponding target densities are substituted in the weight functions.
The full details of the DPF are given in \algRef{alg:DPF}.
Again to simplify the bookkeeping our proposed algorithm always assigns the conditional path to the first particle, and this is enforced during resampling.

\begin{eqnarray}
	w_{1}(\featVecSMC{1}) & = & \targetDist{1} \label{eq:initDPFWeight} \\
	w_{t}(\featVecSMC{t}|\featVecSMC{t-1}) & = & \frac{\targetDist{t}}{\targetDist{t-1}} \label{eq:DPFWeight}
\end{eqnarray}

\begin{algorithm}
	\caption{Conditional resampling for DPF.}
	\begin{algorithmic}[1]
		\Function{ResampleDPF}{$\textbf{w}$, $M$, $P$}
			\State $c$ $\gets$ \Call{findRoot}{$\sum_{i=1}^{P} \text{min}(1, x w_i) - M$} \Comment{Find unique root $x$ to equation on right}

			\State $\ancestorVec \gets \{ 1 \}$ \Comment{First index is conditional path}
			
			\State $j \gets 1$ \Comment{Initialize number of retained particles}
			
			\If{$w_1 \geq \frac{1}{c}$}
				\State $\rawWeight{1}{} \gets w_1$

			\Else
				\State $\rawWeight{1}{} \gets \frac{1}{c}$
			
			\EndIf
			
			\For{$i \in \{2, \ldots, P\}$}
				\If{$w_i \geq \frac{1}{c}$} \Comment{Retain particles with large weights}
					\State $\ancestorVec \gets (\ancestorVec, i)$
					
					\State $\rawWeight{j}{} \gets w_i$

					\State $j \gets j + 1$
				
				\Else \Comment{Resample particles with small weights}
					\State $U \sim \text{Uniform}(\cdot \mid [0, 1])$
					
					\If{$c \, w_{i} \geq U$}					
						\State $\ancestorVec \gets (\ancestorVec, i)$

						\State $\rawWeight{j}{} \gets \frac{1}{c}$

						\State $j \gets j + 1$						

					\EndIf

				\EndIf
				
			\EndFor
			
			\For{$i \in \{1, \ldots, j\}$}
				\State $w_{i}^{new} \gets \frac{\rawWeight{i}{}}{\sum_{l=1}^{j}\rawWeight{l}{}}$ \Comment{Normalize new weights}
			\EndFor
			
			\State \textbf{return} $\ancestorVec$, $\textbf{w}^{new}$, $j$
			
		\EndFunction
	\end{algorithmic}
	\label{alg:ResampleDPF}
\end{algorithm}

\begin{algorithm}
	\caption{Sample a row of the feature allocation using the discrete particle filter update.}
	\begin{algorithmic}[1]
		\Function{DiscreteParticleFilter}{$x_{n}$, $\featVec_{n}$, $\perm$, $\boldsymbol{\rho^{n}}, \testPath$, $M$}
	
		\State $\featVecSMC{T}^{1} \gets \featVec[\perm]$ \Comment{Set conditional path}

		\For{$t \in \{1, \ldots, T-1\}$}
			\State $\featVecSMC{t}^{1}$ $\gets$ $(\featVecSMC{T})_{1:t}$ \Comment{First particle of each generation matches conditional path}
			\State $\featVecSMC{t}^{2}$ $\gets$ $(\featVecSMC{t-1}^{1}, 1 - (\featVecSMC{t}^{1})_{t})$ \Comment{Expand conditional path with alternate feature value for time $t$}
		\EndFor		

		\State $P \gets 2$ \Comment{Initialize number of particles}

		\For{$i \in \{1, 2\}$}
			\State $\rawWeight{1}{i} \gets w_{1}(\featVecSMC{1}^{i})$ \Comment{Initialize incremental importance weights}
		\EndFor
		
		\For{$i \in \{1, 2\}$}
			\State $\normWeight{1}{i} \gets \frac{\rawWeight{1}{i}}{\sum_{j} \rawWeight{1}{j}}$ \Comment{Compute normalized weights}
		\EndFor
			
		\For{$t \in \{2, \ldots, T \}$}
			\If{$P > M$} \Comment{Check if there are too many particles}
				\State $\ancestorVec$, $\normWeightVec{t}$, $P$ $\gets$ \Call{ResampleDPF}{$\normWeightVec{t}$, $M$, $P$}  \Comment{Resample using Algorithm~\ref{alg:ResampleDPF}}
			\Else
				\State $\ancestorVec \gets (1, \ldots, P)$ \Comment{Set ancestor indices to identity map}
			\EndIf

			\State $j \gets 2$ \Comment{Track number of particles}	
			
			\For{$i \in \{2, \ldots, P\}$}
				\For{$z \in \{0, 1\}$}
					\State $j \gets j + 1$
					\State $\featVecSMC{t}^{j} \gets (\featVecSMC{t-1}^{a_i}, z)$ \Comment{Expand unconditional particles}
				\EndFor
			\EndFor
			
			\State $P \gets j$ \Comment{Update number of particles}
	
			\For{$i \in \{1, \ldots, P\}$}
				\State $\rawWeight{t}{i} \gets \normWeight{t-1}{a_i} w_{t}(\featVecSMC{t}^{i}|\featVecSMC{t-1}^{a_i})$ \Comment{Update incremental importance weights}
			\EndFor
			
			\For{$i \in \{1, \ldots, P\}$}
				\State $\normWeight{t}{i} \gets \frac{\rawWeight{t}{i}}{\sum_{j} \rawWeight{t}{j}}$ \Comment{Compute normalized weights}
			\EndFor
						
		\EndFor
		
		\State $\featVec \sim \sum_{i=1}^{P} \normWeight{T}{i} \delta_{\featVecSMC{T}^{i}}(\cdot)$ \Comment{Sample updated feature allocation}
		
		\State $\featVec \gets \featVec[\perm^{-1}]$ \Comment{Reorder sampled feature allocation vector by inverse of $\perm$}
		
		\State \textbf{return} $\featVec$		
		
		\EndFunction		
	\end{algorithmic}
	\label{alg:DPF}
\end{algorithm}

%% file: results.tex
\section{Results}

We first demonstrate the potential slow mixing of the standard Gibbs sampler on a toy dataset and illustrate how the row wise Gibbs updates can alleviate this problem.
Next we explore how to tune the parameters of the PG and DPF samplers.
We then compare the behaviour of the Gibbs sampler and our proposed methods on a number of synthetic datasets.

We have compared the performance of the Gibbs, Row Gibbs (RG), Particle Gibbs (RG) and Discrete Particle Filter (DPF) using three models.
The first model we tested with was the Linear Gaussian (LG) model, which has been widely used in the IBP literature \citep{griffiths2011indian}.
The second model we considered was the Latent Feature Relational Model (LFRM) proposed by \cite{miller2009nonparametric}.
The final model we consider is a modified version of the PyClone model used for inferring population structure from admixed data in cancer genomics \citep{roth2014pyclone}.
The original PyClone model clusters sets of mutations which appear in a similar proportion of cells.
We have modified this model to use feature allocations to indicate which cell populations have each mutation.
Full details of the models and the updates used for parameters are in the Section~\ref{sec:models}.

When comparing methods we applied the Friedman test to see if there were any significant difference in performance between the methods (p-value $<$ 0.001).
If the Friedman test was significant we then applied the post-hoc Nemenyi test with a Bonferroni correction to all pairs of models to determine which models showed significantly different performance from each other (p-value $<$ 0.001) \citep{demvsar2006statistical}.
All statements of significance are with respect to this test.
Because the samplers have different computational complexity per iteration, we report the results using wall clock time instead of per iteration.
This ensures a fair comparison, as for example, we can perform many more updates using the Gibbs sampler than the PG sampler in a given time interval.
We report the relative log density when comparing methods to better represent how far away from convergence the samplers are.
Let $\hat{\ell}$ be the log density of the data under the true parameters used for simulation and $\ell$ the observed log density.
The relative log density is given by $\frac{\ell - \hat{\ell}}{\hat{\ell}}$.

Code implementing the samplers and models is available online at \url{https://github.com/aroth85/pgfa}.
All experiments were done using version 0.2.2 of the software.
Code for performing the experiments is available online at \url{https://github.com/aroth85/pgfa_experiments}.

\subsection{Row updates improve mixing}

\begin{figure}[h]
\includegraphics[scale=0.8]{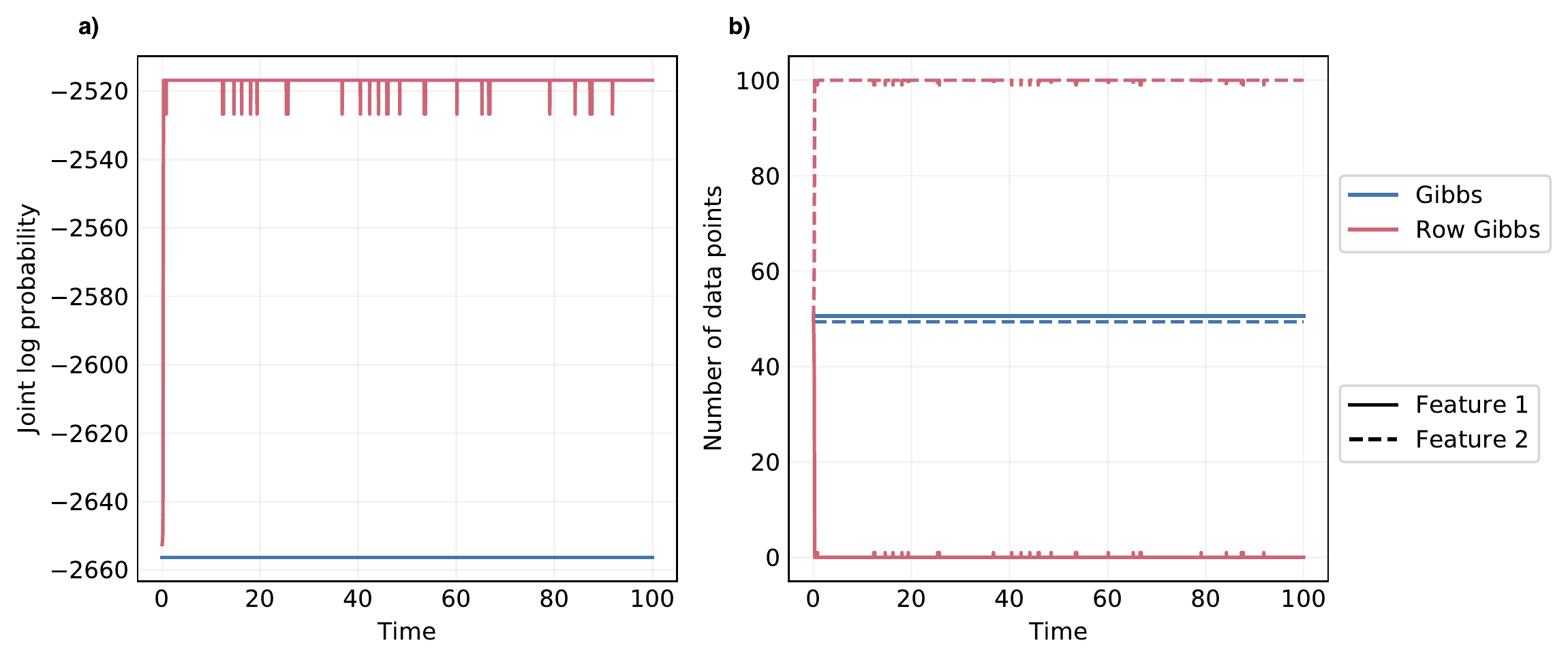}
\caption{
	Comparison of element wise Gibbs to row Gibbs sampler.
	\textbf{a}) Log joint probability of the samplers over time.
	\textbf{b}) Number of data points assigned to each feature over time.
	Lines for the Gibbs sampler are jittered away from 50 for visibility.
}
\label{fig:g_vs_rg}
\end{figure}

To illustrate the potential benefits of using row wise updates, we first consider a simple pedagogical example.
We simulated $N=100$ data points from the linear Gaussian model with $D=1$, $K=2$, $\tau_{v}=0.25$, and $\tau_{x}=25$.
We set the value of the feature parameters $V$ to be 100 for both features with half of the data points using the first feature and half using the second feature.
For inference we used the FBB prior with $K=2$, $a=0.5$ and $b=1$.
This prior distribution for the feature allocation heavily favours the configuration where all data points use one feature and the other is not used.
Because both features have identical values, there is no difference is likelihood for a data point to use one feature or the other.
Thus, if a sampler is mixing efficiently it should quickly assign all data points to one feature and none to the other.
We ran both the element wise Gibbs and row wise Gibbs samplers for 100 seconds recording the value of the log joint probability and number of data points that used each feature at each iteration.
We set all model parameters except the feature allocation to their true values, and did not update them in contrast to the remaining experiments where the feature parameters are updated.
We show the trace of the log joint probability in \figRef{fig:g_vs_rg} \textbf{a}).
The element wise Gibbs sampler (blue) is clearly trapped in a local mode from initialization and cannot move away from the initial configuration.
This is due to the need for the element wise Gibbs sampler to traverse a region of low probability to use the other feature.
Specifically, a data point must use both features or neither feature in one iteration before it can then only use the other feature in the next iteration.
\figRef{fig:g_vs_rg} \textbf{b}) supports this hypothesis as we see that the number of data points using each feature never changes over the course of sampling.
In contrast the row wise Gibbs sampler (red) rapidly increases the joint probability \figRef{fig:g_vs_rg} \textbf{a}) and moves all data points to a single feature \figRef{fig:g_vs_rg} \textbf{b}).
This contrived example clearly illustrates the potential for slow mixing that element wise updates can cause and that row wise updates can solve the problem.
We will see that this behaviour is a general phenomenon of the element wise Gibbs sampler, even when the initialization is not constructed to be adversarial as in this case.

\subsection{Setting tuning parameters}

The PG and DPF samplers have a number of tuning parameters which affect performance.
We explored the impact these parameters have on performance using synthetic data generated from the LG model.
We generated four datasets and four sets of initial parameter values.
For all combinations of datasets and initial parameters we performed five random restarts of the sampler.
Thus we executed 80 chains for each method considered, all with the same data and parameter initialization.
Data was simulated from the LG model using the FBB prior with $\alpha=2$, $\tau_v=0.25$, $\tau_x=25$, $D=10$, $K=20$ and $N=100$.
These parameters were chosen to generate datasets where we would expect the sampler to converge to the true parameter values used for simulation.
We randomly assigned 10\% of the data matrix to be missing and used these entries to compute root mean square reconstruction error (RMSE).
For each experiment we varied a single tuning parameter, setting the remaining parameters to default values, namely an annealing power of 1.0, a number of particles of 20, a resampling threshold of 0.5, and test paths consisting of a vector of zeros.

\subsubsection{Number of particles}

The first parameter we explore is the number of particles.
For standard SMC algorithms a large number of particles are typically used, as this parameter ultimately controls the quality of the Monte Carlo approximation.
In contrast to standard SMC, the Particle Gibbs framework is less sensitive to the number of particles.
This is primarily a result of the fact many conditional SMC (cSMC) moves can be used within a sampling run, in contrast to the one shot approach of SMC.
For our particular problem the length of the cSMC runs also tends to be short because we rarely expect very large numbers of features to be used in the model.
As a result, our updates will also be less sensitive to path degeneracy.

We benchmarked the PG and DPF algorithms using varying number of particles (\figRangeRef{fig:pg_num_particles_trace}{fig:num_particles_boxplot}).
Both algorithms appear to be relatively insensitive to the number of particles used.
Using 50 and 100 particles leads to significantly worse performance for both the PG (\statTabsRef{pg_num_particles}) and DPF samplers (\statTabsRef{dpf_num_particles}) than using fewer particles after the algorithms have run for 10 seconds. 
However, after 1000 seconds there were no significant differences between the runs with different numbers of particles for either method.
One surprising feature is that runs using as few as two particles still work well.
We caution this observation may not hold for other models or larger numbers of features, and more particle may be required.
We also note that it is possible to parallelize these samplers across particles which would allow for more particles to be used, though we did not investigate this.

\subsubsection{Resampling threshold}

We use an adaptive resampling scheme for the PG algorithm, whereby resampling only occurs if the relative effective sample size (ESS) falls below a specified threshold.
The DPF algorithm does not require this tuning parameter as the resampling mechanism is deterministic.
\figsRef{fig:pg_resample_threshold_trace}{fig:pg_resample_threshold_boxplot} shows the results of the benchmark experiment.
The performance of the PG algorithm was insensitive to the value of this parameter with the exception of using a threshold of 1.0 which corresponds to always resampling.
Always resampling performed significantly worse (\statTabsRef{pg_resample_threshold}) than several other thresholds at all time points.
Somewhat surprisingly when the resampling threshold is 0.0, that is never resampling, the sampler still performed well.

\subsubsection{Annealing power}

As discussed in the methods we can use a sequence of target distribution which anneals the data likelihood.
In principle this allows the method to defer resampling away particles with low data likelihood at early stages.
We explore the impact of the annealing parameter in \figRangeRef{fig:pg_annealing_power_trace}{fig:annealing_power_boxplot}.
The PG sampler using no annealing, that is setting the power to zero, performed significantly worse (\statTabsRef{pg_annealing_power}).
The actual value of the annealing power seemed to be less important provided it was larger than zero.
The DPF sampler was generally insensitive to this parameter.
The only significant difference observed was between using a power of 0.0 and 3.0 after 10 seconds (\statTabsRef{dpf_annealing_power}) and this difference disappears for later times.
This is likely due to the fact all possible paths from early time steps are included in by the DPF sampler, and are not resampled away.

\subsubsection{Test path}\label{sec:testPath}

In order to evaluate the data likelihood term in the target distributions, we must instantiate the values of the feature allocation vector that have not been updated yet.
We consider several strategies for doing so:
\begin{itemize}
\item Conditional: Use the value of the conditional path.
\item Ones: Set the value of all features to one.
\item Random: Draw the value of the feature vector uniformly at random.
\item Two stage: Run an unconditional SMC sampler using the conditional path to draw a test path.
\item Unconditional: Similar to two stage but using zeros as the test path for the first pass unconditional SMC.
\item Zeros: Set the value of all features to zero.
\end{itemize}
The Conditional and Two Stage strategies do not lead to valid Gibbs updates due to the dependency on the conditional path.
However, we include them in this analysis as they could be used during a burnin phase.
After burnin, another strategy which does lead to a valid Gibbs update could be used.
The Two Stage and Unconditional strategies both use a pilot run of unconditional SMC.
This increases run time, and introduces additional tuning parameters.
For the purpose of this experiment, we set the tuning parameters of both the SMC and cSMC passes to the same values.

\figRangeRef{fig:pg_test_path_trace}{fig:test_path_boxplot} show the results of the experiment. 
For the PG sampler the Ones and Random test paths performed significantly worse than other approaches.
At early time points the Conditional and Zeros strategies were the best, but at later time points the Two Stage and Unconstrained approaches were not significantly worse (\statTabsRef{pg_test_path}).
For the DPF algorithm the Conditional, Random, and Zeros methods significantly outperformed other approaches after 10 seconds (\statTabsRef{dpf_test_path}).
Both the Conditional and Zeros methods significantly outperformed the Random method at this time.
For later time points no methods had significantly different performance.
This result suggests that the simple approach of using a test path of zeros is effective, though there may still be some benefit of using the Conditional strategy for burnin.
This experiment also suggests that the PG sampler is sensitive to this parameter, whereas the DPF sampler is quite robust.

\subsubsection{Summary}

Based on these results we used the following parameter values for subsequent experiments.

\begin{itemize}
	\item Annealing power - 1.0
	\item Number of particles - 20
	\item Resample threshold - 0.5
	\item Test path - Zeros
\end{itemize}

These were not necessarily the optimal parameters based on the experiments, but were reasonably close to optimal.
Note we use the Zeros test path strategy to ensure we have a valid Markov Chain kernel targeting the correct distribution.

\subsection{Method comparison}

To compare the performance of our proposed approaches to the standard Gibbs sampler we generated synthetic data from three feature allocation models.
For all comparisons we ran 80 chains for each sampler as in the tuning experiments.
We simulated data with parameter values which should lead to easily identifiable solutions and thus we would expect the samplers to converge to a distribution concentrated on the parameters used for simulation. 

\subsubsection{Linear Gaussian model}

\begin{figure}
	\includegraphics[scale=0.55]{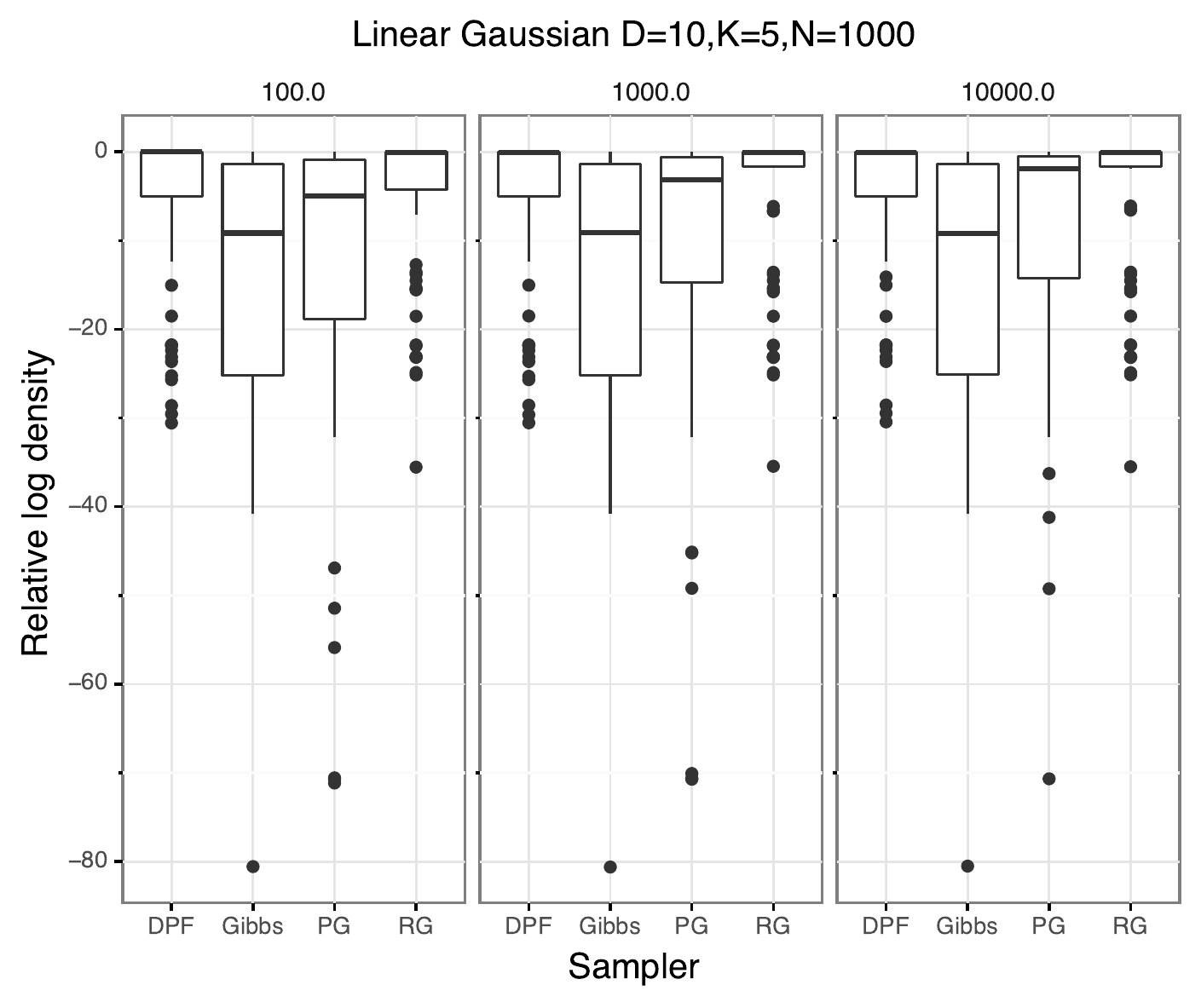}
	\includegraphics[scale=0.55]{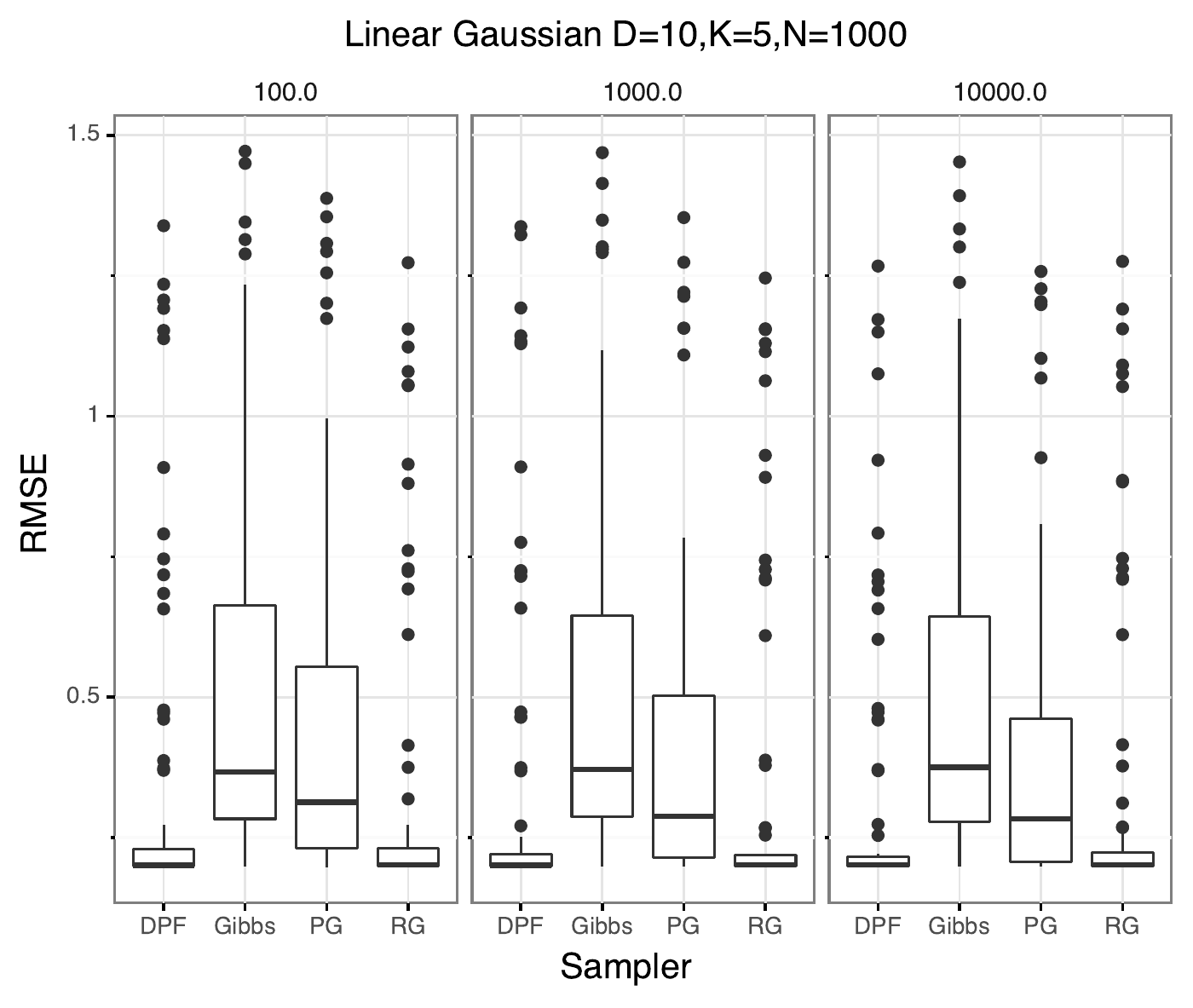}
	\vfill
	\includegraphics[scale=0.55]{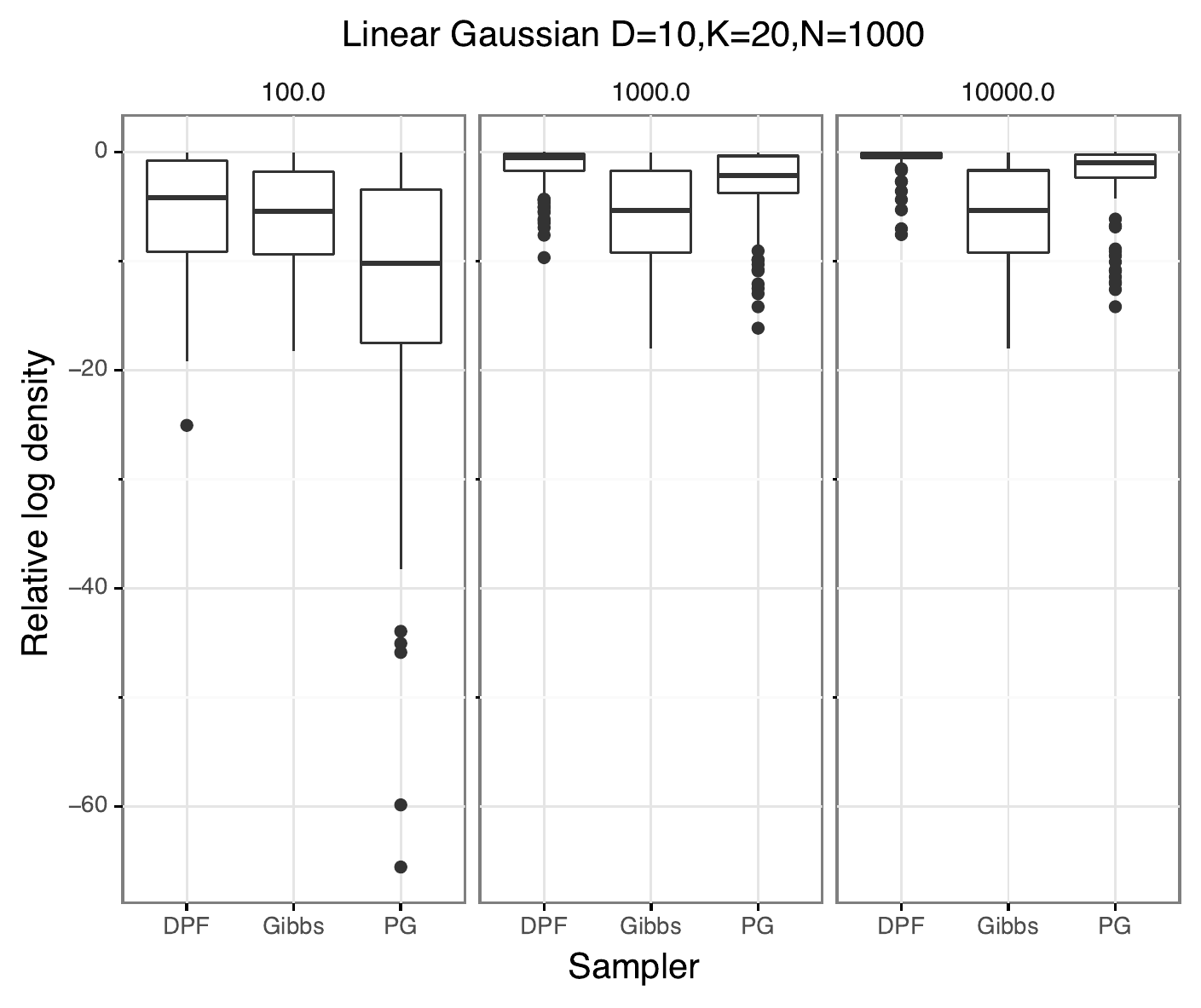}
	\includegraphics[scale=0.55]{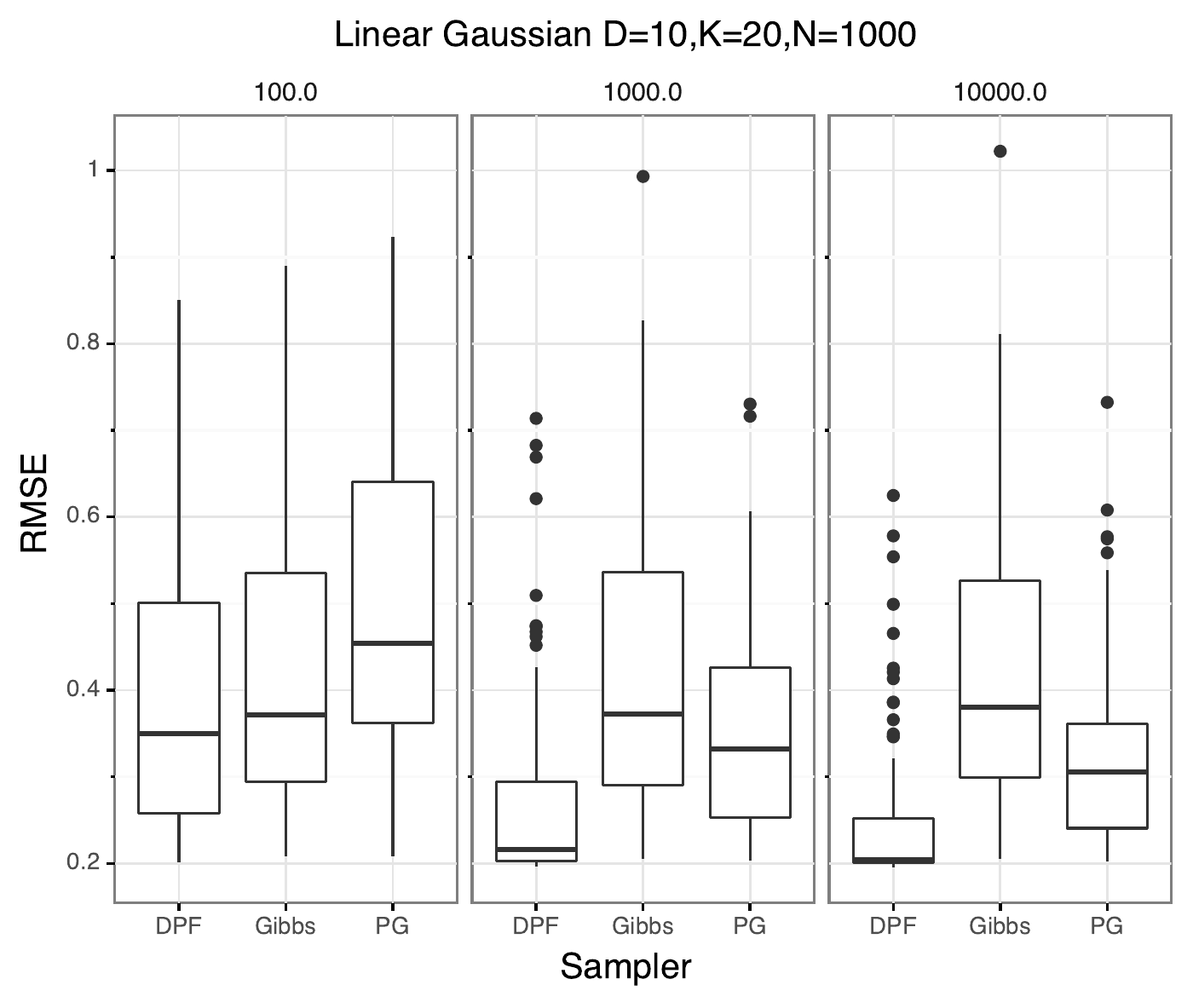}
	\vfill
	\includegraphics[scale=0.55]{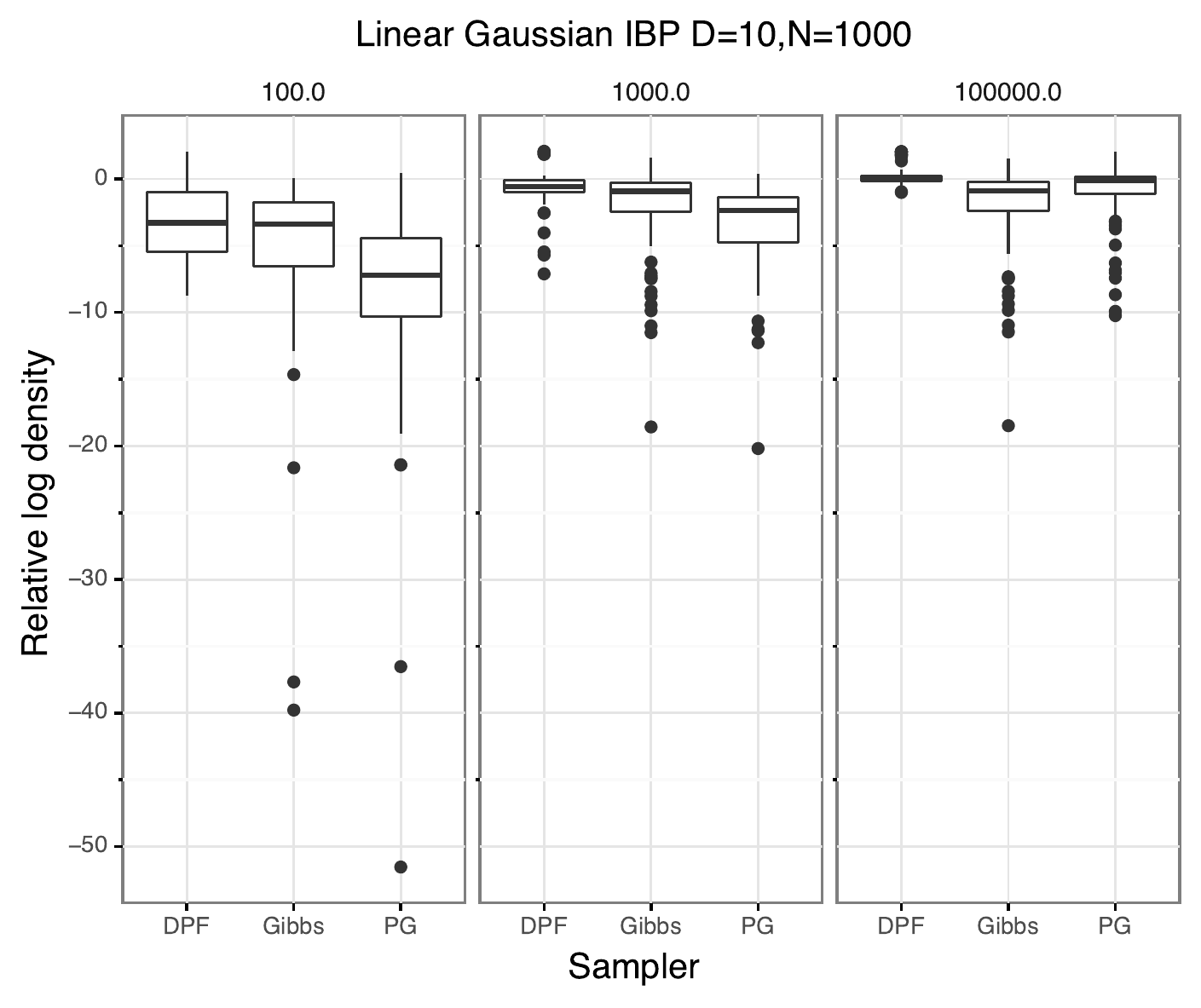}
	\includegraphics[scale=0.55]{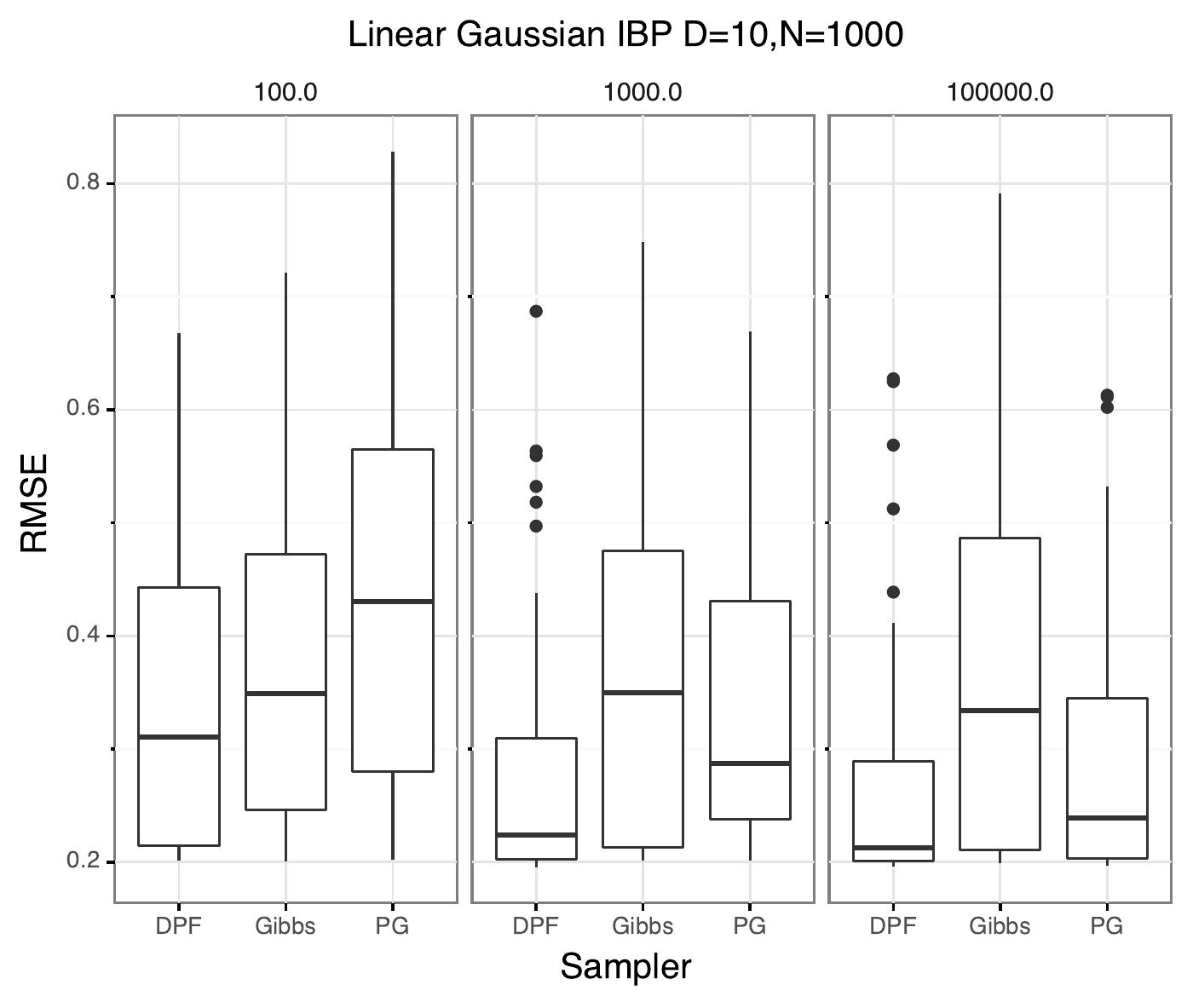}
	\caption{
		Performance of different samplers using synthetic data from the LG model.
		The box plots represent the distribution of values from 80 random starts of each parameter setting.
		We show the values of the relative log density (left, higher is better) and root mean square error reconstruction of missing values (right, lower is better).
	}
	\label{fig:lg_boxplot}
\end{figure}

We generated datasets using two sets of model parameters.
The first dataset was simulated using the FBB prior and $K=5$ and the second was simulated using the FBB model with $K=20$.
We fit the second dataset using both the FBB prior with $K=20$ and the IBP prior.
For both datasets we simulated $N=1000$ data points from the linear Gaussian model with $\alpha=2$, $\tau_{v}=0.25$, and $\tau_{x}=25$.

The results of these experiments are shown in \figRef{fig:lg_boxplot} and \figRangeRef{fig:lg_finite_5_trace}{fig:lg_ibp_20_trace}.
For the first experiment with $K=5$ it was computationally feasible to use the RG sampler.
Because we use 20 particles for the DPF algorithm it is equivalent to the RG sampler in this case.
The RG sampler serves as the gold standard for the DPF and PG methods in this experiment.
For the $K=5$ dataset the RG sampler significantly outperforms the Gibbs sampler, supporting the results of our initial toy data experiment (\statTabsRef{lg_finite_5}).
The DPF and RG samplers do not perform significantly different as expected, and outperform the other two approaches.
The PG sampler does not significantly outperform the Gibbs sampler.

For the second dataset, fitting the FBB prior model ($K=20$) the DPF sampler does not significantly outperform the Gibbs sampler after 100 seconds (\statTabsRef{lg_finite_20}).
However, for longer runs the performance advantage of the DPF sampler becomes significant.
At the earliest time point the Gibbs sampler significantly outperforms the PG sampler, but the situation reverse at later time points.

The results are somewhat different for the third experiment fitting the IBP model.
In this case we see that the Gibbs sampler outperformed the PG sampler at early time points (\statTabsRef{lg_ibp_20}).
As the samplers were run longer the PG sampler began to outperform the Gibbs sampler.
The DPF sampler outperforms both approaches.
One explanation for the better performance of the Gibbs sampler over the PG sampler is that the Gibbs sampler can propose more moves to alter the dimensionality of the model in the same time period.
Thus during the burnin phase the Gibbs sampler can more efficiently move the model to the correct number of features which improves performance.
However, the fact the DPF sampler outperforms both, suggests that the ability to perform efficient updates on the non-singleton columns dominates this effect.

\subsubsection{Latent Feature Relational Model}

We next explored performance using the LFRM model described by \cite{miller2009nonparametric}.
As for the LG experiment, we generated datasets using two sets of model parameters.
We fit the second dataset using both the FBB prior with $K=20$ and the IBP prior.
We again executed 80 runs for all samplers using the same strategy as previous experiments.
We simulated $N=100$ data points with parameters $\alpha=2$ and $\tau=0.25$ from the non-symmetric LFRM model.
We randomly assigned 5\% of the data matrix to be missing.
In addition to the relative log density we report the reconstruction error of the model for the entire data matrix, both observed and missing values.

The result of these experiments are shown in \figRangeRef{fig:lfrm_finite_5_trace}{fig:lfrm_boxplot}.
The RG and DPF methods significantly outperformed the other two methods for the K=5 experiment in terms of relative log density (\statTabsRef{lfrm_finite_5}).
However, the difference in reconstruction error was not significant. 
There was no significant difference between samplers for the other two runs (\tabRangeRef{tab:friedman_lfrm_finite_20}{tab:nemenyi_lfrm_ibp_20}).

\subsubsection{PyClone model}

\begin{figure}
	\includegraphics[scale=0.55]{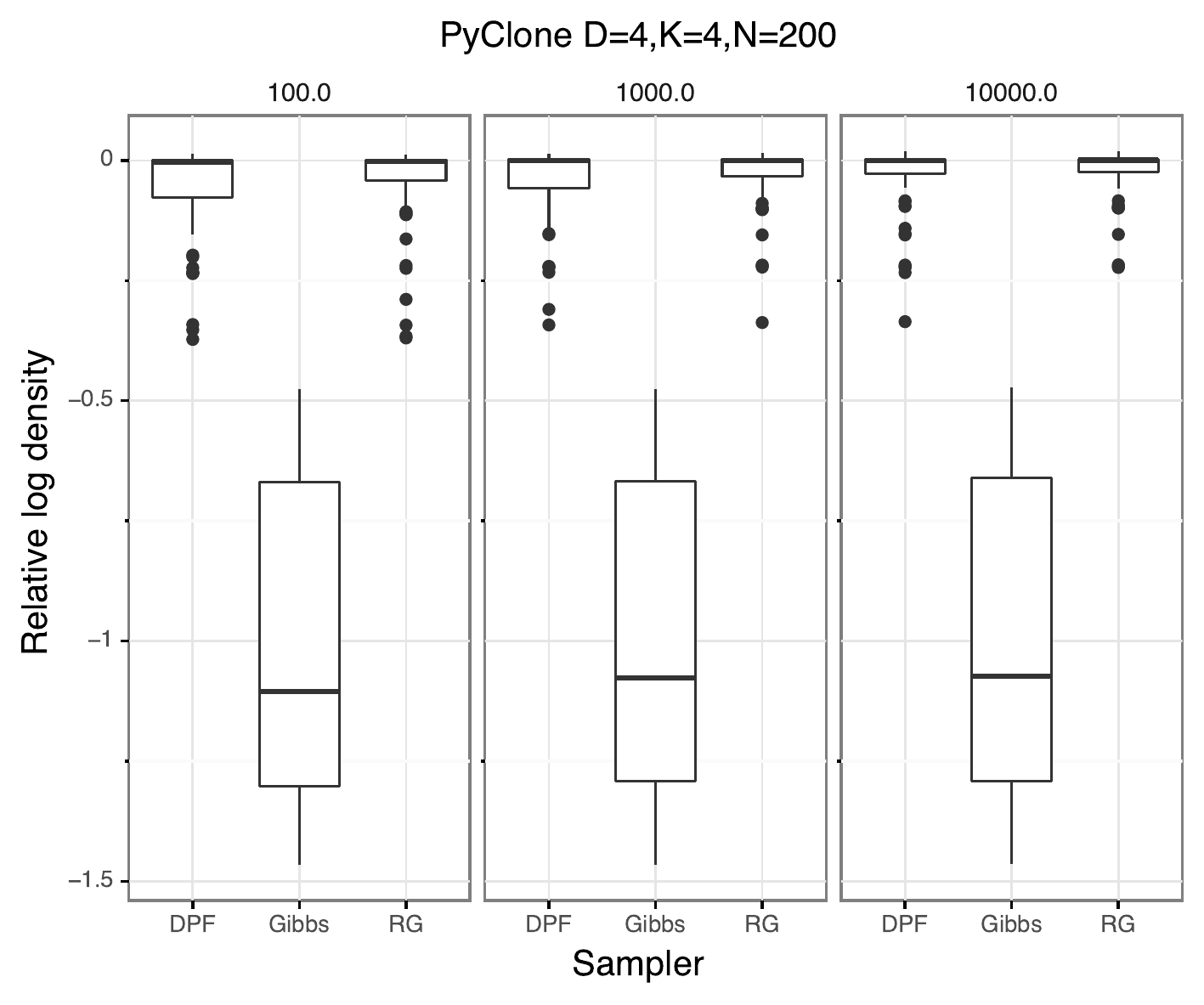}
	\includegraphics[scale=0.55]{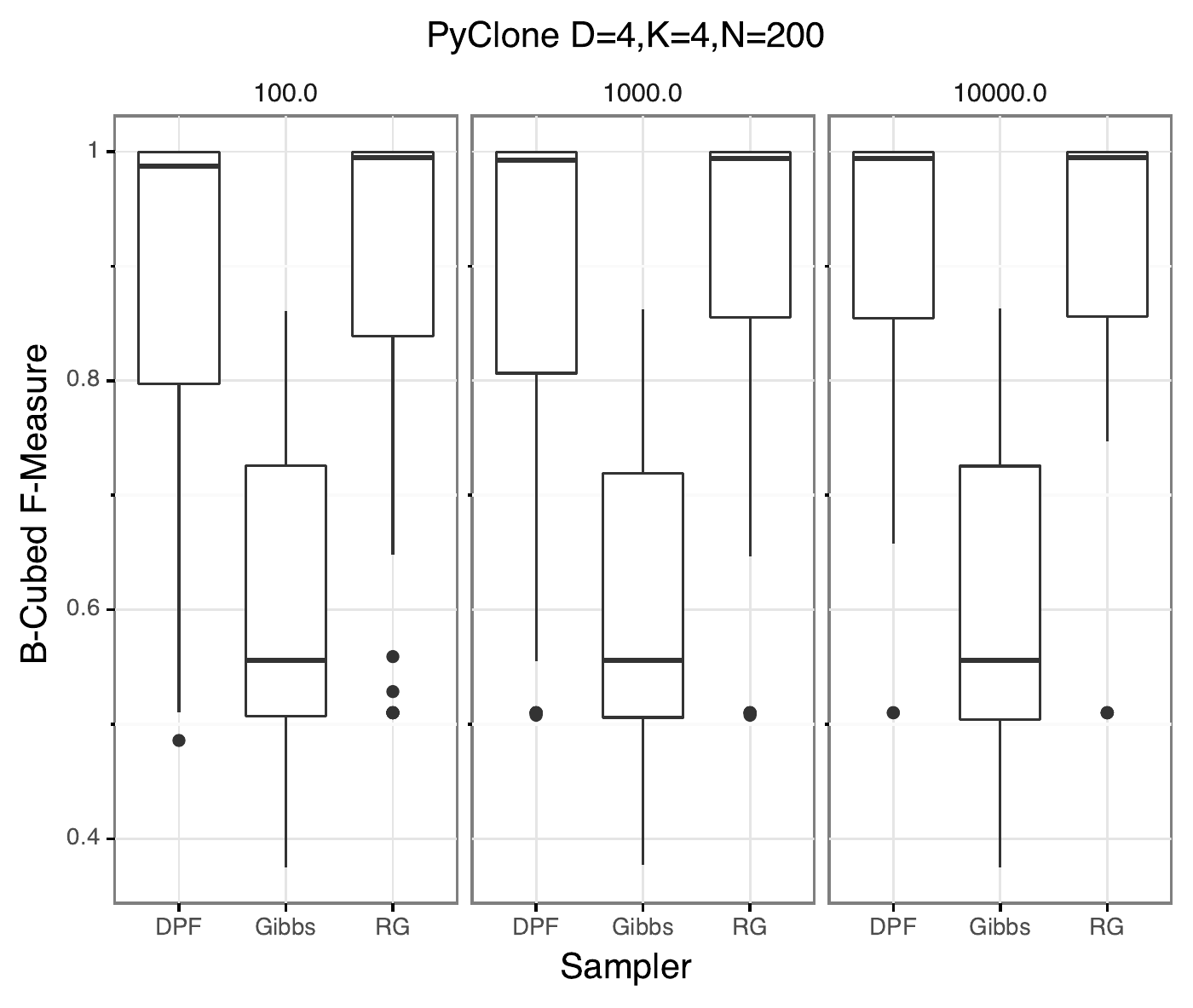}
	\vfill
	\includegraphics[scale=0.55]{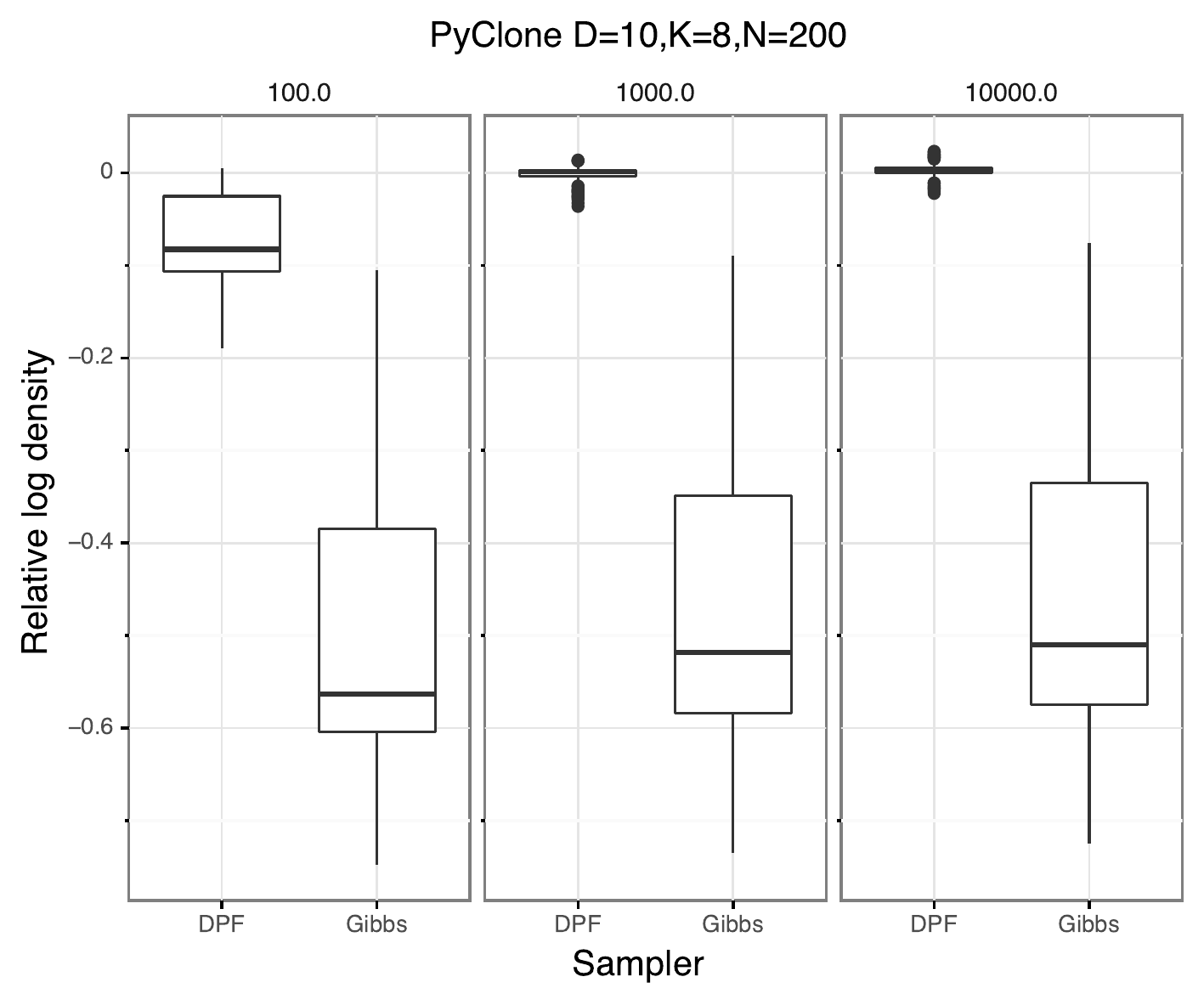}
	\includegraphics[scale=0.55]{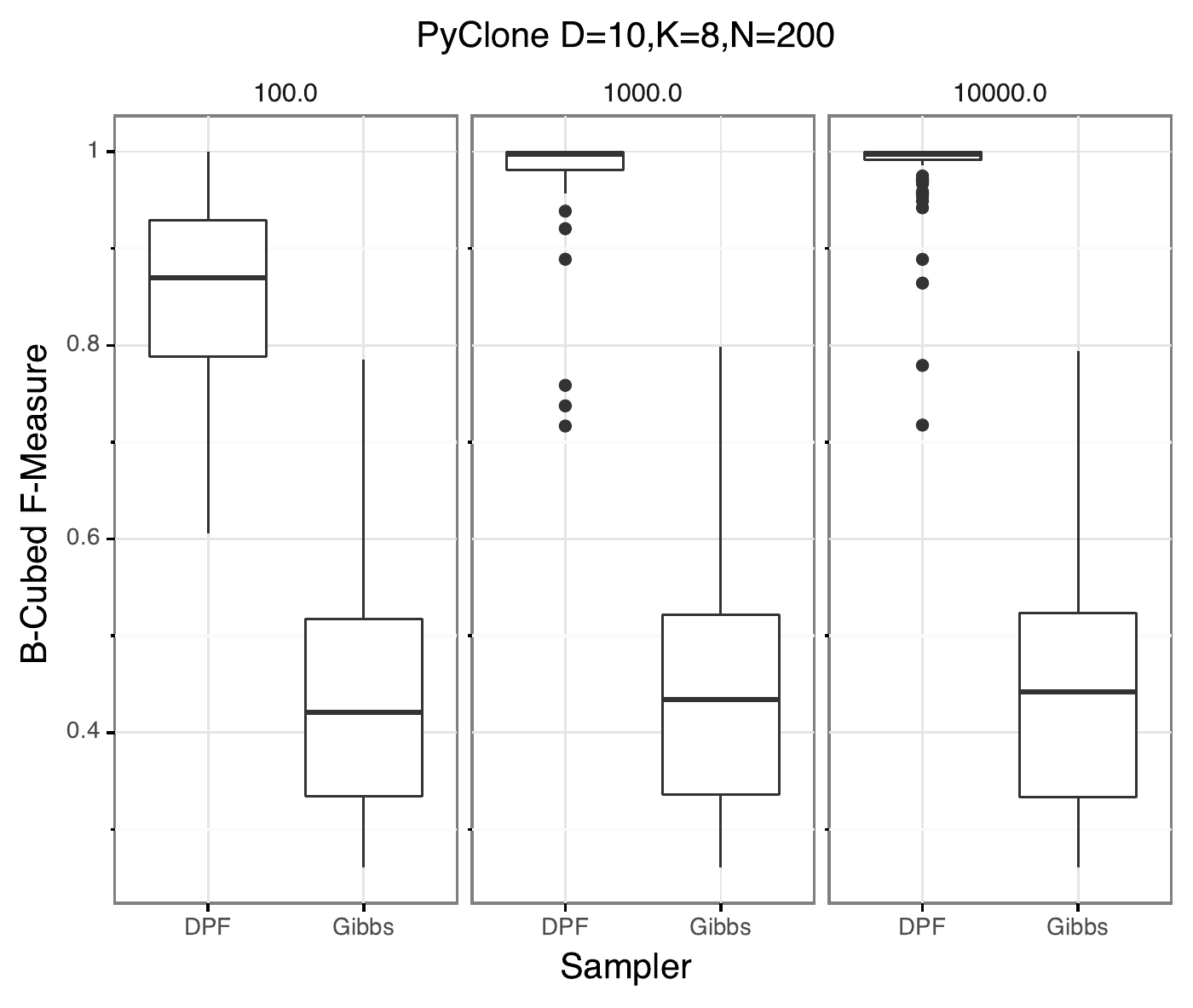}
	\vfill
	\includegraphics[scale=0.55]{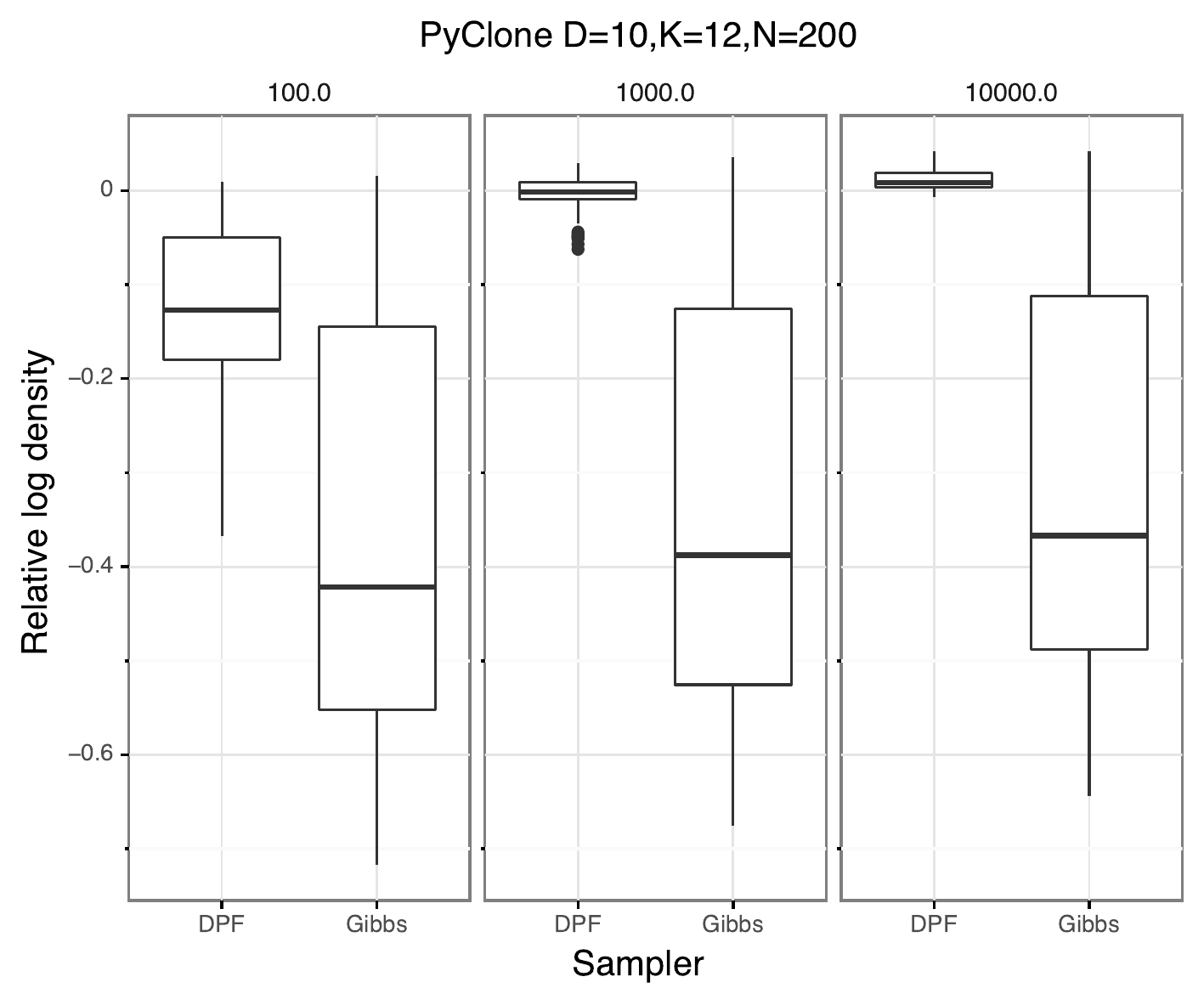}
	\includegraphics[scale=0.55]{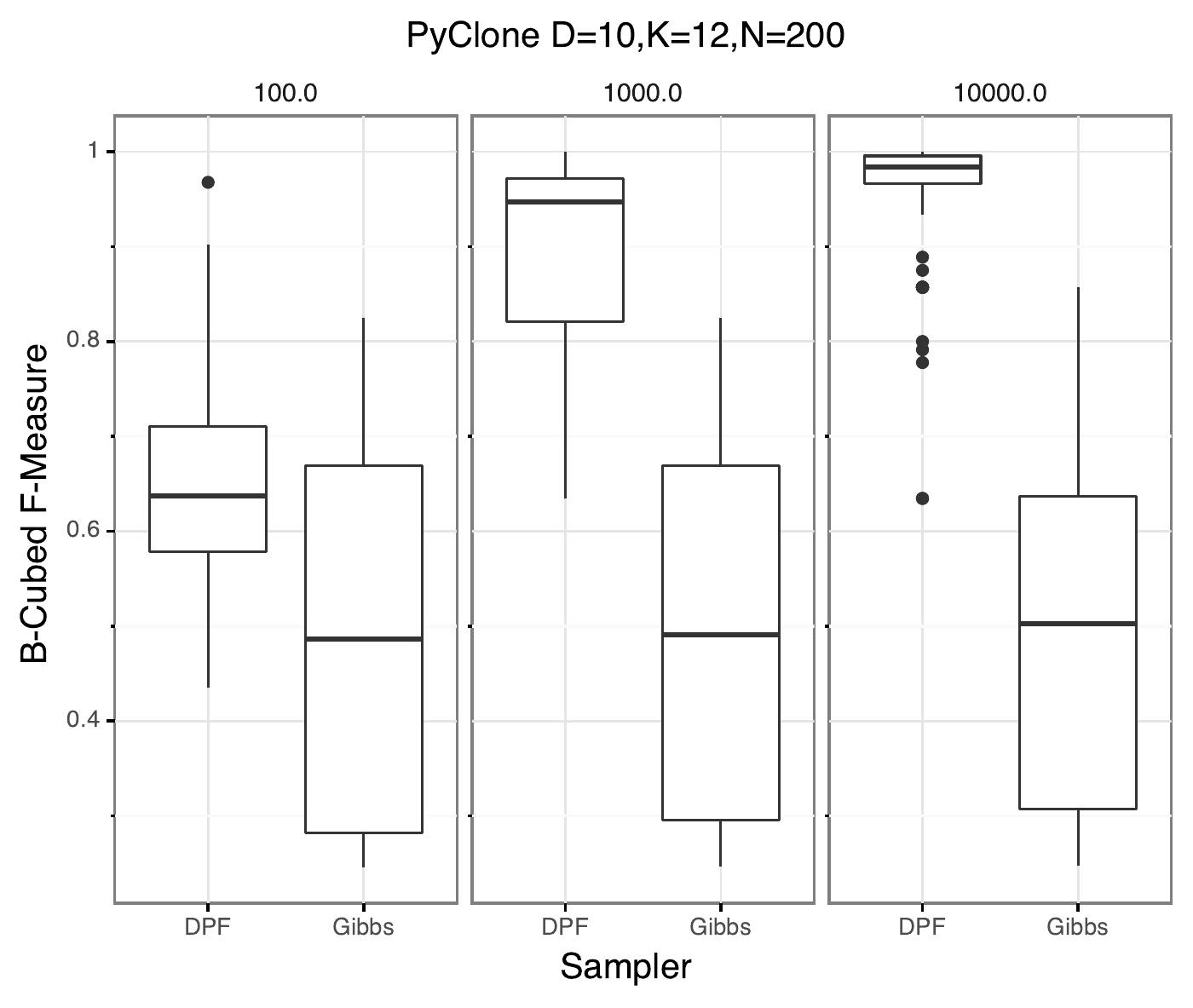}
	\caption{
		Performance of different samplers using synthetic data from the PyClone model.
		The box plots represent the distribution of values from 80 random starts of each parameter setting.
		We show the values of the relative log density (right) and B-Cubed F-measure (left).
	}
	\label{fig:pyclone_boxplot}
\end{figure}

The final model we tested with was a modified version of the PyClone model described in \cite{roth2014pyclone}.
The modifications are described in the supplement. 
We simulated three datasets using the FBB prior with $\alpha=2$, $a_V=b_V=1$ with $N=200$ data points.
For the first dataset we set $D=4$ and $K=4$, the second we set $D=10$ and $K=8$ and the third $D=10$ and $K=12$.
We did not fit the model using the IBP prior as we could not develop an efficient proposal for updating singleton entries.
In addition to the relative log density we computed the B-Cubed F-Measure \citep{amigo2009comparison}.
The B-Cubed metric is a measure of feature allocation accuracy analogous to the V-Measure metric \citep{rosenberg2007v} used to evaluate clustering algorithms.
We focused on feature allocation accuracy as the features are interpretable quantities that we wish to infer in this application.

The results of the experiments are shown in \figRef{fig:pyclone_boxplot} and \figRangeRef{fig:pyclone_finite_4_trace}{fig:pyclone_finite_12_trace}.
Note that we exclude the PG method from these figures as the performance was so poor as to obscure the scales of the plots.
For the first dataset the RG and DPF sampler both outperformed the other approaches (\statTabsRef{pyclone_finite_4}).
The PG approach performed significantly worse than all other approaches including the Gibbs sampler.
The two other datasets presented similar trends, with the DPF sampler outperforming both approaches and the PG sampler performing the worst (\tabRangeRef{tab:friedman_pyclone_finite_8}{tab:nemenyi_pyclone_finite_12}).
The performance of the Gibbs sampler did not improve from times 1000 to 10000.
Both the PG and DPF samplers show improved performance as sampling is run for longer.
This suggests that the Gibbs sampler is potentially trapped in the vicinity of a local optima, which it cannot escape from.

%% file: discussion.tex
\section{Discussion}

In this work we have developed several methods for updating an entire row of a feature allocation matrix.
Our results suggest that such samplers can significantly improve performance compared to the widely used single entry Gibbs sampler.
Directly implementing row wise Gibbs updates is intractable for more than a small number of features due to the exponential number of feature allocations.
We overcome this limitation by using the PG methodology to develop an algorithm which scales linearly in the number of features.
When coupled with the DPF framework we obtain significantly better performance than the standard Gibbs sampler.
The use of the DPF framework appears to be critical, as the standard PG sampler did not always perform well.
In particular, the performance of the PG sampler when applied to the PyClone model was significantly worse than the standard Gibbs sampler.
However, the DPF approach significantly outperformed both the Gibbs and PG methods.
Furthermore, when applied to models such as the LFRM where the standard Gibbs sampler performs well, our approach does not perform significantly worse.
This suggests that despite the increased computational complexity of the PG framework, there is little downside to employing this approach.
Taken together our results suggest the DPF algorithm is a computationally efficient and generally applicable approach for performing Bayesian inference for feature allocation models.
Our algorithm is applicable to both the parametric FBB and non-parametric IBP model.

We have focused on developing row wise updates for the feature allocation matrix.
When applied to the parametric FBB model these updates can significantly improve performance.
However, when applied to the non-parametric models using the IBP prior we did not see consistent improvement.
We believe a major problem in the non-parametric regime is the updates for the singleton features.
The most general approach of using MH updates with proposals from the priors seems to lead to very slow mixing.
While this is an issue for the Gibbs sampler as well, the low computational cost of updating non-singleton entries allows this sampler to perform more singleton updates.
We believe that the development of efficient schemes to update the columns in a single move will be particularly useful.
This has already been explored to some extent in \cite{fox2014joint}, where split-merge moves are used as proposals for an MH update.
It should be possible to further improve upon these split-merge style moves using the PG framework, in a similar way to what has been done in the Bayesian clustering literature \citep{bouchard2017particle}.
Such updates would complement the approach we have developed in this work.

We have not exploited the potential for performing parallel computation that is offered by the PG framework.
In particular we could parallelize any loops over particles in Algorithm~\ref{alg:PG} which could potentially yield significant speed-ups.
It has been noted by \cite{whiteley2010efficient} and \cite{lindsten2014particle} that the use of backward or ancestor sampling can significantly reduce the effect of path degeneracy for SMC models.
These approaches could naturally be combined with our method, and could allow for the use of fewer particles.

%% file: appendix.tex
\renewcommand\thefigure{S\arabic{figure}}
\renewcommand\thetable{S\arabic{table}}

\section{Appendix}

\input{models}

\subsection{Supplementary figures}

\begin{figure}
	\includegraphics[scale=1.0]{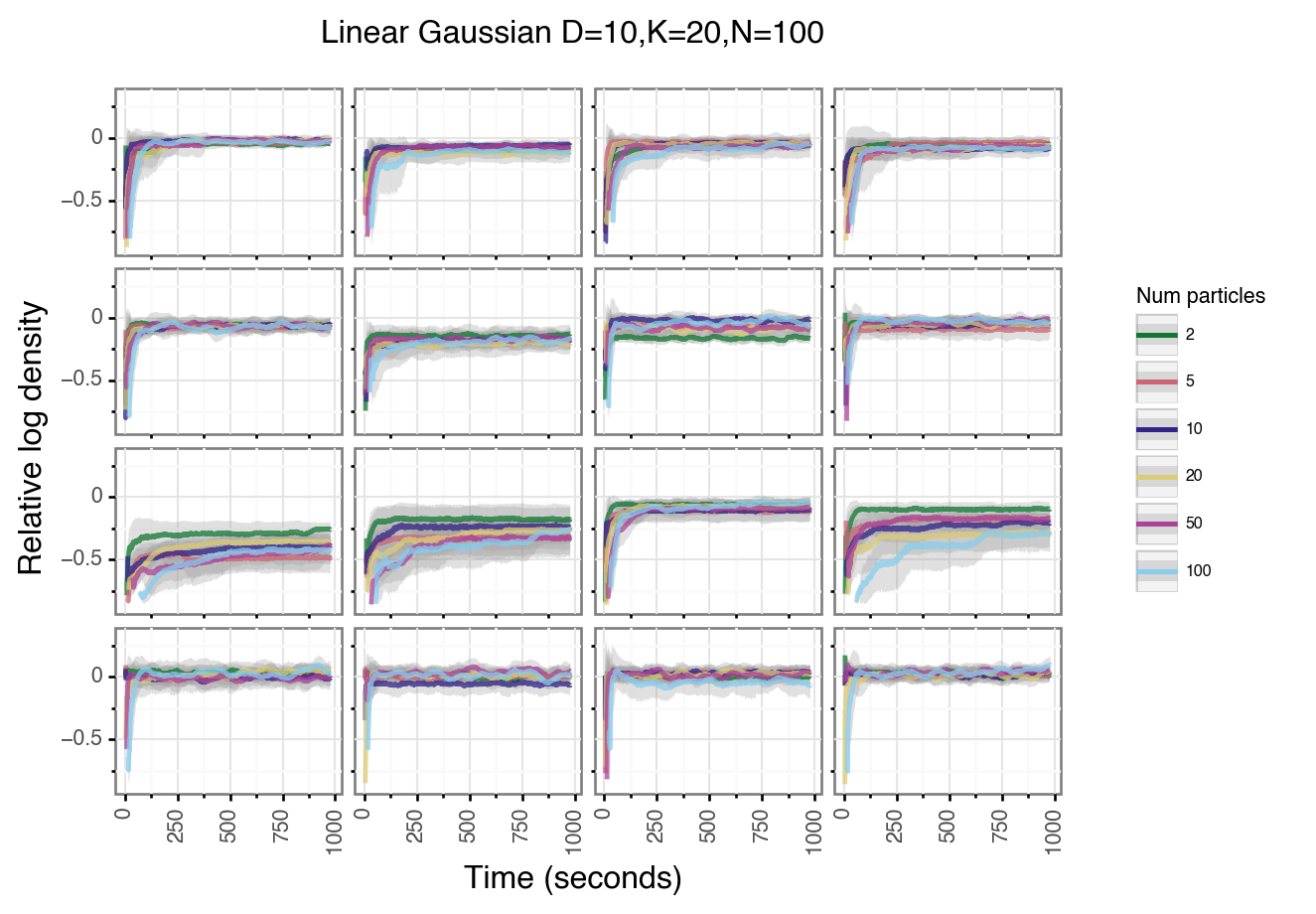}
	\caption{\traceTuningCaption{PG}{number of particles}}
	\label{fig:pg_num_particles_trace}
\end{figure}

\begin{figure}
	\includegraphics[scale=1.0]{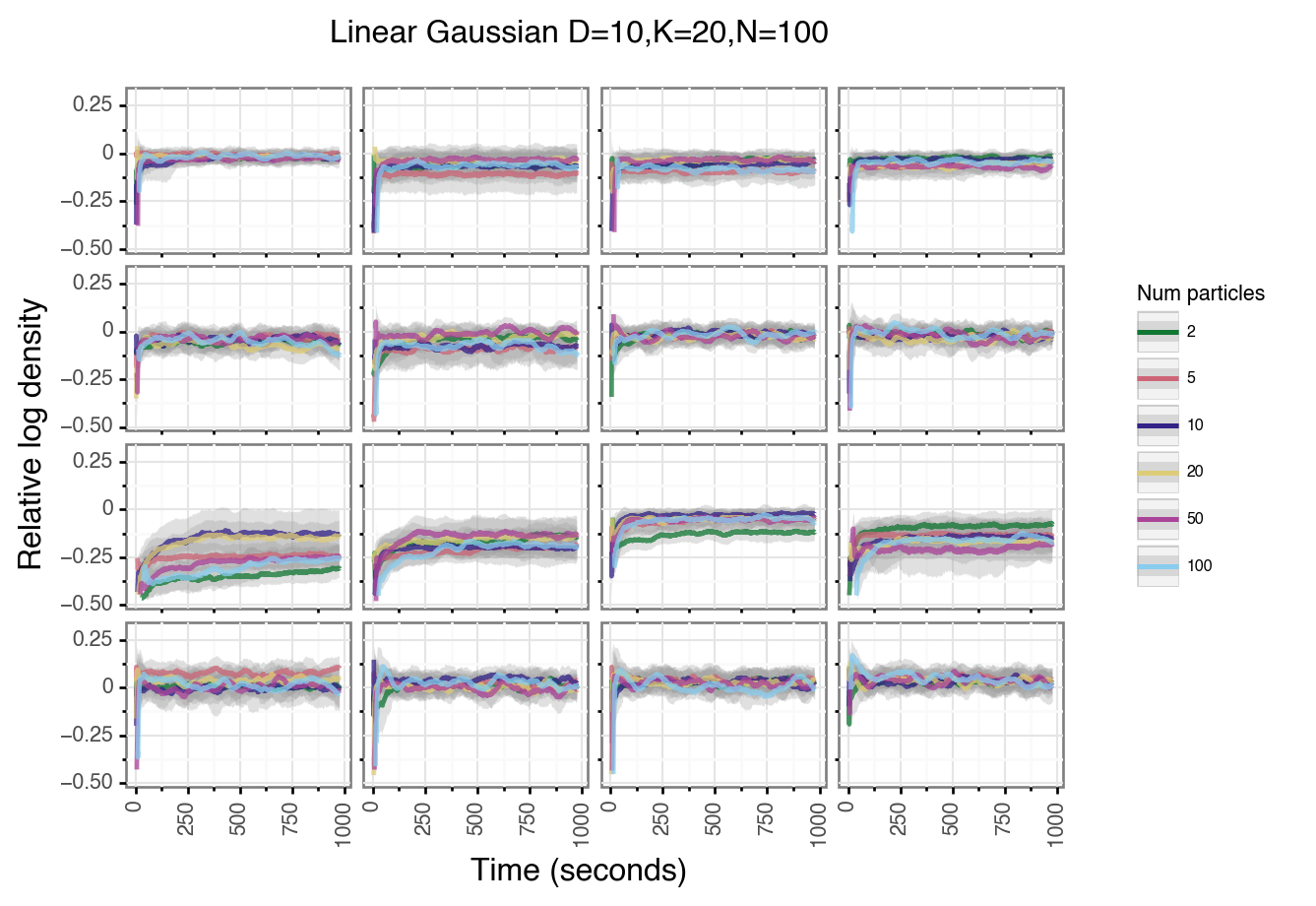}
	\caption{\traceTuningCaption{DPF}{number of particles}}
	\label{fig:dpf_num_particles_trace}
\end{figure}

\begin{figure}[h]
	\includegraphics[scale=0.6]{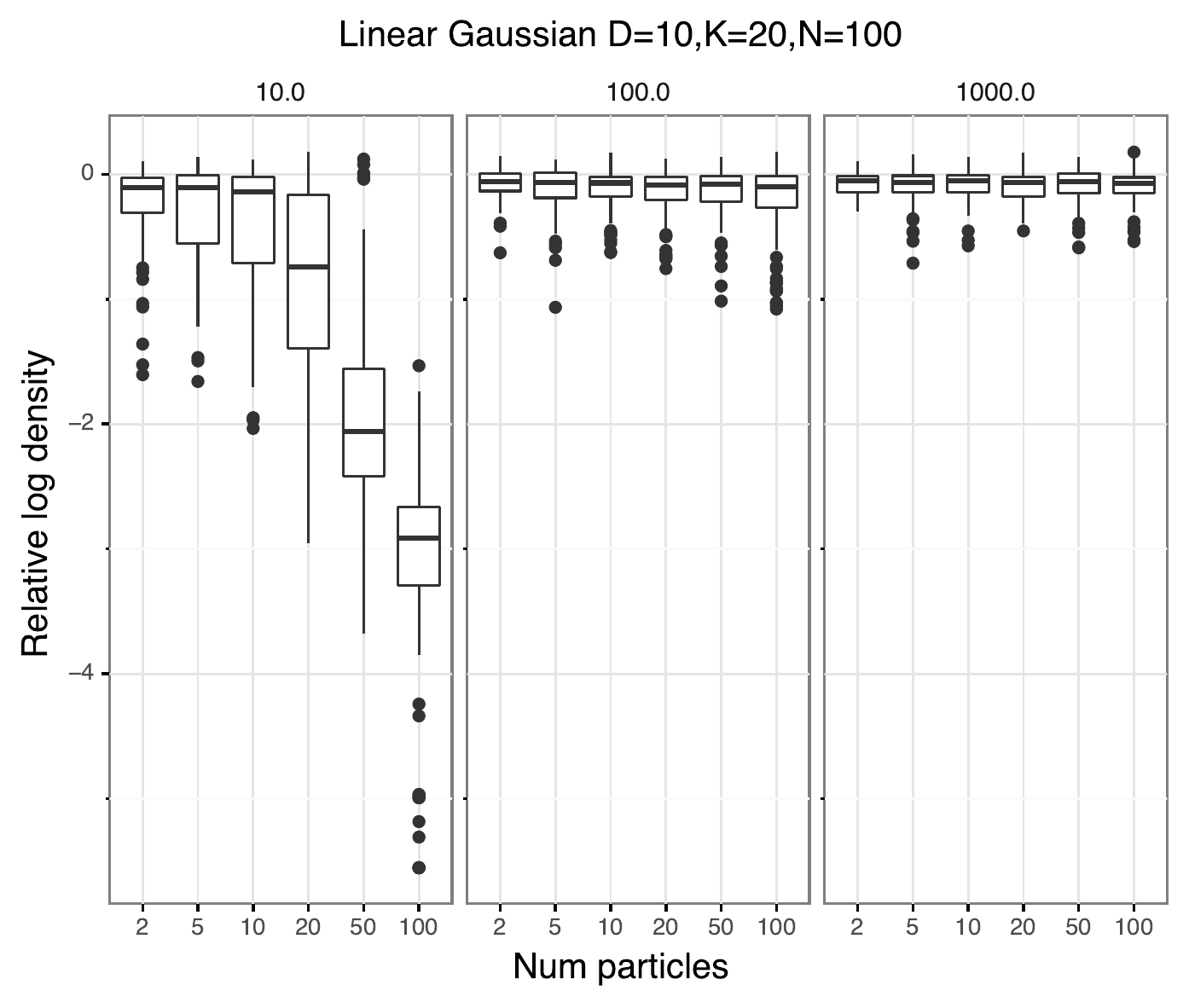}
	\includegraphics[scale=0.6]{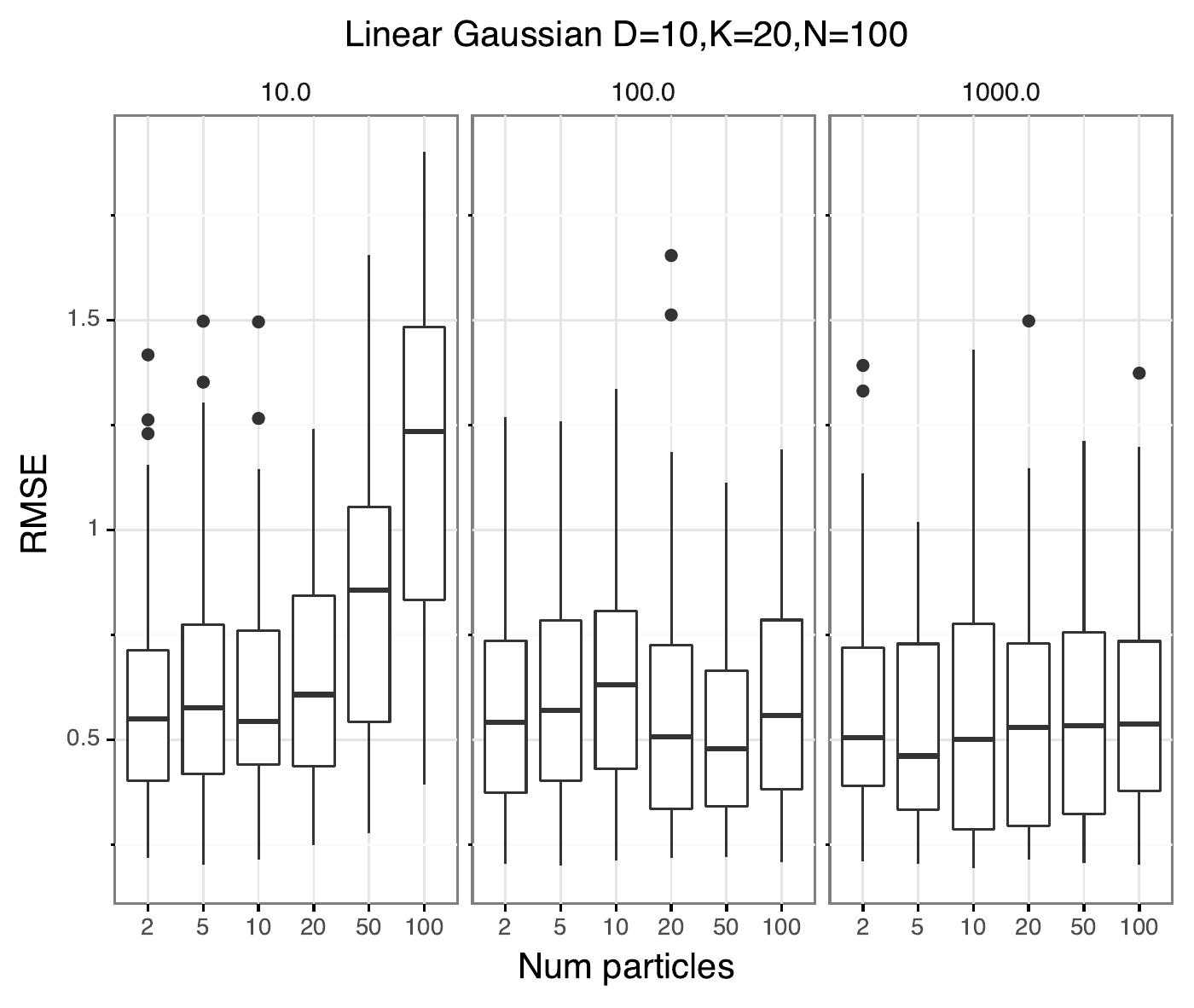}
	\vfill
	\includegraphics[scale=0.6]{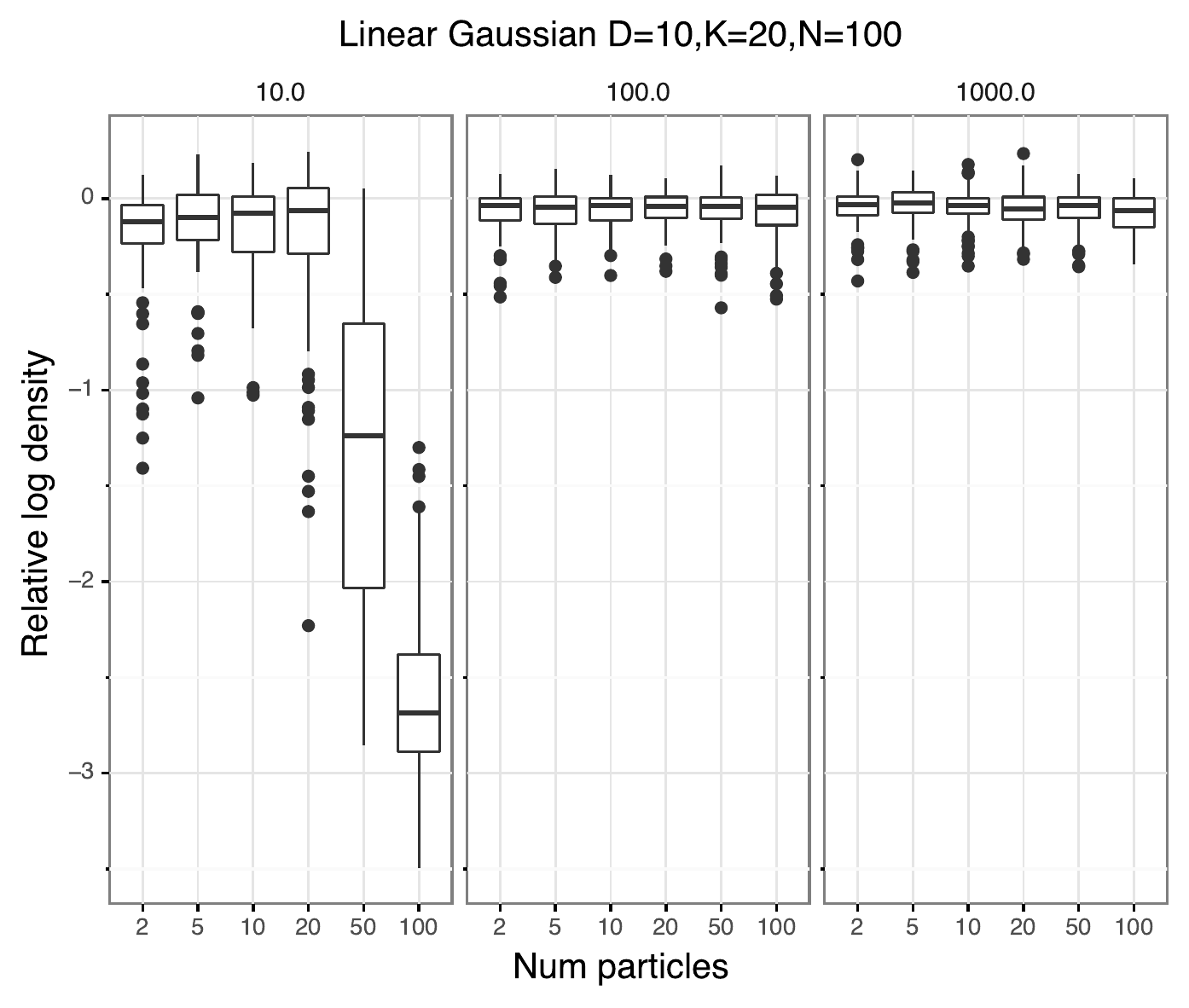}
	\includegraphics[scale=0.6]{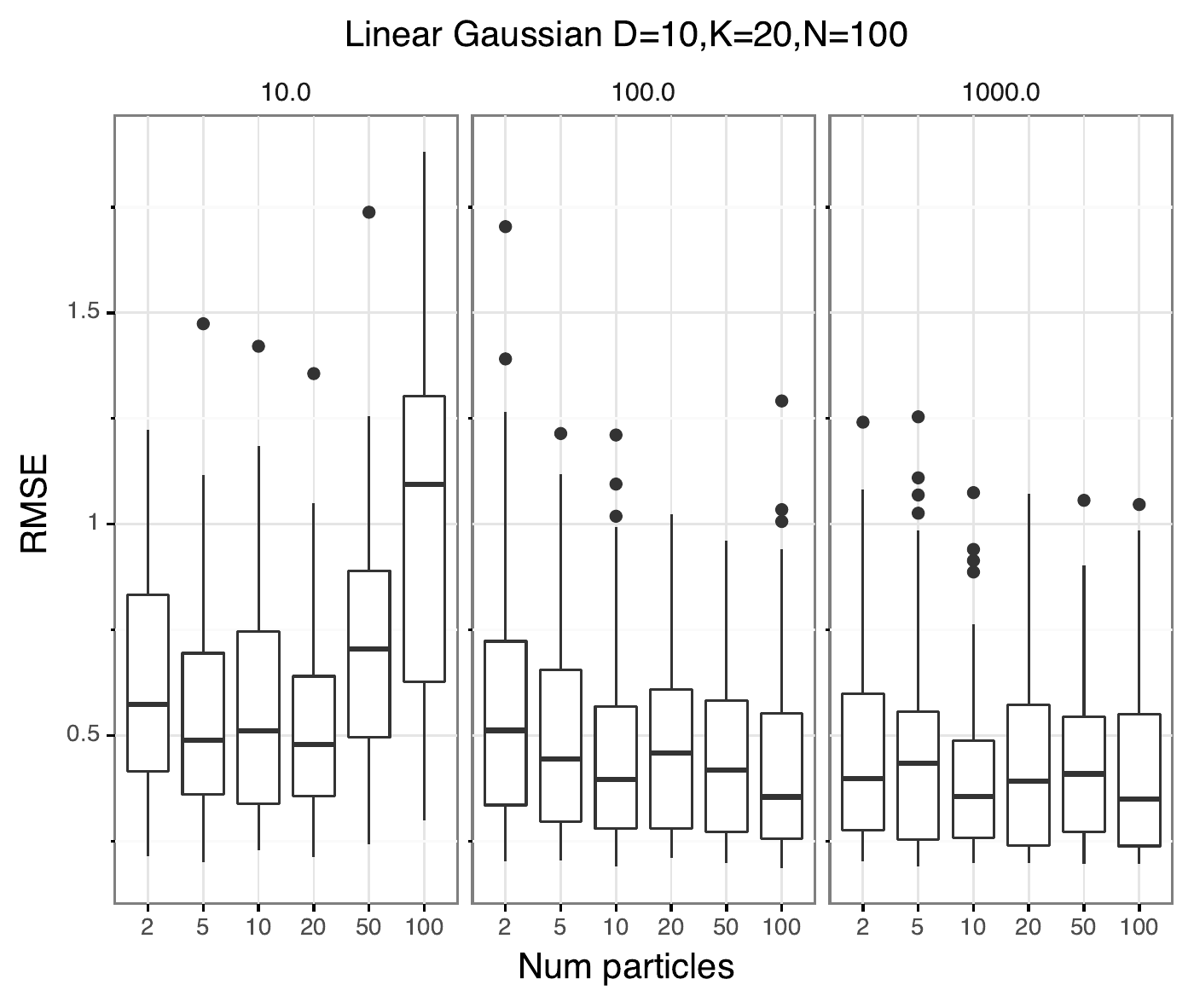}
	\caption{
		Performance of the PG (top) and DPF (bottom) samplers as function of the number of particles.
		The box plots represent the distribution of values from 80 random starts of each parameter setting.
		We show the values of the relative log density (right) and root mean square error reconstruction of missing values (left).
	}
	\label{fig:num_particles_boxplot}
\end{figure}

\begin{figure}
	\includegraphics[scale=1.0]{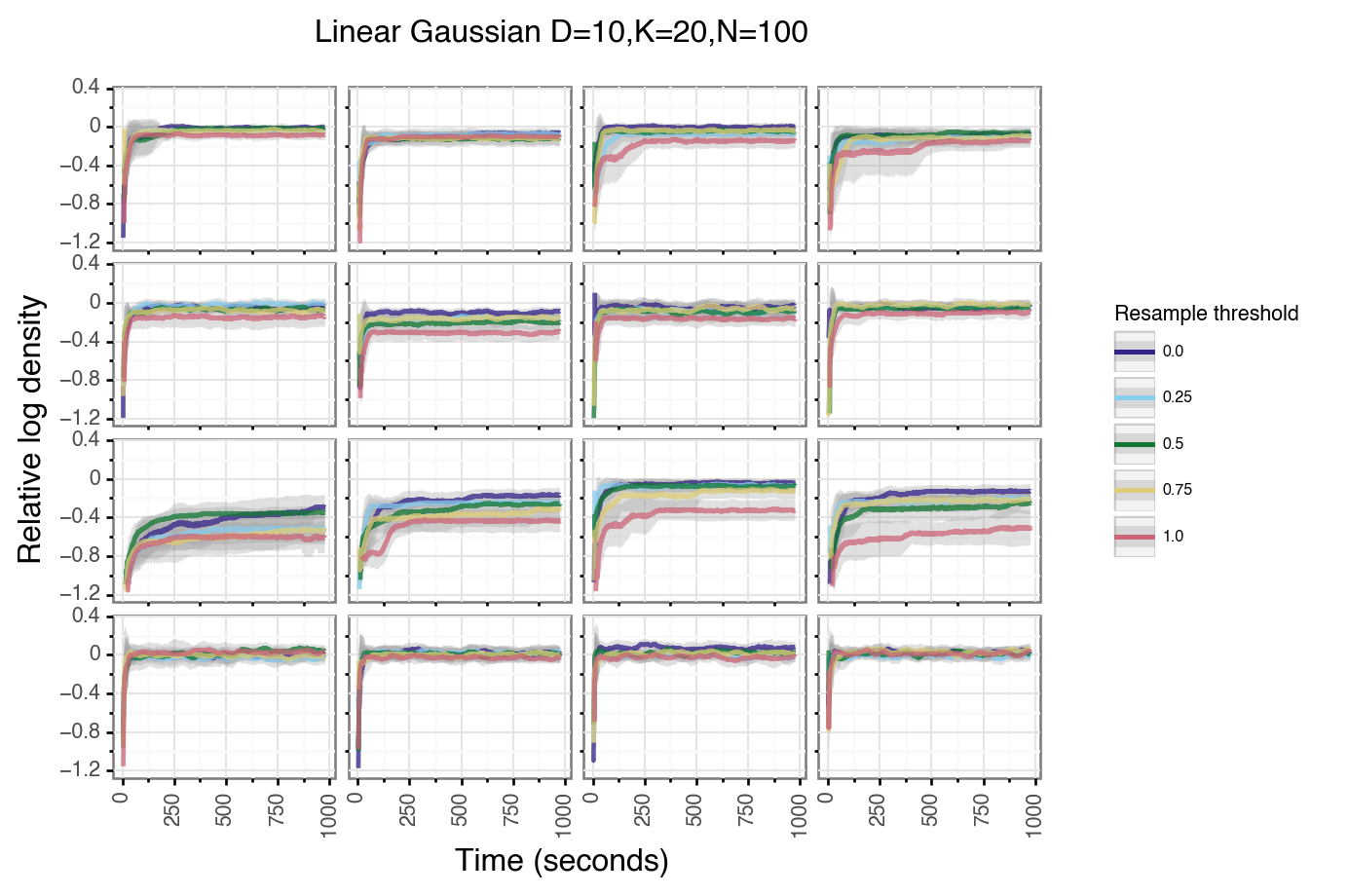}
	\caption{\traceTuningCaption{PG}{resampling thresholds}}
	\label{fig:pg_resample_threshold_trace}
\end{figure}

\begin{figure}
	\includegraphics[scale=0.6]{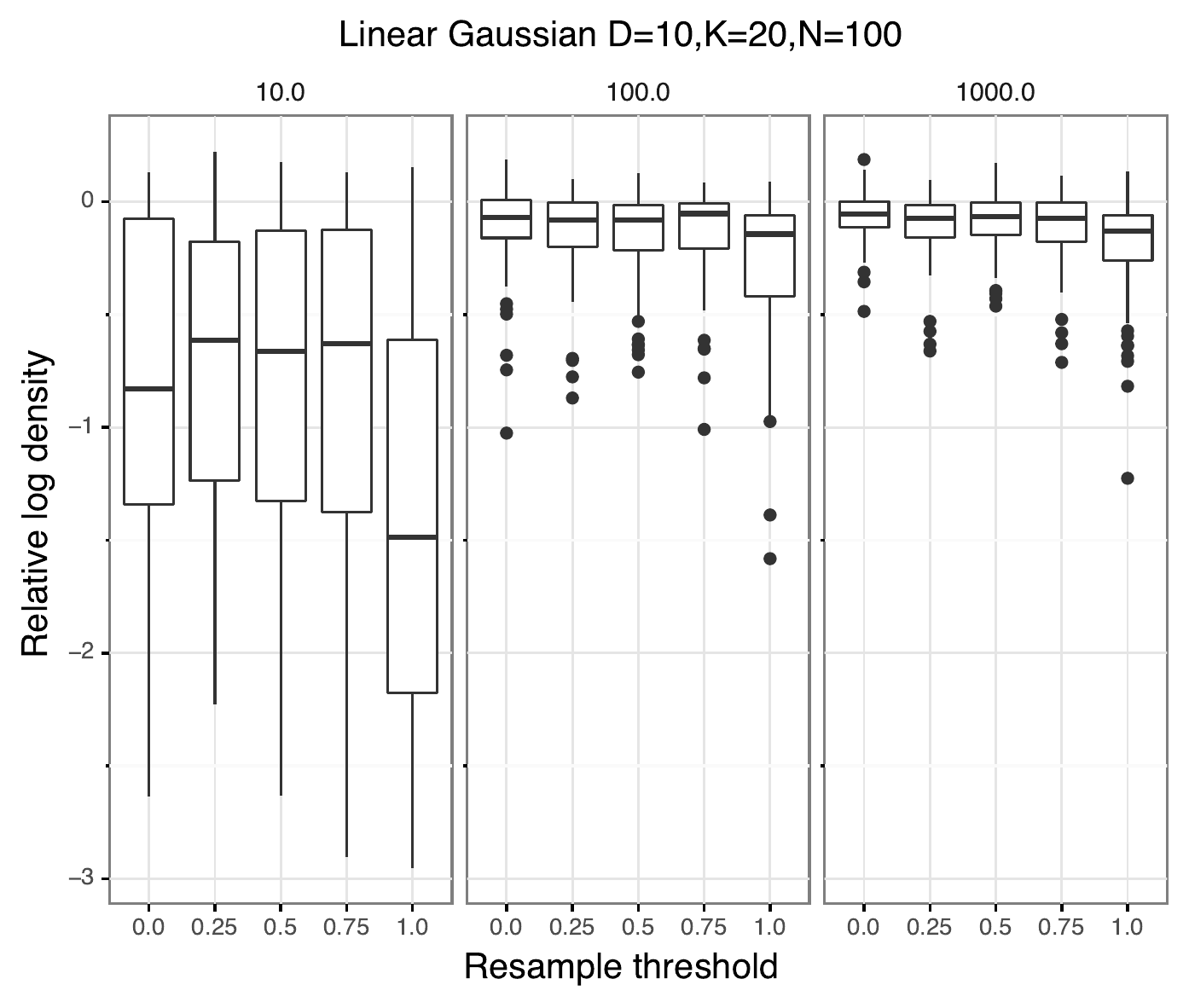}
	\includegraphics[scale=0.6]{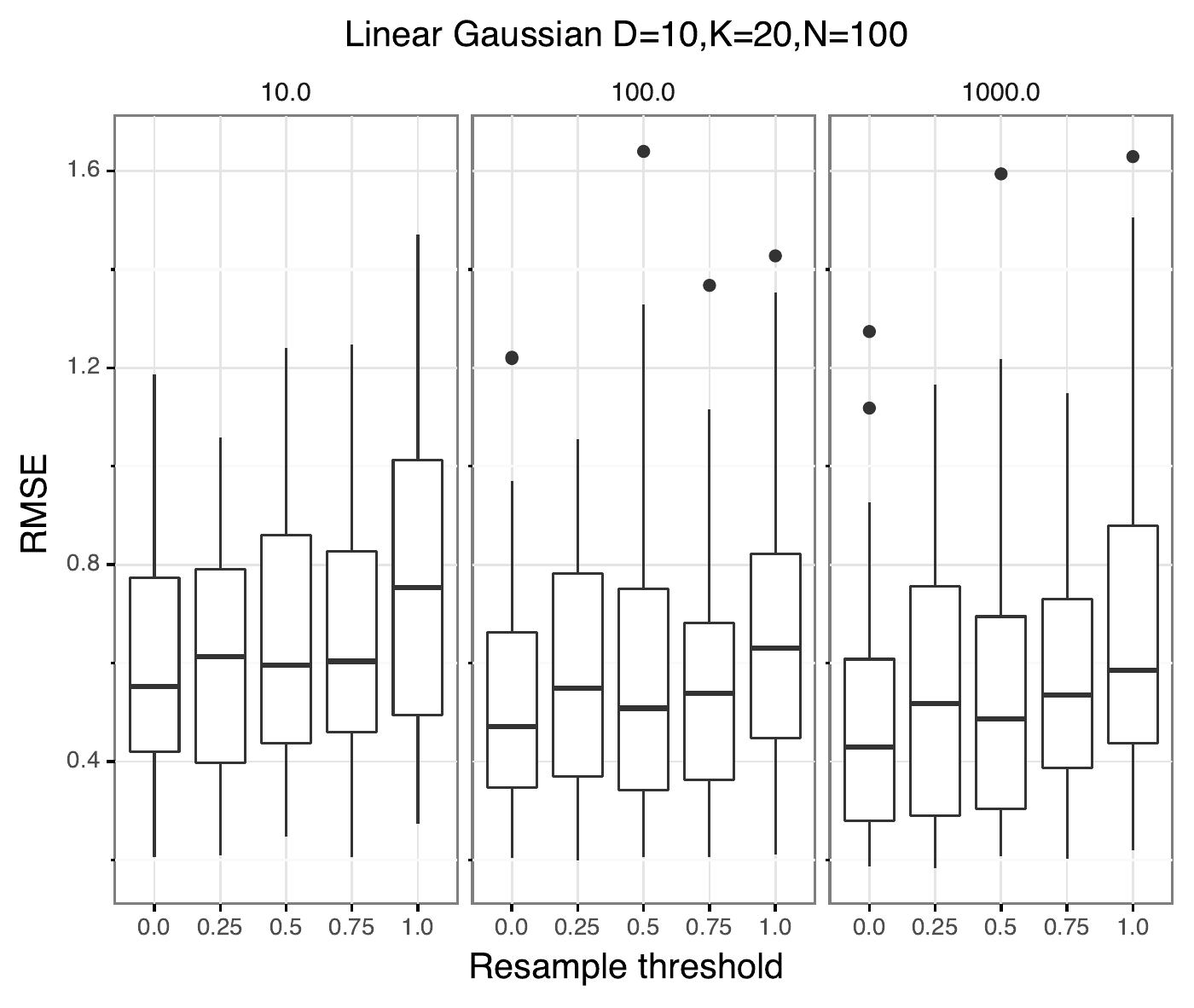}
	\caption{
		Performance of the PG samplers as function of the resampling threshold.
		The box plots represent the distribution of values from 80 random starts of each parameter setting.
		We show the values of the relative log density (right) and root mean square error reconstruction of missing values (left).
	}
	\label{fig:pg_resample_threshold_boxplot}
\end{figure}

\begin{figure}
	\includegraphics[scale=1.0]{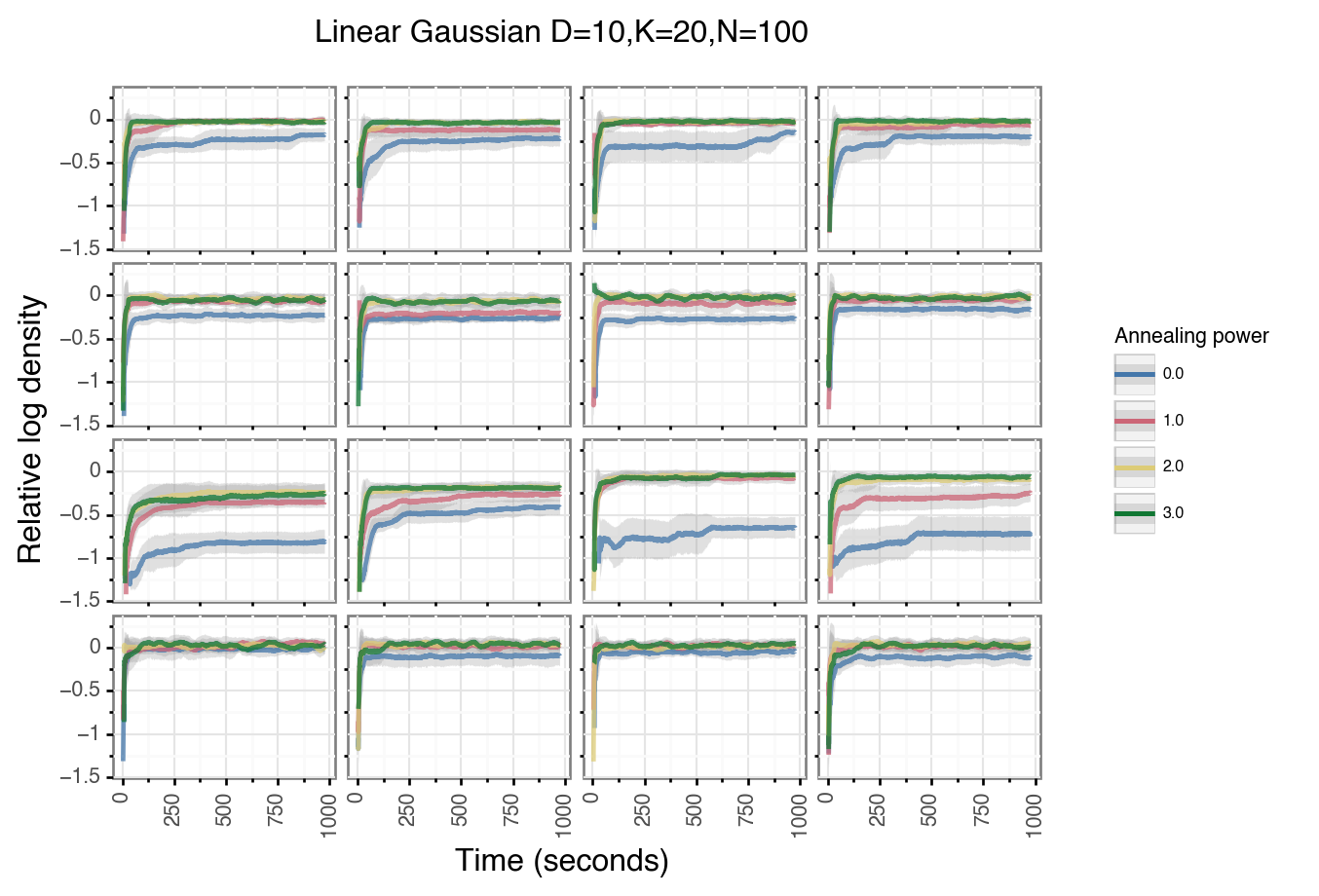}
	\caption{\traceTuningCaption{PG}{annealing powers}}
	\label{fig:pg_annealing_power_trace}
\end{figure}

\begin{figure}
	\includegraphics[scale=1.0]{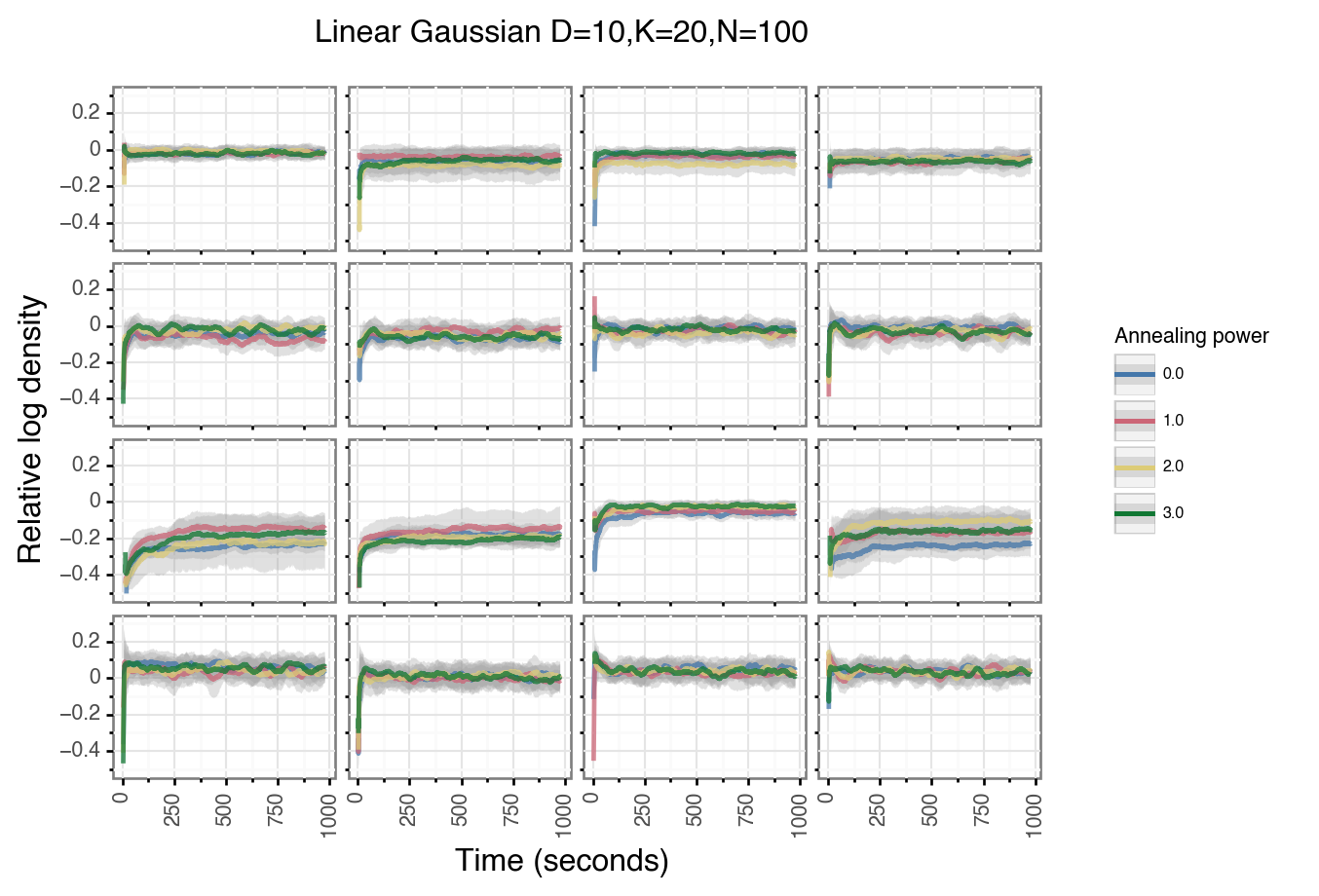}
	\caption{\traceTuningCaption{DPF}{annealing powers}}
	\label{fig:dpf_annealing_power_trace}	
\end{figure}

\begin{figure}
	\includegraphics[scale=0.6]{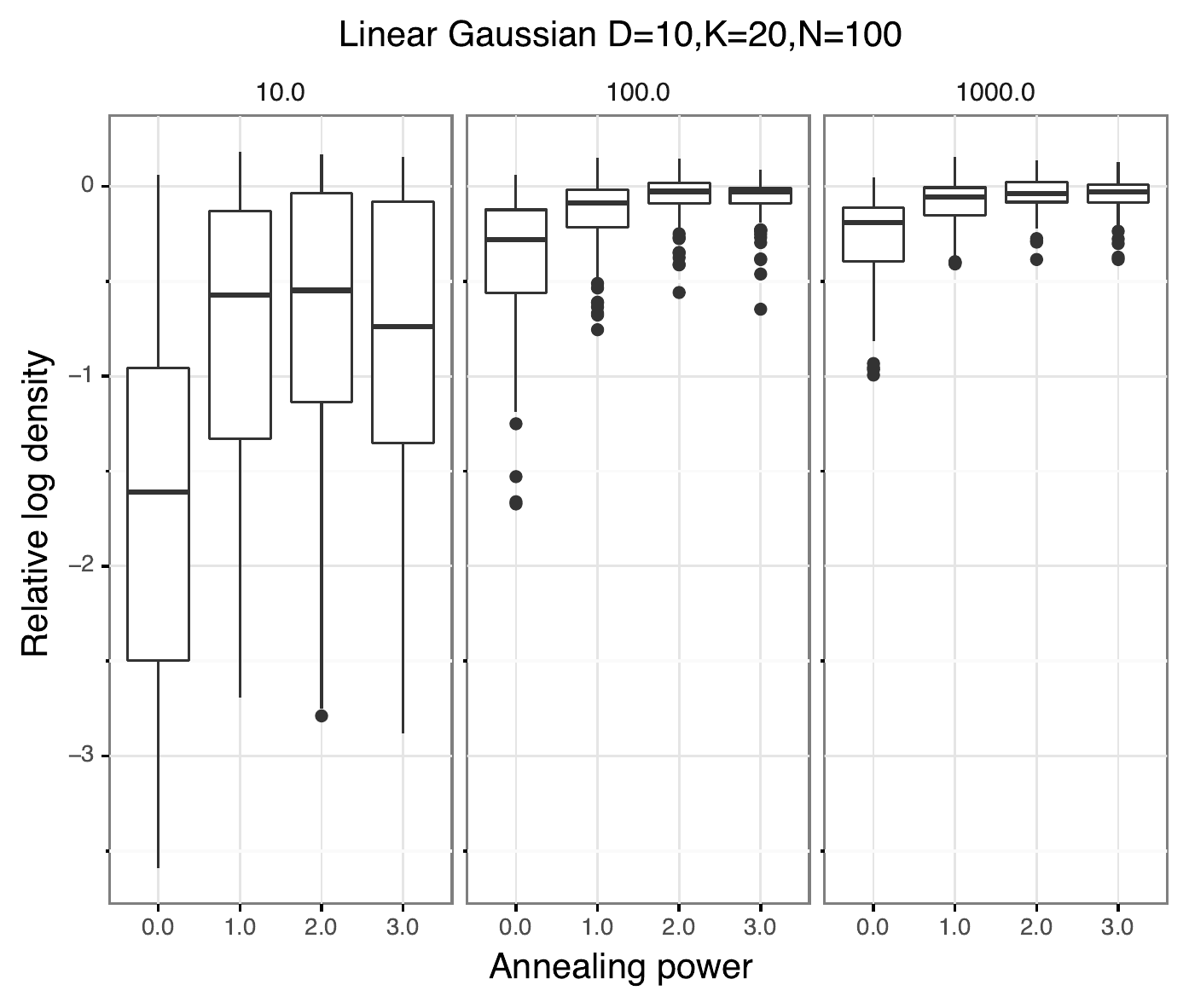}
	\includegraphics[scale=0.6]{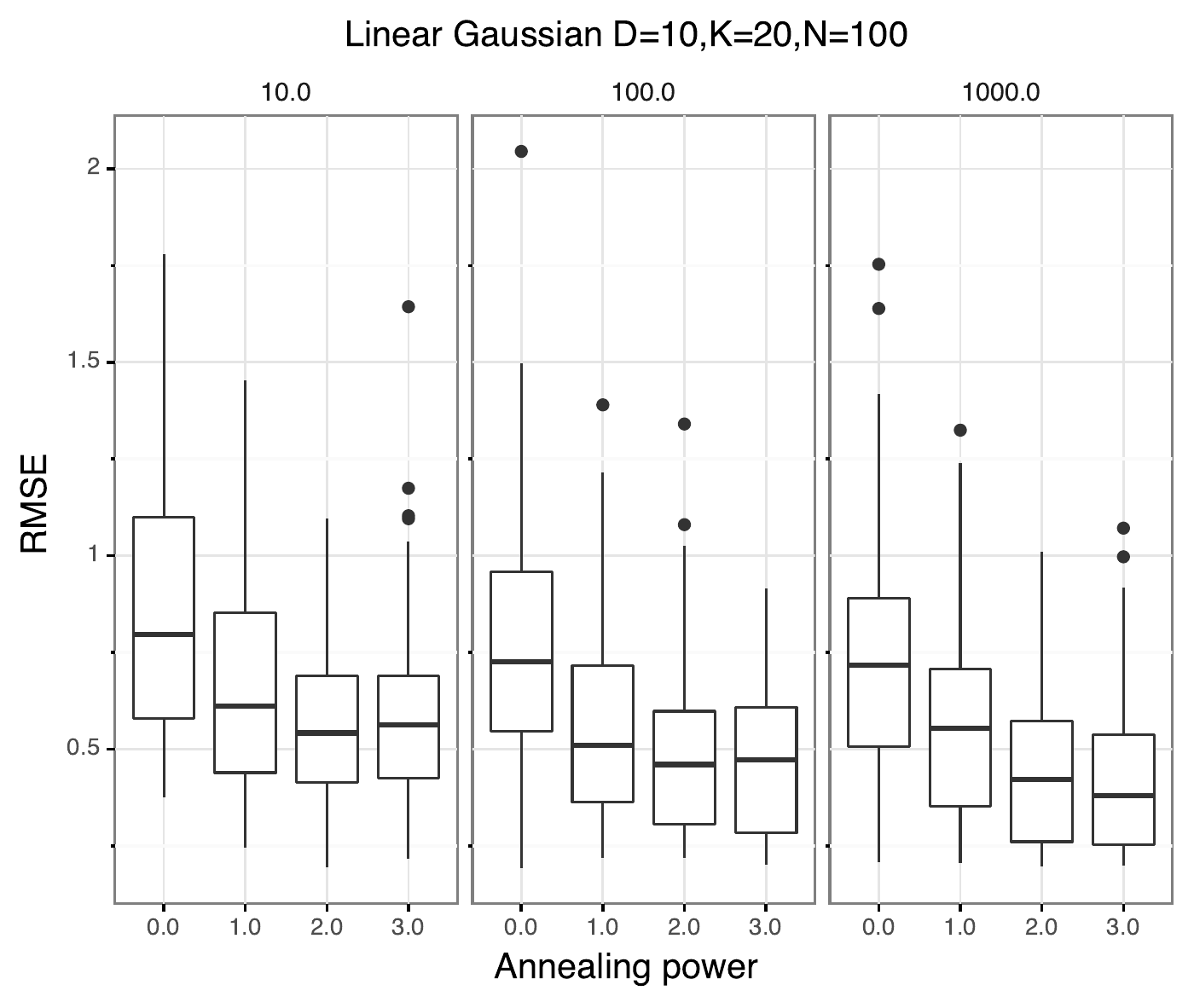}
	\vfill
	\includegraphics[scale=0.6]{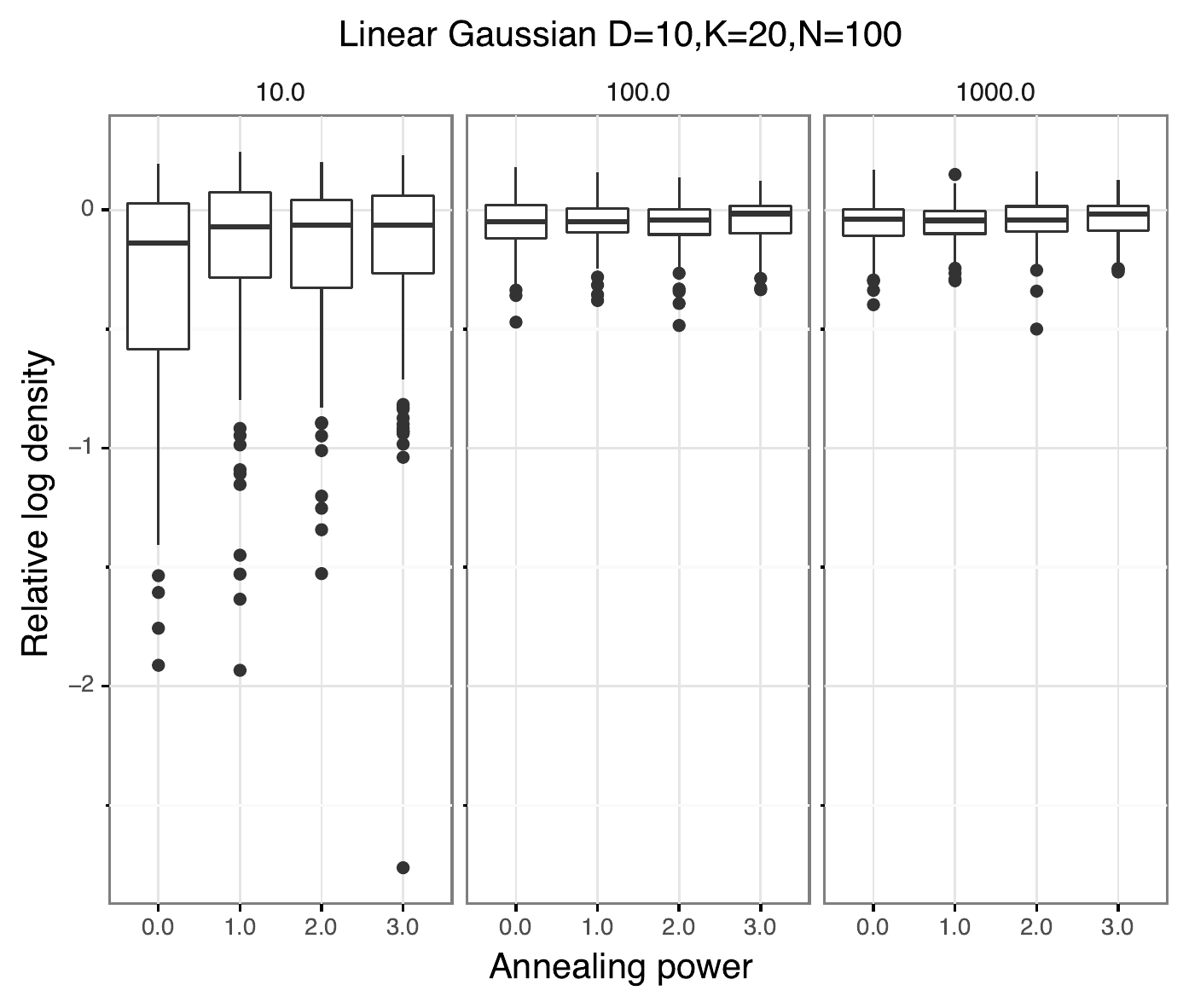}
	\includegraphics[scale=0.6]{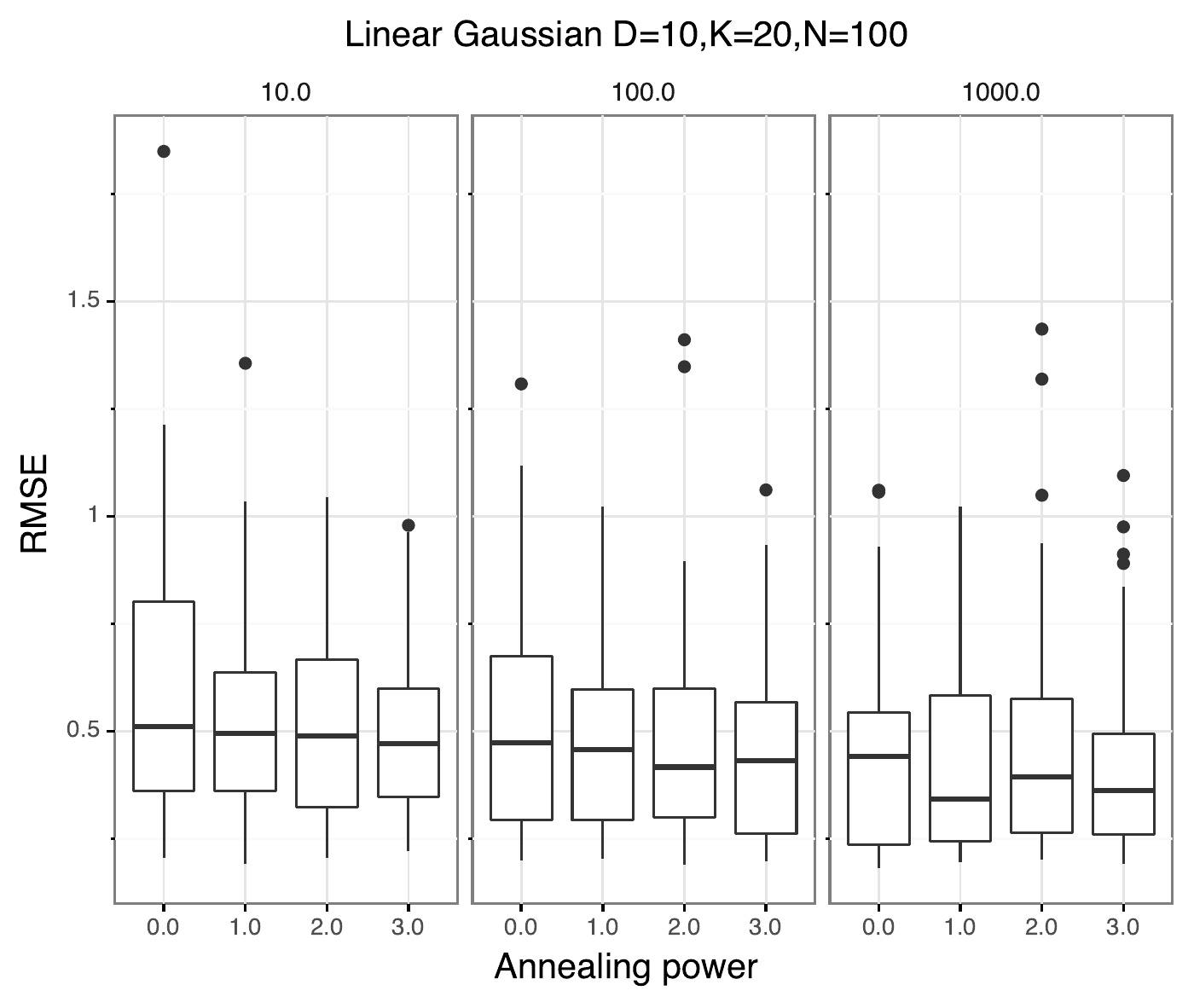}
	\caption{
		Performance of the PG (top) and DPF (bottom) samplers as function of the annealing power of the intermediate target distribution.
		The box plots represent the distribution of values from 80 random starts of each parameter setting.
		We show the values of the relative log density (right) and root mean square error reconstruction of missing values (left).
	}
	\label{fig:annealing_power_boxplot}
\end{figure}

\begin{figure}
	\includegraphics[scale=1.0]{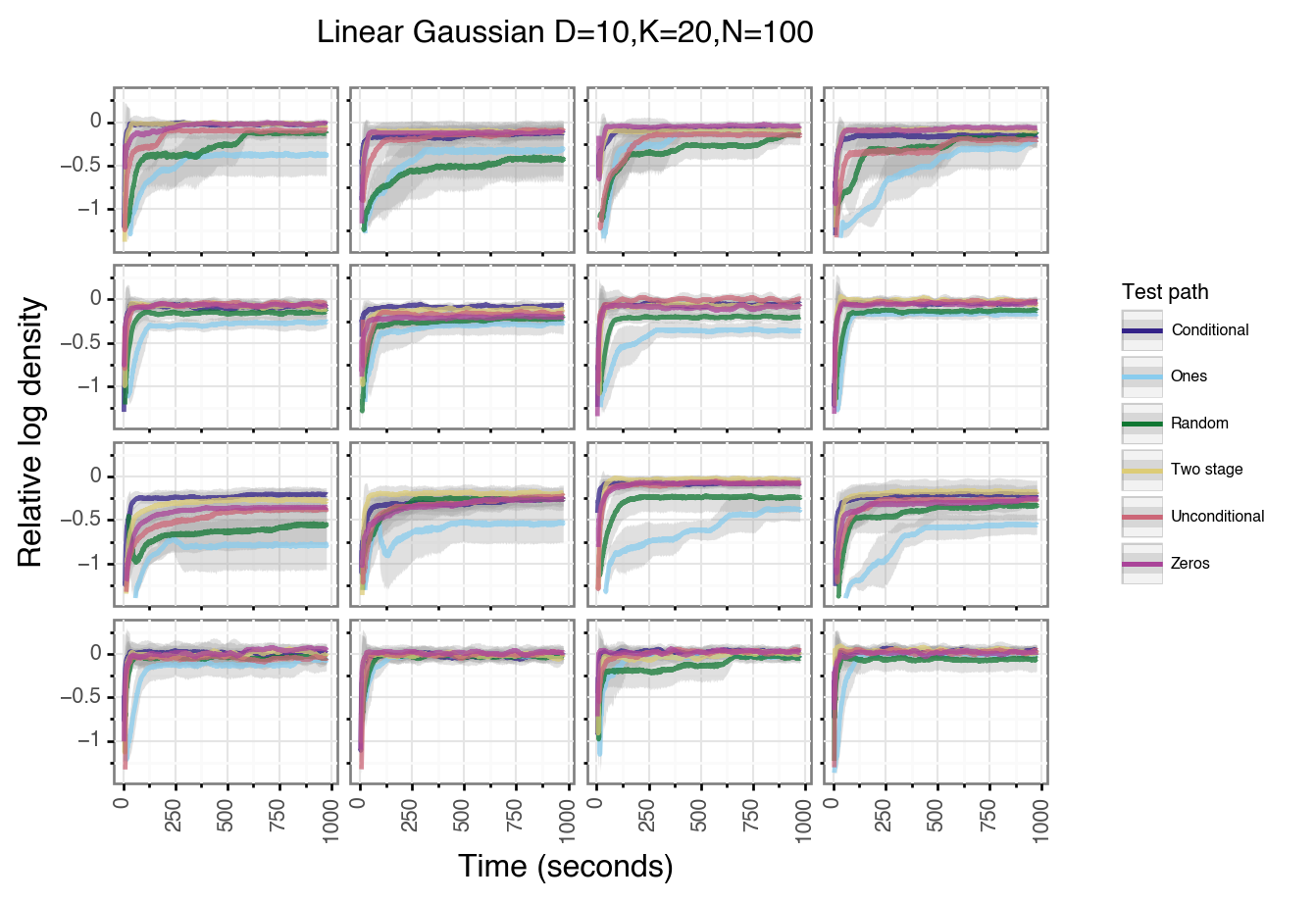}
	\caption{\traceTuningCaption{PG}{test paths}}
	\label{fig:pg_test_path_trace}
\end{figure}

\begin{figure}
	\includegraphics[scale=1.0]{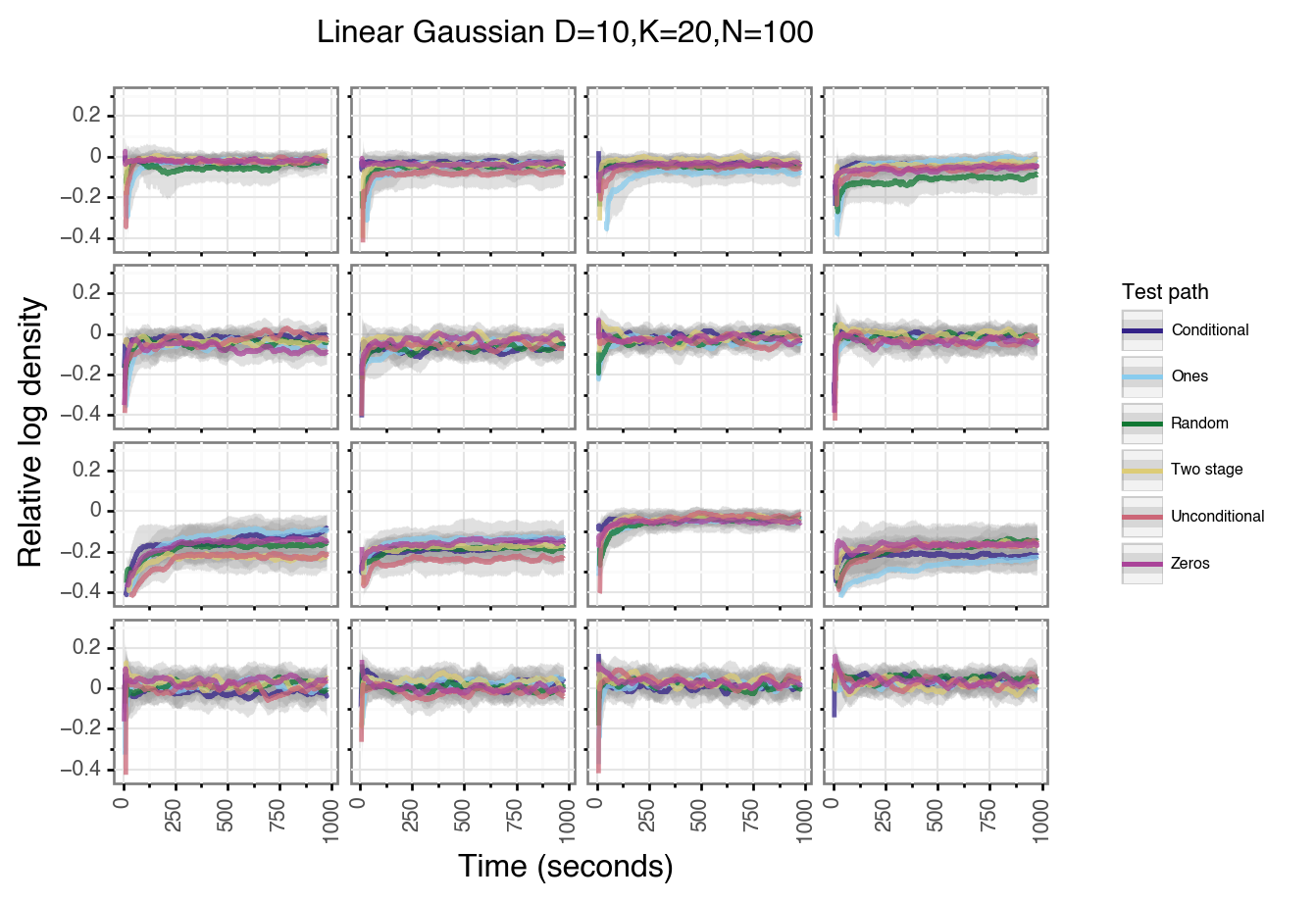}
	\caption{\traceTuningCaption{DPF}{test paths}}
	\label{fig:dpf_test_path_trace}
\end{figure}

\begin{figure}
	\includegraphics[scale=0.6]{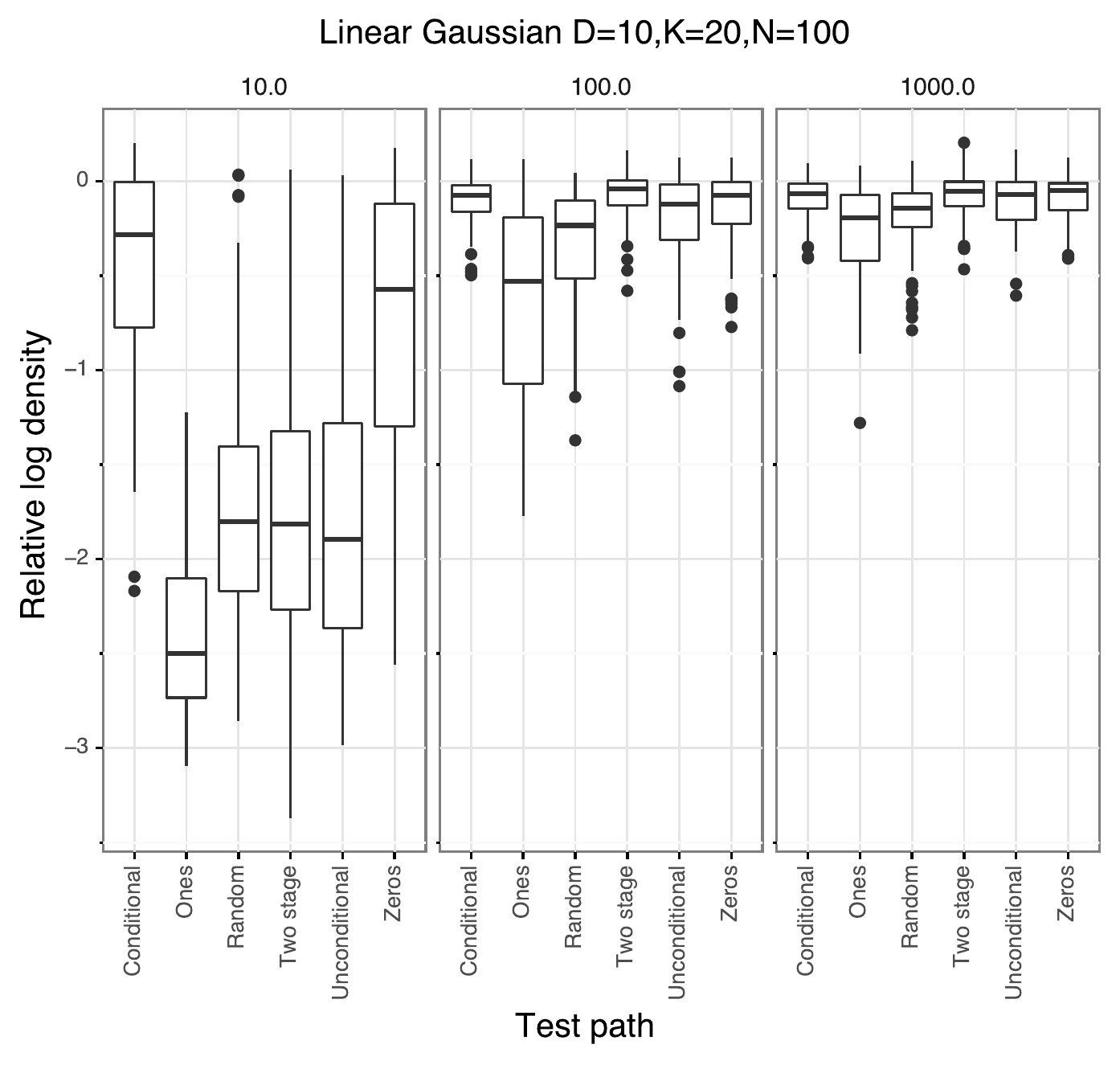}
	\includegraphics[scale=0.6]{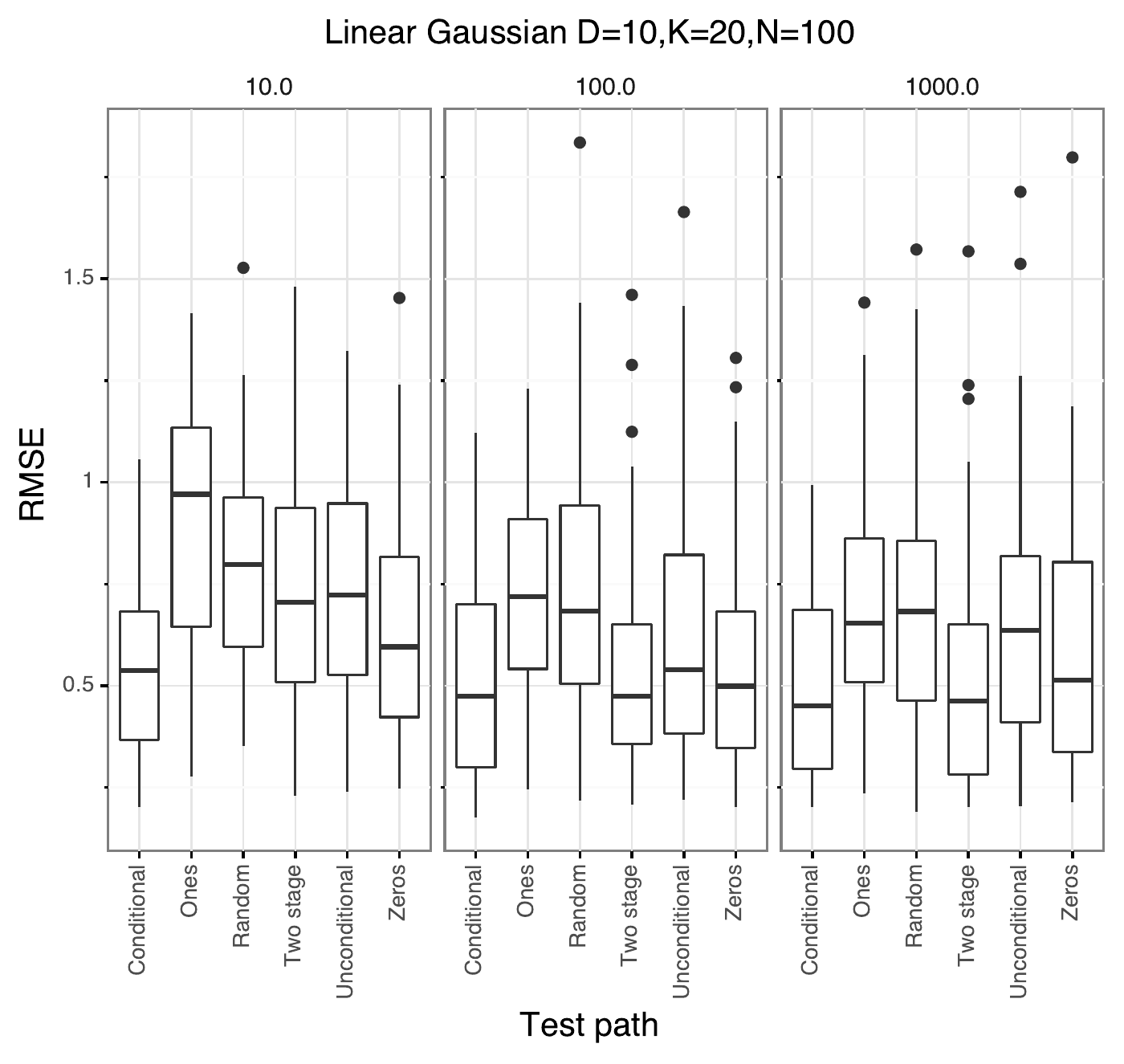}
	\vfill
	\includegraphics[scale=0.6]{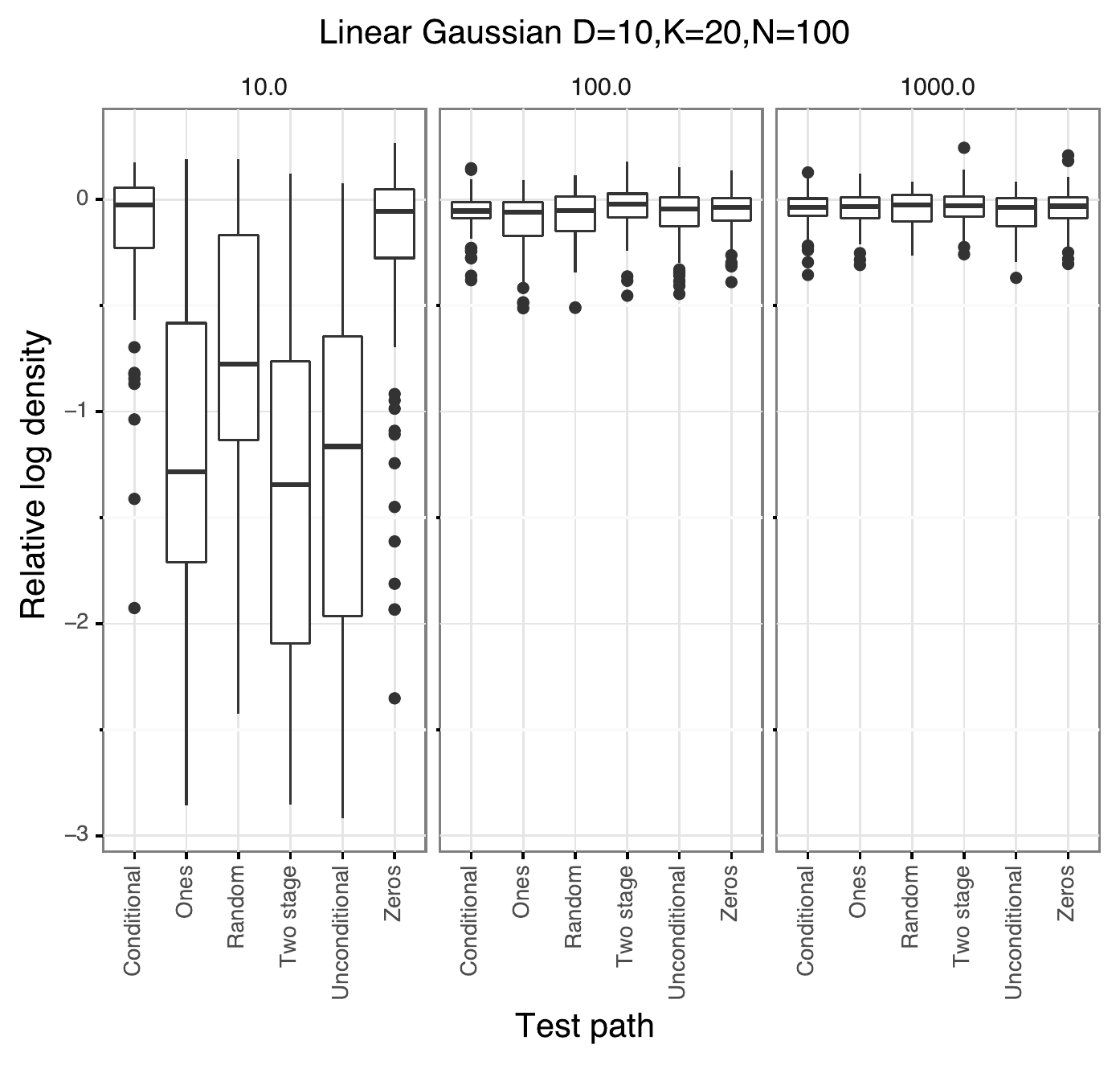}
	\includegraphics[scale=0.6]{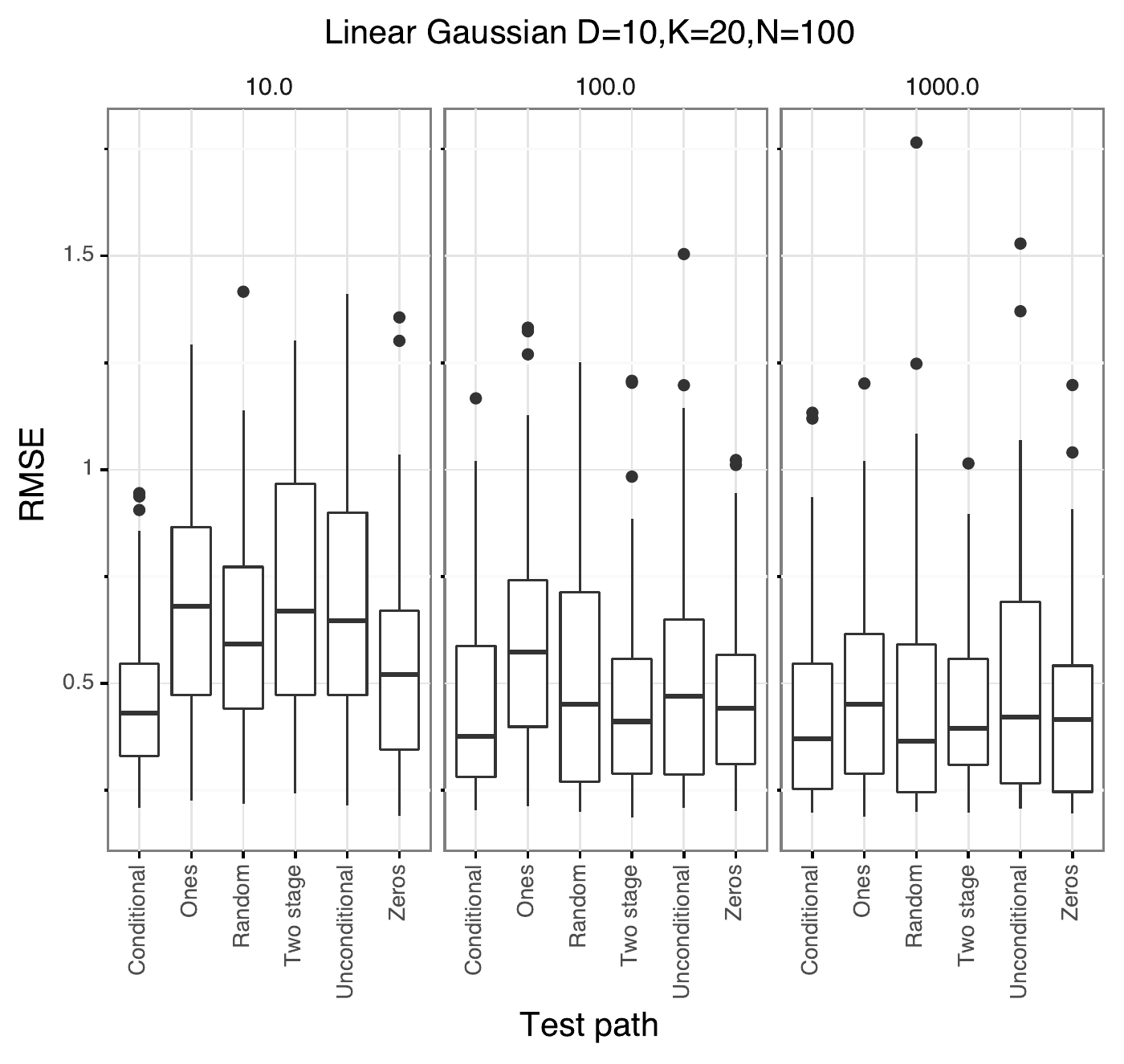}
	\caption{
		Performance of the PG (top) and DPF (bottom) samplers as function of the test path used to evaluate the data likelihood.
		The box plots represent the distribution of values from 80 random starts of each parameter setting.
		We show the values of the relative log density (right) and root mean square error reconstruction of missing values (left).
	}
	\label{fig:test_path_boxplot}
\end{figure}

\begin{figure}
	\includegraphics[scale=1.0]{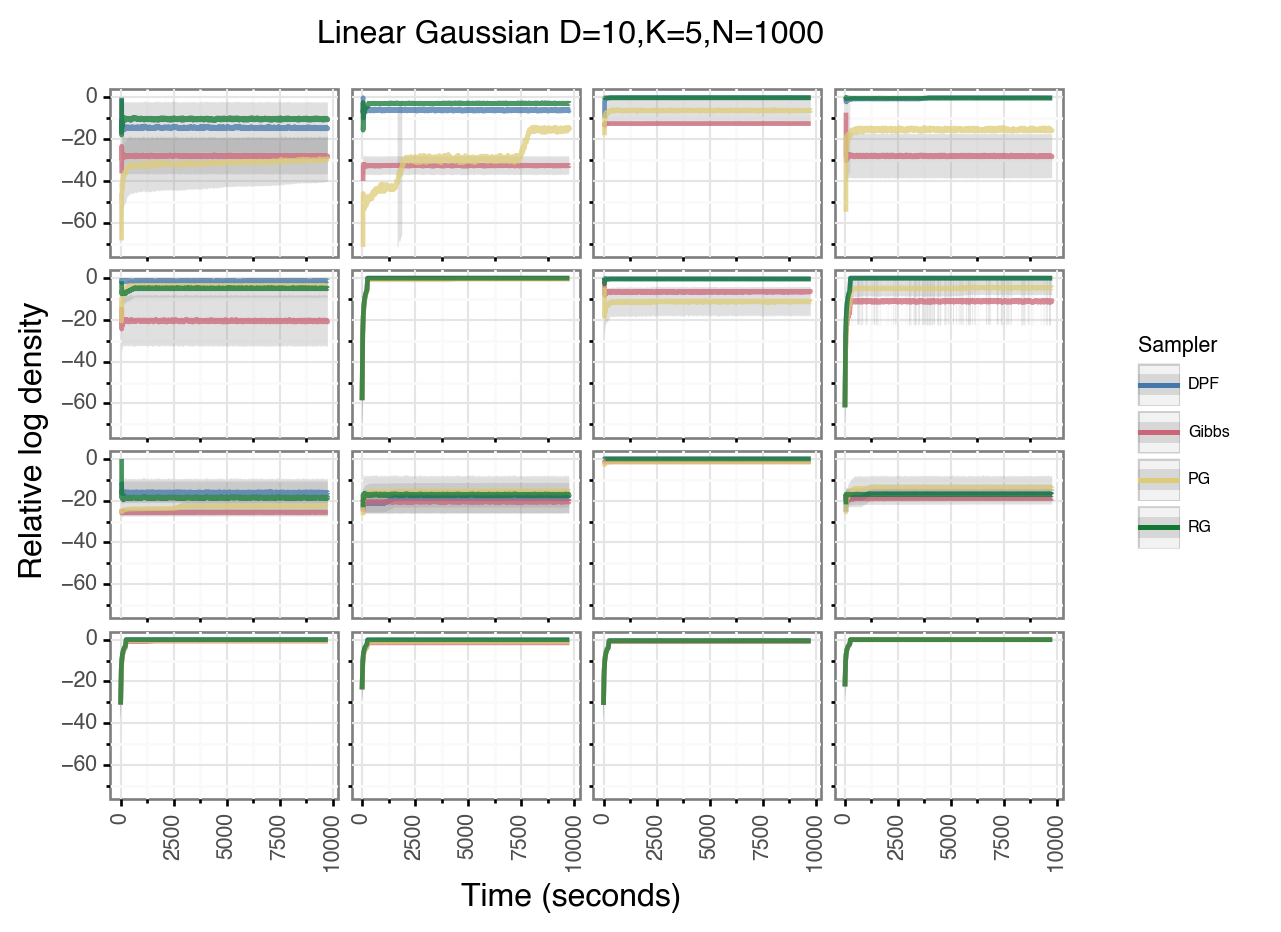}
	\caption{\traceComparisonCaption{linear Gaussian}{K=5}}
	\label{fig:lg_finite_5_trace}
\end{figure}

\begin{figure}
	\includegraphics[scale=1.0]{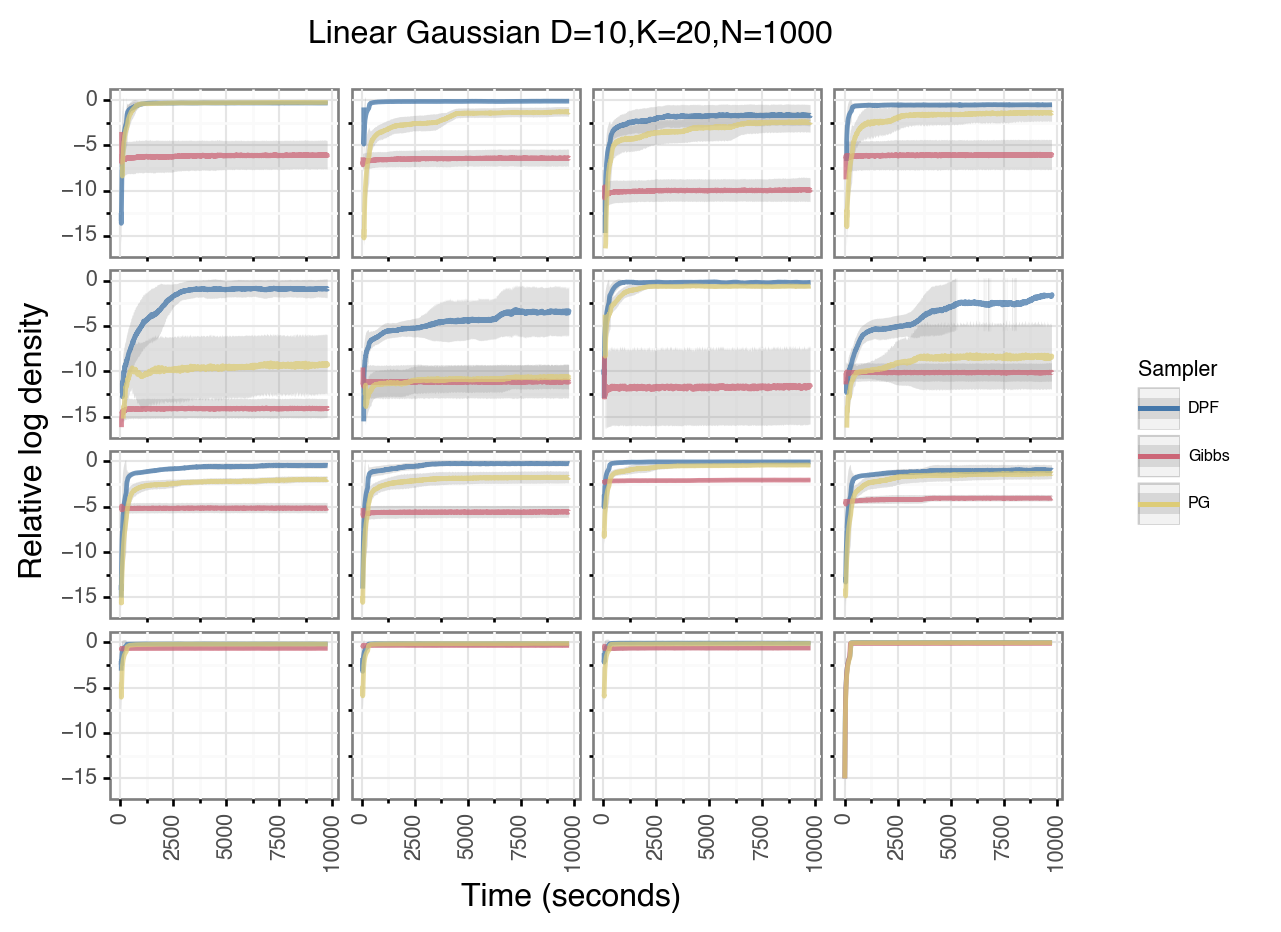}
	\caption{\traceComparisonCaption{linear Gaussian}{K=20}}
	\label{fig:lg_finite_20_trace}
\end{figure}

\begin{figure}
	\includegraphics[scale=1.0]{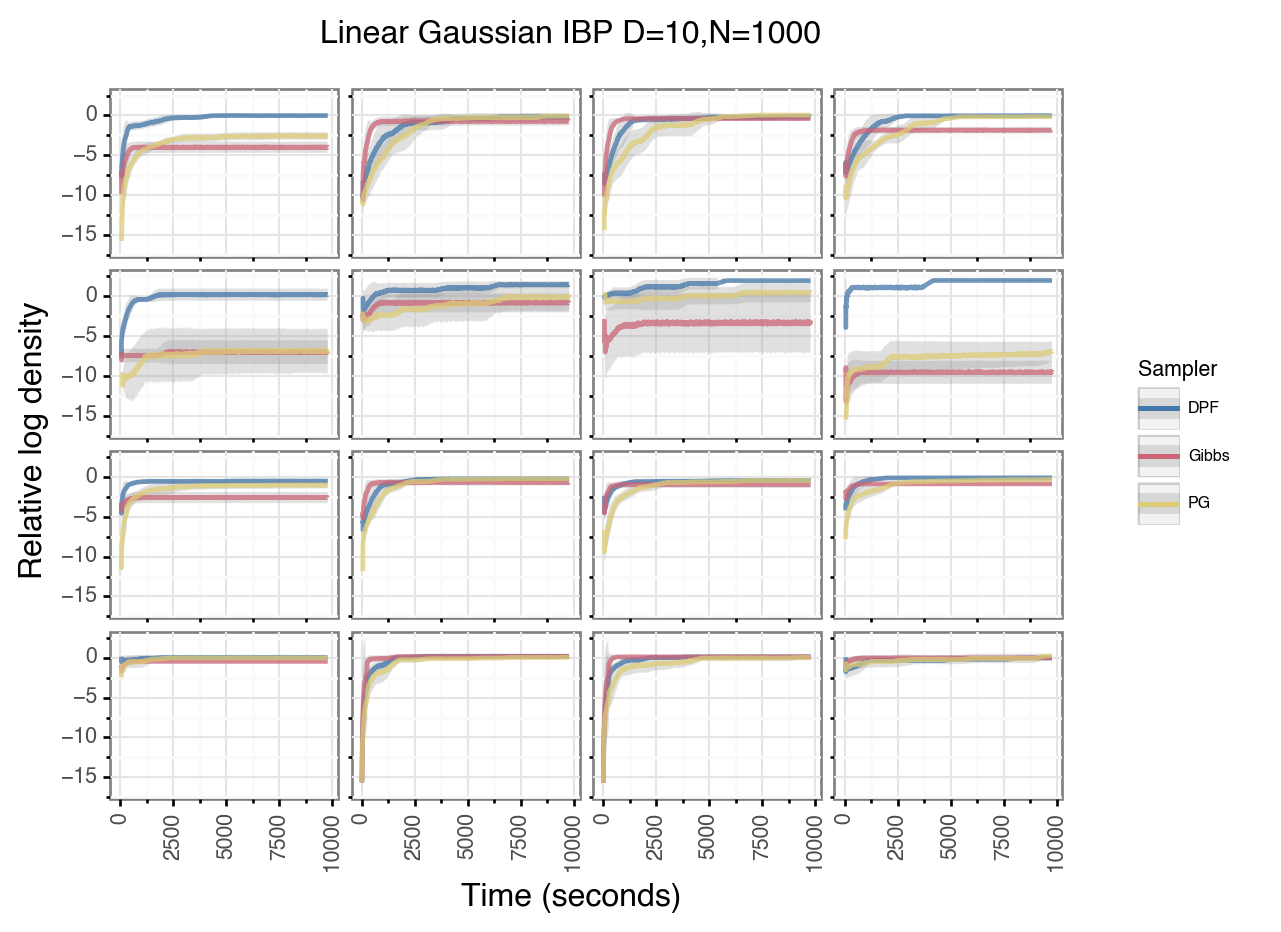}
	\caption{\traceComparisonCaption{linear Gaussian}{K=20 using an IBP prior}}
	\label{fig:lg_ibp_20_trace}
\end{figure}


\begin{figure}
	\includegraphics[scale=1.0]{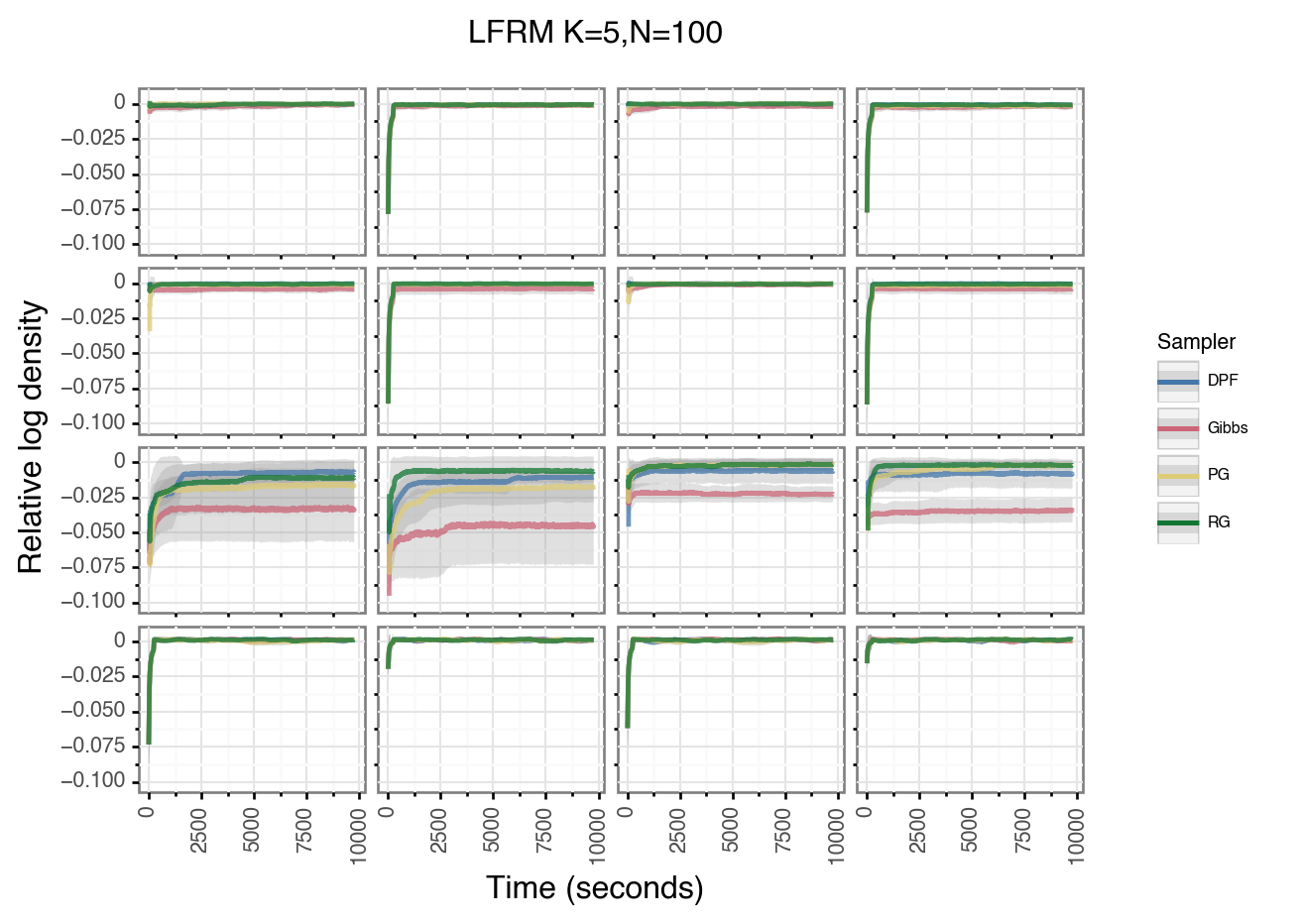}
	\caption{\traceComparisonCaption{non-symmetric LFRM}{K=5}}
	\label{fig:lfrm_finite_5_trace}
\end{figure}

\begin{figure}
	\includegraphics[scale=1.0]{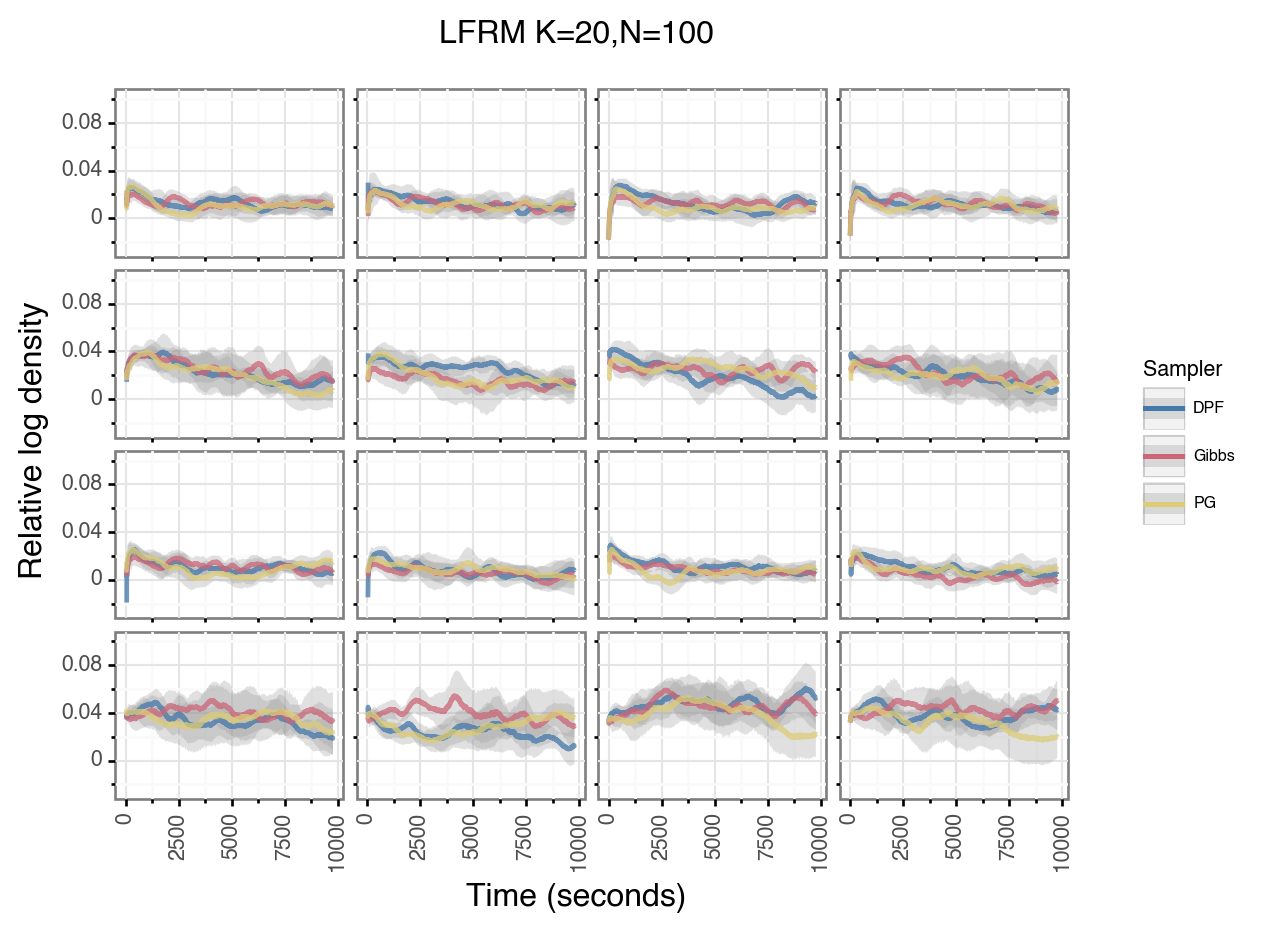}
	\caption{\traceComparisonCaption{non-symmetric LFRM}{K=20}}
	\label{fig:lfrm_finite_20_trace}
\end{figure}

\begin{figure}
	\includegraphics[scale=1.0]{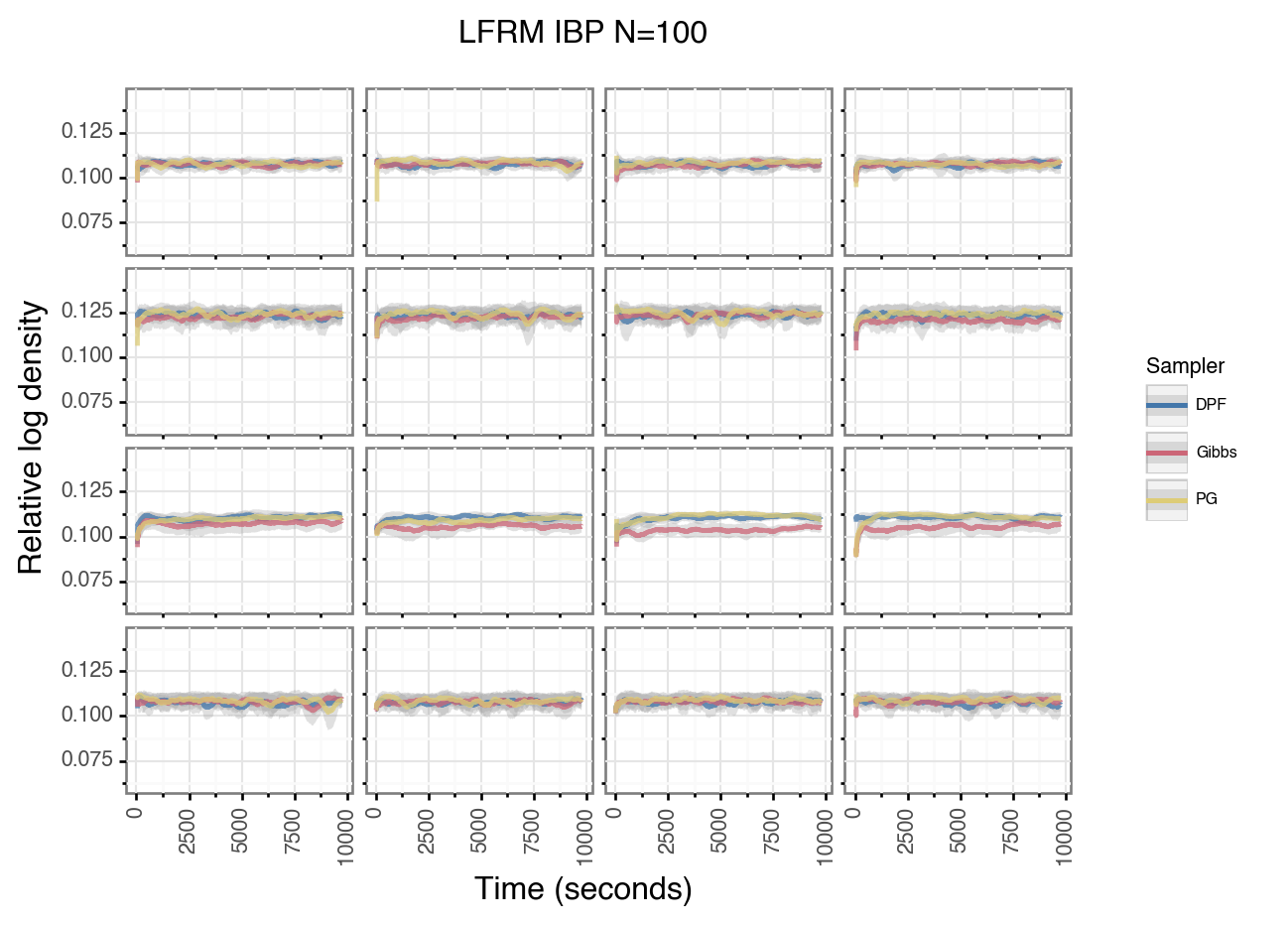}
	\caption{\traceComparisonCaption{non-symmetric LFRM}{K=20 using an IBP prior}}
	\label{fig:lfrm_ibp_20_trace}
\end{figure}

\begin{figure}
	\includegraphics[scale=0.55]{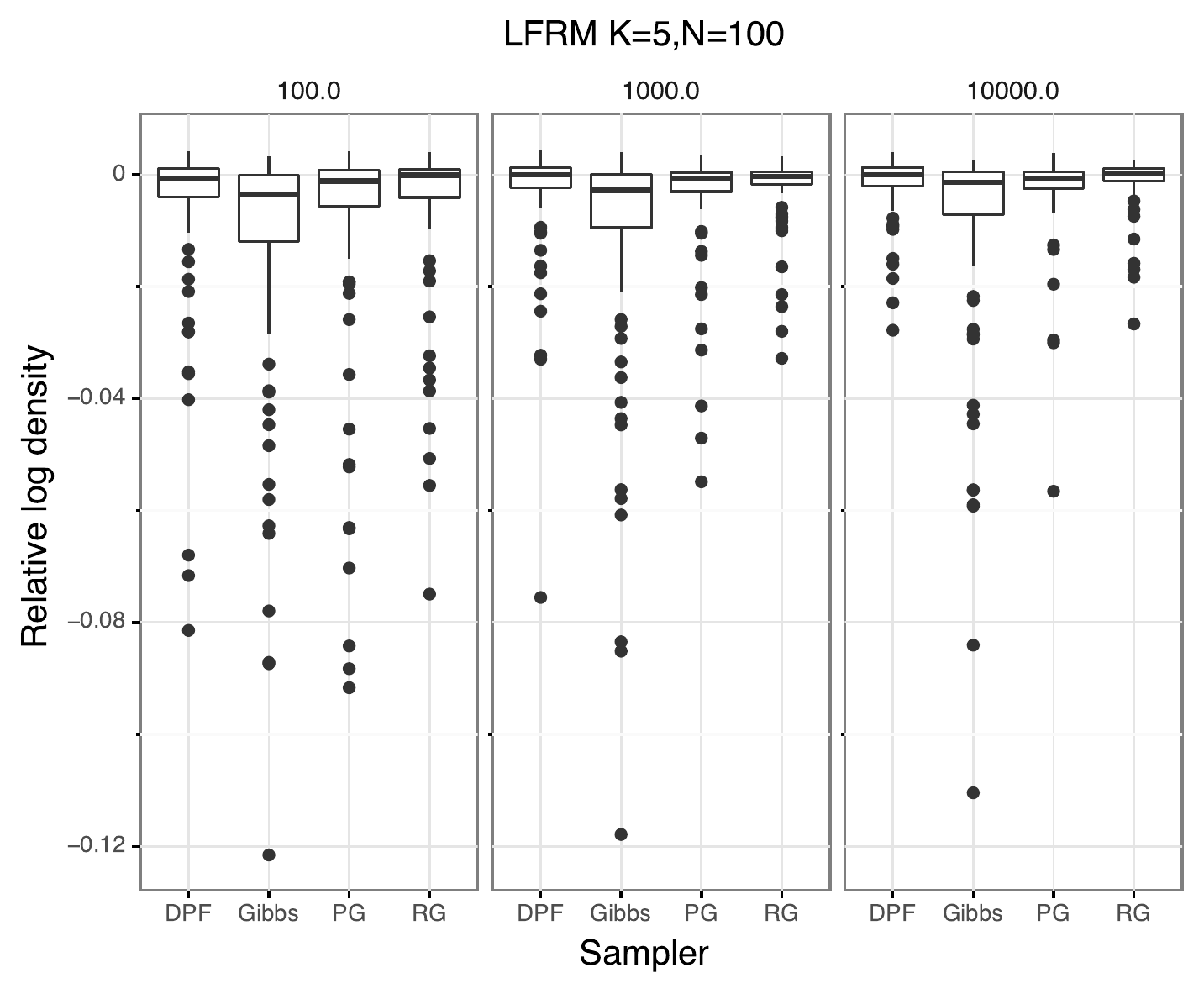}
	\includegraphics[scale=0.55]{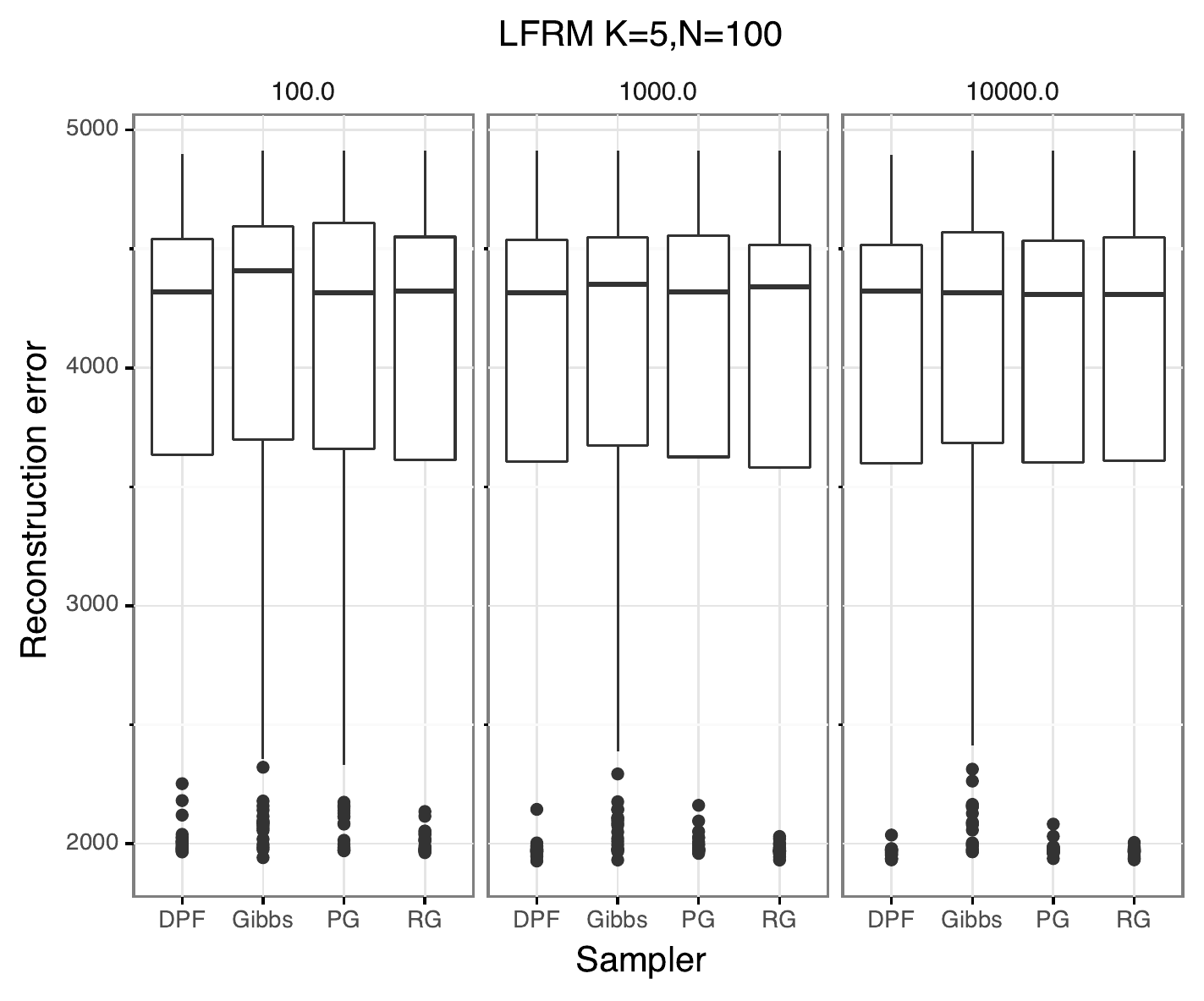}
	\vfill
	\includegraphics[scale=0.55]{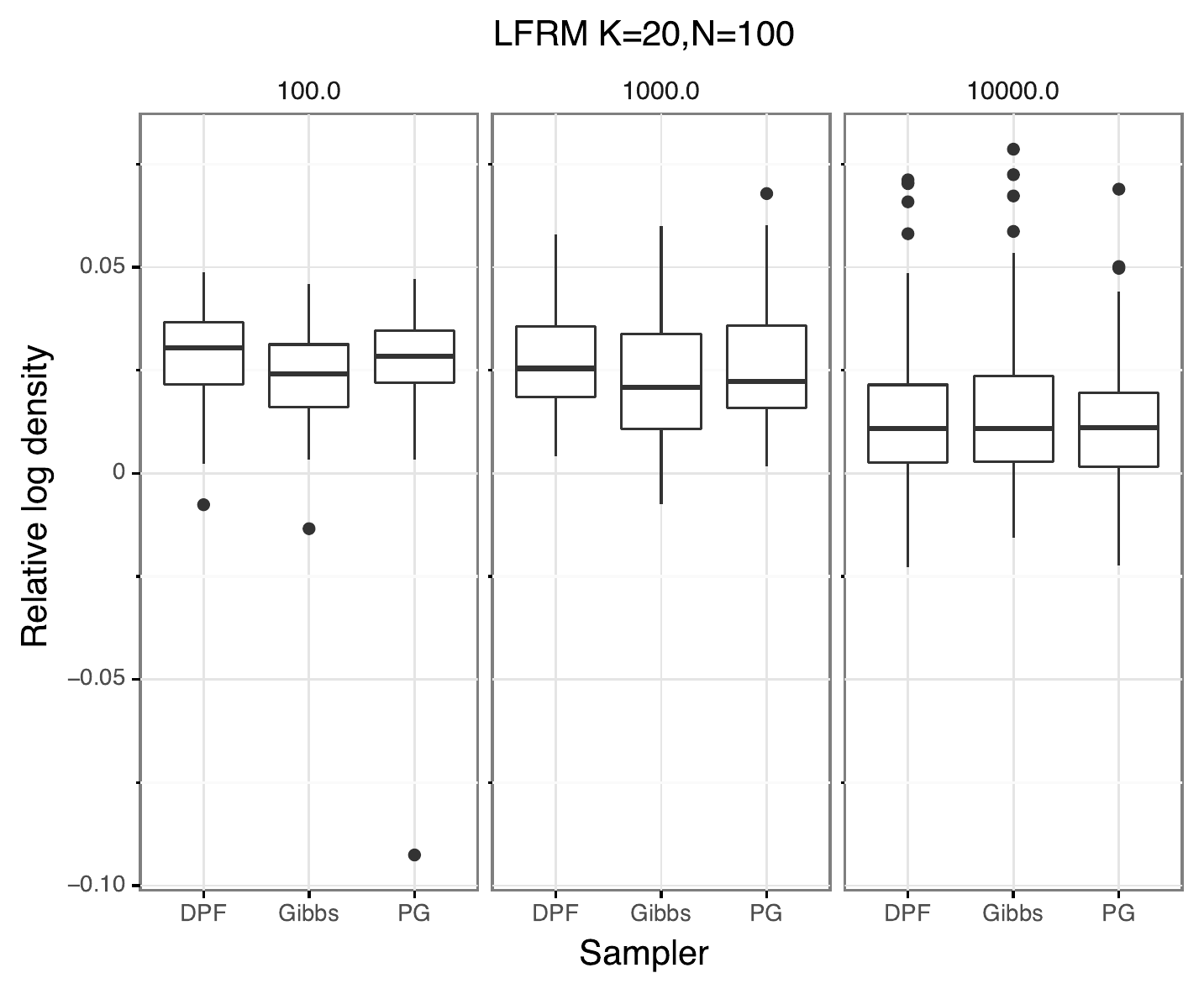}
	\includegraphics[scale=0.55]{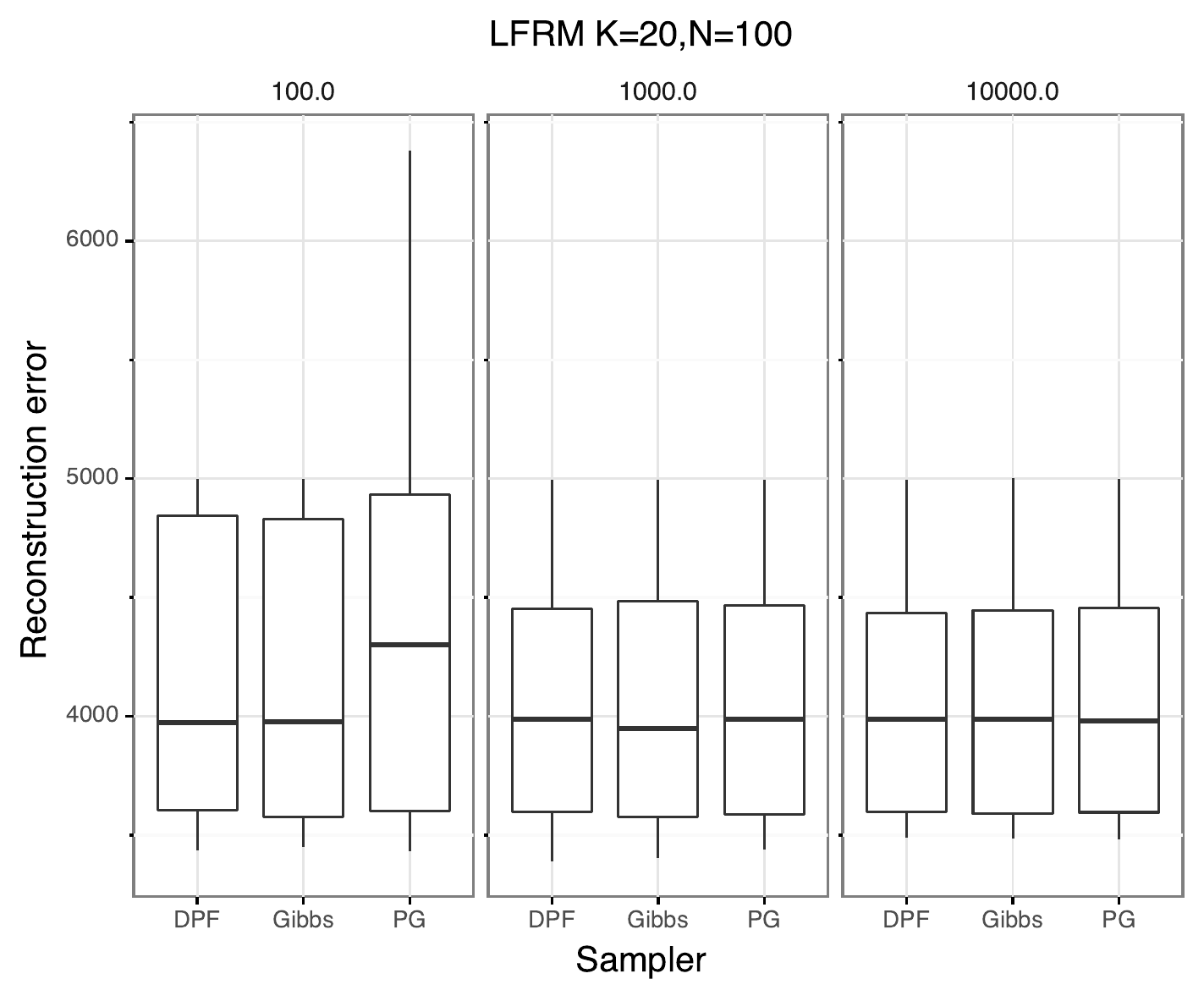}
	\vfill
	\includegraphics[scale=0.55]{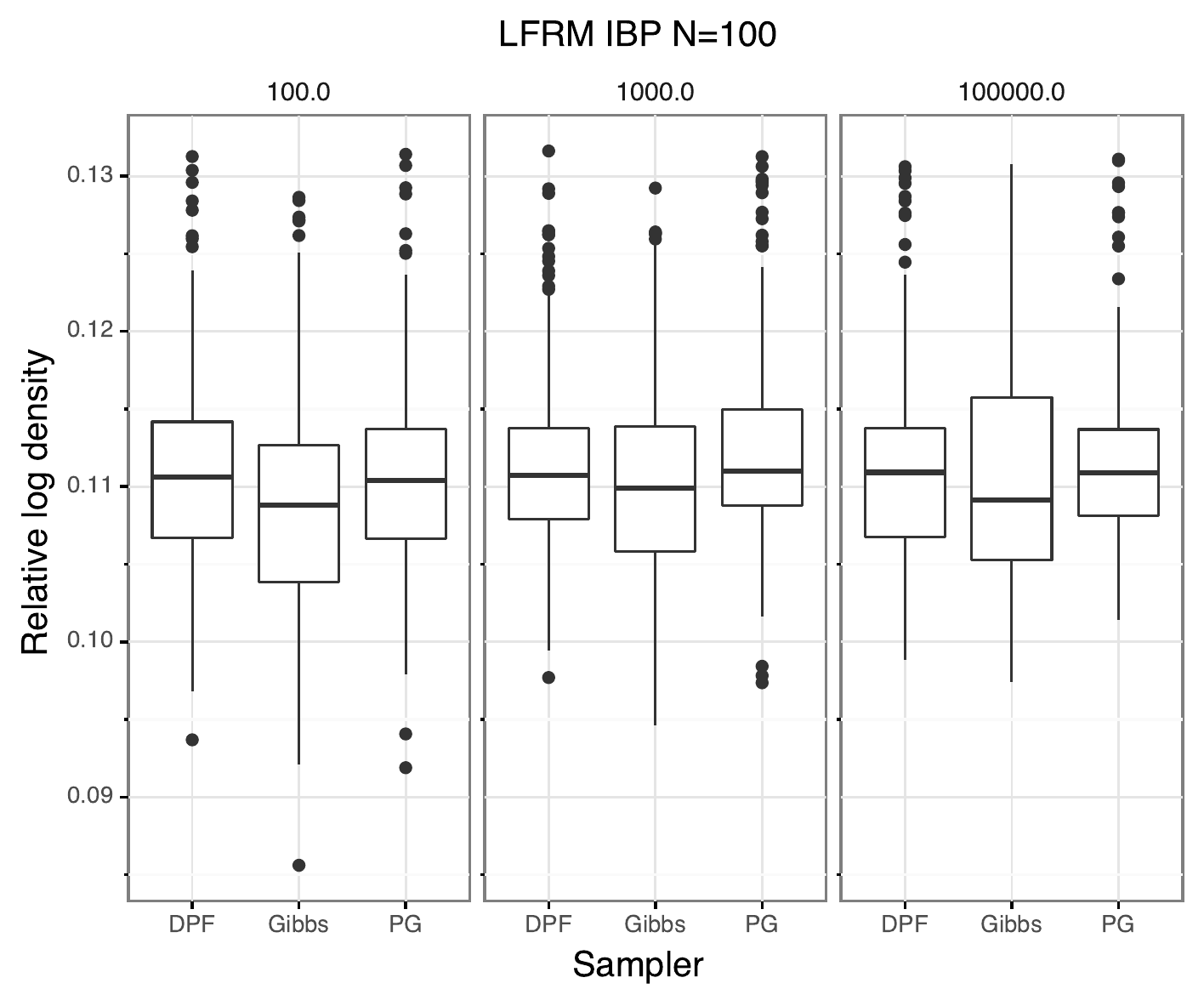}
	\includegraphics[scale=0.55]{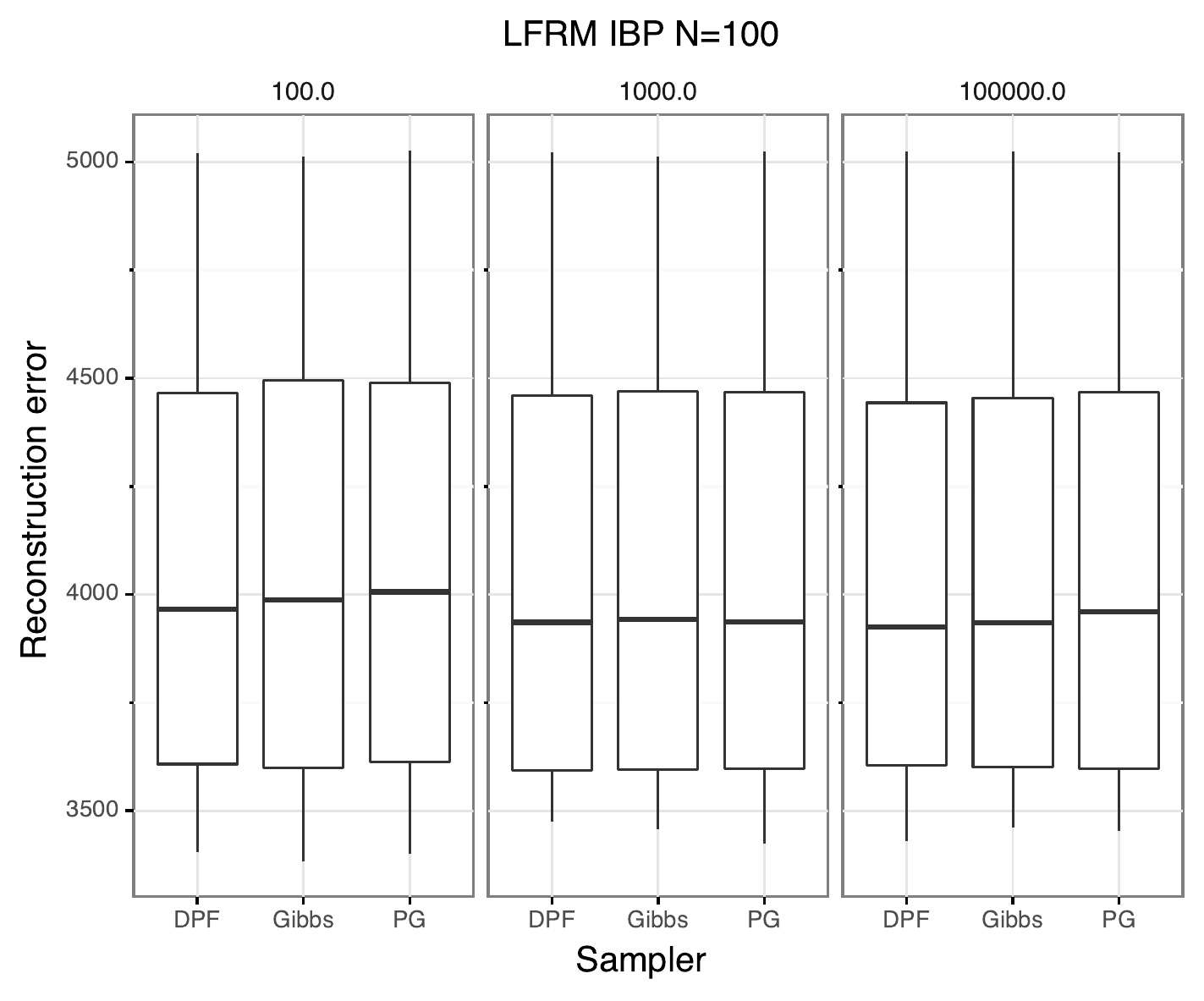}
	\caption{
		Performance of different samplers using synthetic data from the LFRM model.
		The box plots represent the distribution of values from 80 random starts of each parameter setting.
		We show the values of the relative log density (right) and reconstruction error (left).
	}
	\label{fig:lfrm_boxplot}
\end{figure}

\begin{figure}
	\includegraphics[scale=1.0]{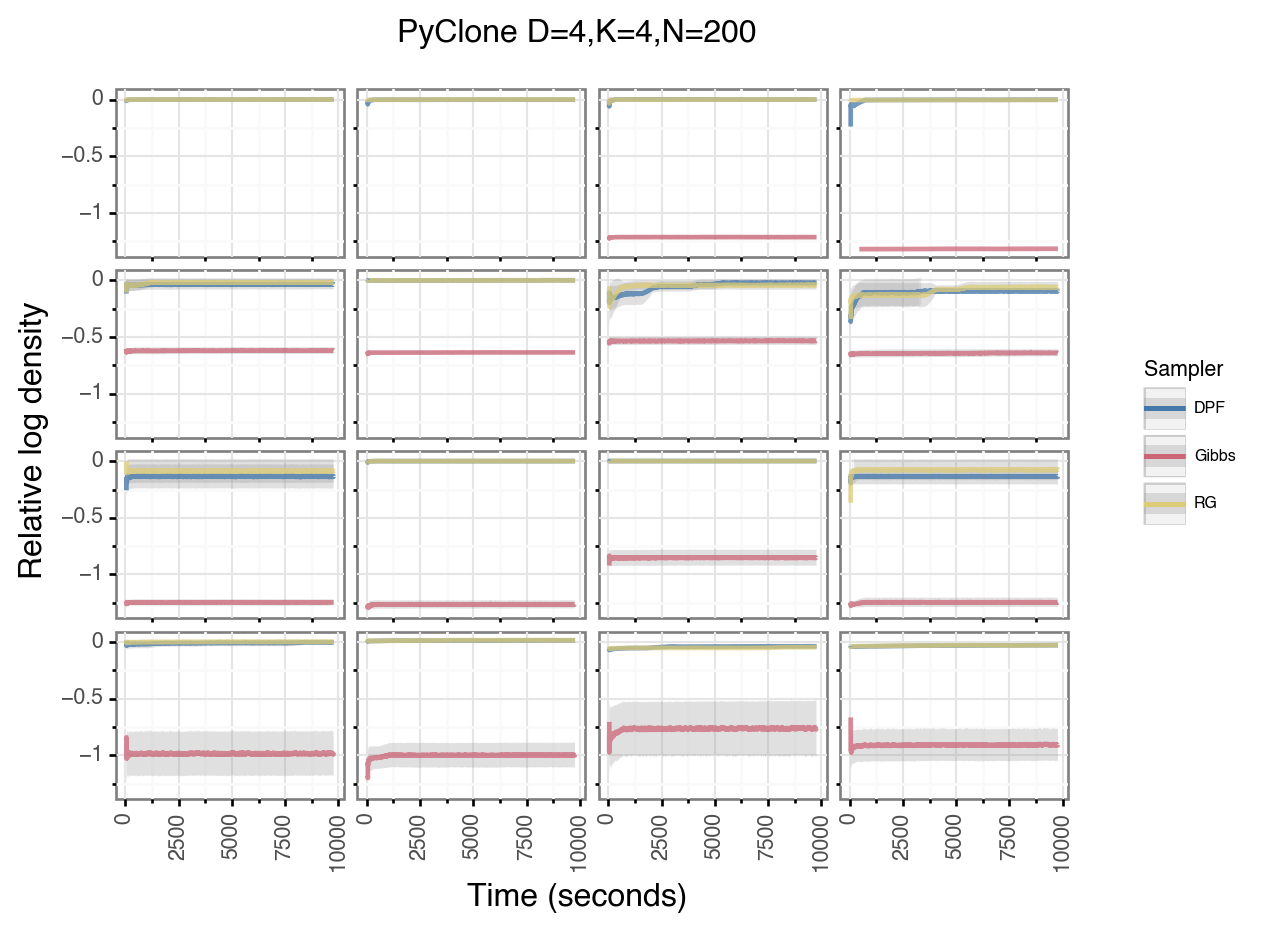}
	\caption{\traceComparisonCaption{PyClone}{D=4 and K=4}}
	\label{fig:pyclone_finite_4_trace}
\end{figure}

\begin{figure}
	\includegraphics[scale=1.0]{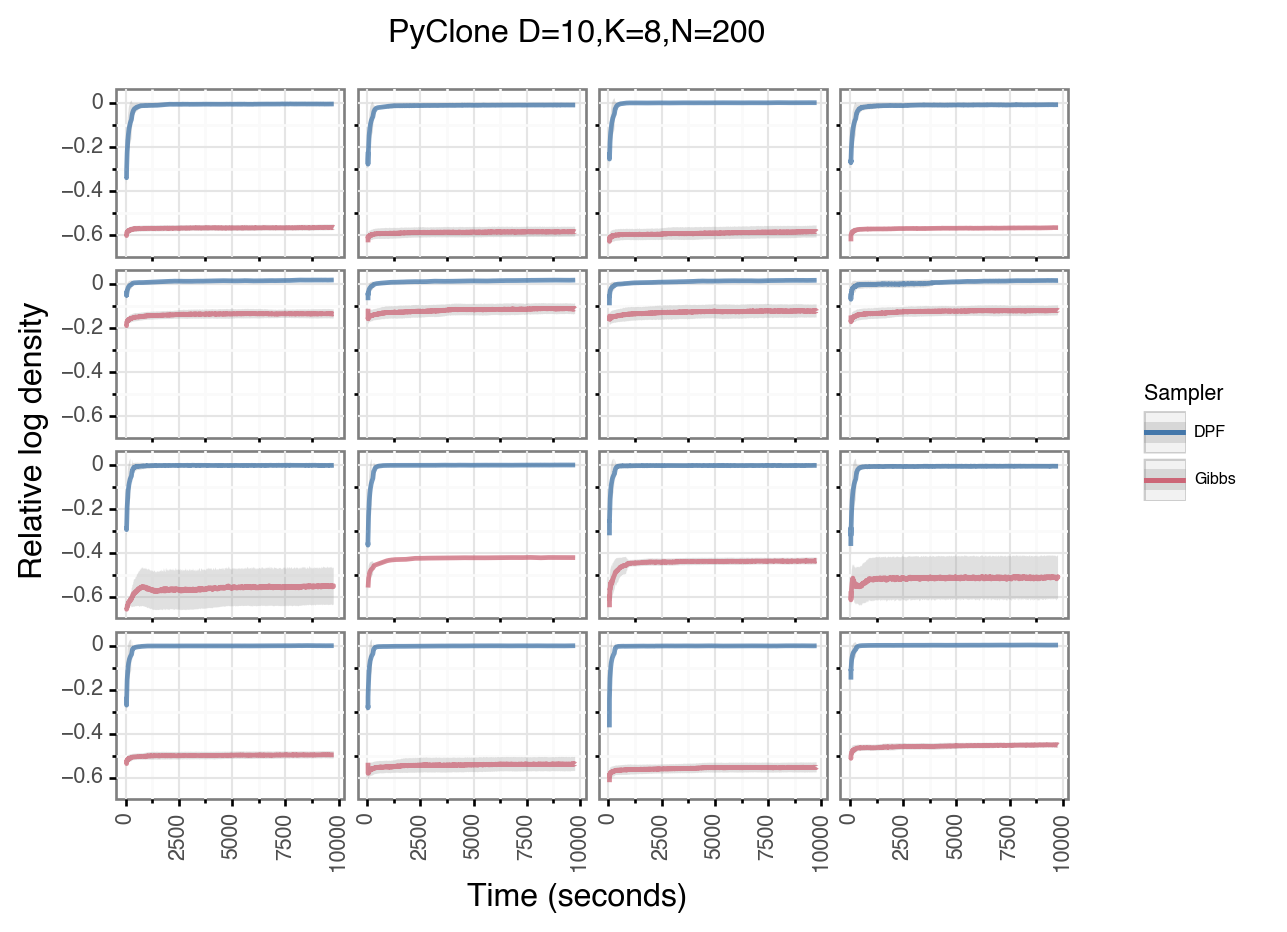}
	\caption{\traceComparisonCaption{PyClone}{D=10 and K=8}}
	\label{fig:pyclone_finite_8_trace}	
\end{figure}

\begin{figure}
	\includegraphics[scale=1.0]{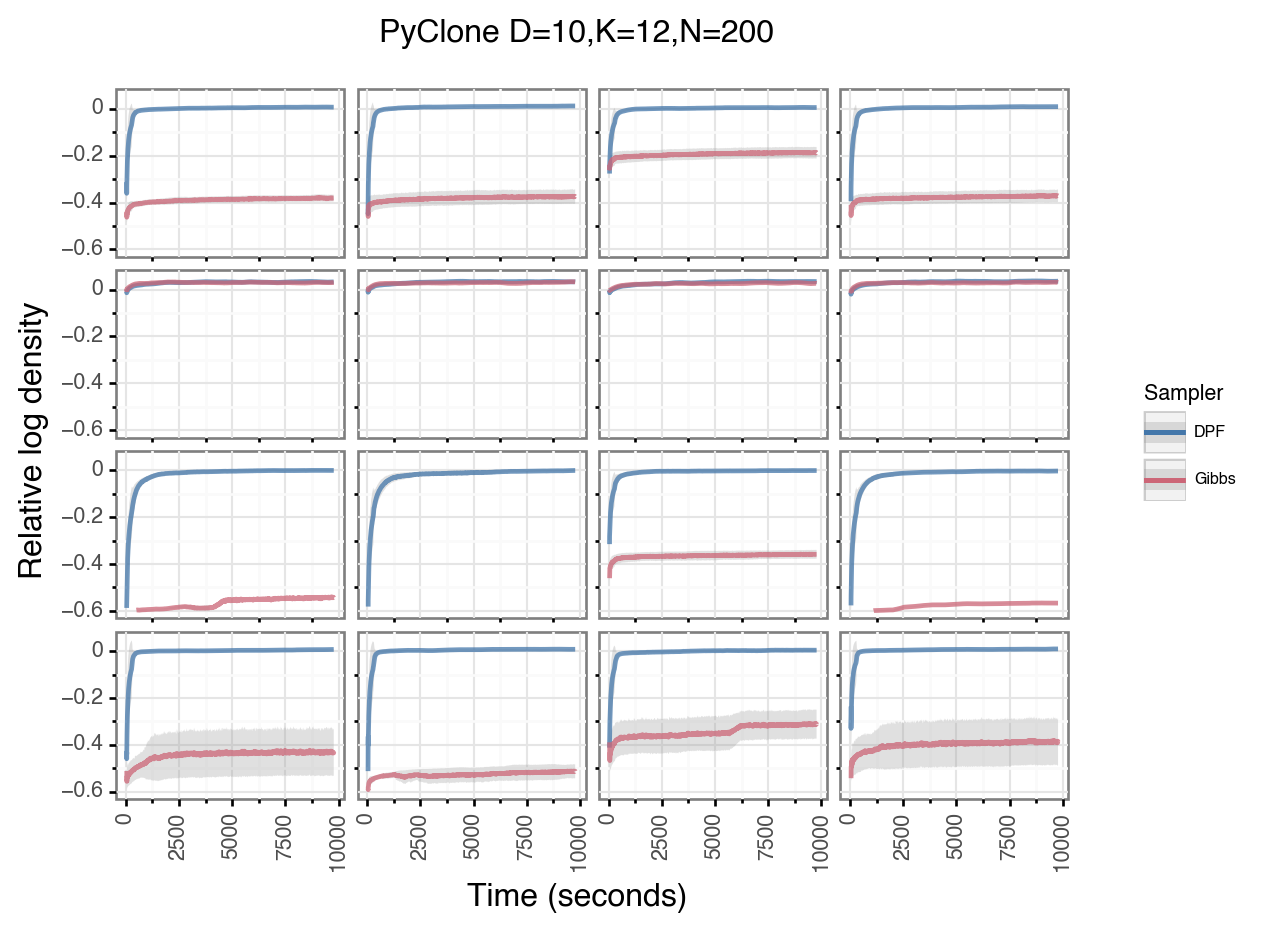}
	\caption{\traceComparisonCaption{PyClone}{D=8 and K=12}}
	\label{fig:pyclone_finite_12_trace}	
\end{figure}


\clearpage

\subsection{Supplementary tables}

\input{tables/latex/supp_tables}

%% file: models.tex
\subsection{Models}\label{sec:models}

We describe the three models we used for performance comparisons.
We use the notation $Z \mid \alpha \sim \text{FAM}(\cdot \mid \alpha)$ to describe sampling from a feature allocation distribution, either the FBB or IBP priors.
The number of features $K$ is implicitly determined by the number of columns of $Z$.
We place a $\text{Gamma}(\cdot \mid 1, 1)$ prior on the hyper-parameter $\alpha$ and use a random walk Metropolis-Hastings kernel to update the variable.

When referring to the Normal distribution we use the mean/precision parametrization.
When referring to the Gamma distribution we use the shape/rate parametrization.

\subsubsection{Linear Gaussian}

The linear Gaussian model has been widely used, particularly in the IBP literature (see \cite{griffiths2011indian} for example).
One reason for the model's popularity is that it is possible to marginalize the feature parameters, so a collapsed sampler can be developed.
In this work we do not exploit this, and instead work with uncollapsed model.
The hierarchical model is as follows:
\begin{eqnarray*}
	Z \mid \alpha & \sim & \text{FAM}(\cdot \mid \alpha) \\
	\tau_v \mid a_v, b_v & \sim & \text{Gamma}(\cdot \mid a_v, b_v) \\
	S_v & = & \tau_v \, \mathbf{I}_D \\
	\tau_x \mid a_x, b_x & \sim & \text{Gamma}(\cdot \mid a_x, b_x) \\
	S_x & = & \tau_x \, \mathbf{I}_D \\
	\boldsymbol{v}_k \mid \tau_v & \sim & \text{Normal}(\cdot \mid \mathbf{0}, S_v) \\
	\boldsymbol{x}_n \mid \{\boldsymbol{v}_k\}_{k=1}^{K}, \tau_x, \boldsymbol{z}_n & \sim & \text{Normal}(\cdot \mid \sum_{k=1}^{K} z_{n k} \boldsymbol{v}_k , S_x)
\end{eqnarray*}
We use a Gibbs kernels to update $\boldsymbol{v}_k$, $\tau_v$ and $\tau_x$.
When using the IBP prior we use a collapsed Metropolis-Hastings step to update the singletons \citep{doshi2009accelerated}.

\subsubsection{Linear Feature Relational Model}

The LFRM model was proposed by \cite{miller2009nonparametric}.
The observed data is a binary matrix $X \in \{0, 1\}^{N \times N}$ which encodes interactions between entities.
It could for example be used to model relationships on a social network.
The model posits that an underlying set of features encoded by $Z$ governs whether the entries in $X$ are on or off.
The hierarchical model is as follows:
\begin{eqnarray*}
	Z \mid \alpha & \sim & \text{FAM}(\cdot \mid \alpha) \\
	\tau \mid a, b & \sim & \text{Gamma}(\cdot \mid a, b) \\
	v_{k l} \mid \tau & \sim & \text{Normal}(\cdot | 0, \tau) \\
	x_{i j} \mid \{ v_{k l} \}, Z & \sim & \text{Bernoulli} \left( \cdot \mid \sigma \left ( \sum_{k=1}^{K} \sum_{l=1}^{K} z_{i k} z_{j l} v_{k l} \right) \right)
\end{eqnarray*}
where $\sigma(x) = \frac{1}{1 + e^{-x}}$.
Note that the model can be symmetric so that $v_{k l} = v_{l k}$ or non-symmetric by letting these parameters vary independently.
We use random walk Metropolis-Hastings kernels to update $\tau$ and $v_{k l}$.
When using the IBP prior we use a Metropolis-Hastings kernel with proposals from the prior to update the singletons.

\subsection{PyClone}

The original PyClone model was proposed by \cite{roth2014pyclone}.
The model assumes we observe data $a_{n m}, b_{n m} \in \mathbb{N}$ which represent the number of sequencing reads without and with mutation $n$ in sample $m$.
We refer to $d_{n m} = a_{n m} + b_{n m}$ as the sequencing depth.
We refer to the proportion of cells with mutation $n$ in sample $m$, $\phi_{n m}$, as the cellular prevalence.
We can model the probability of observing $b_{n m}$ reads with the mutation in the samples by a density $g(b_{n m} \mid d_{n m}, \phi_{n m}, *)$ where $*$ indicates other quantities which are not relevant to the discussion.
In the original PyClone model $\boldsymbol{\phi}_n$ is assumed to be sampled from a Dirichlet process so that mutations appearing at similar cellular prevalences are clustered.
This corresponds to the biological assumption mutations appear within sub-populations of cells, and that the cellular prevalence is the sum of the proportion of cells in the sub-populations containing the mutation.
We can alter this model to explicitly identify which sub-populations have the mutation using a feature allocation model.
Let $f_{k m}$ be the proportion of cells from population $k$ in sample $m$.
We use the feature allocation vector $\boldsymbol{z}_n$ for mutation $n$ to encode which sub-populations have the mutation.
The cellular prevalence is then given by $\phi_{n m} = \sum_{k=1}^{K} z_{n k} f_{k m}$.
Substituting this into the observation density $g$ gives the new model.
The hierarchical model is as follows:
\begin{eqnarray*}
	Z \mid \alpha & \sim & \text{FAM}(\cdot \mid \alpha) \\
	v_{k m} \mid a_v, b_v & \sim & \text{Gamma}(\cdot \mid a_v, b_v) \\
	f_{k m} \mid v_{k m} & = & \frac{v_{k m}}{\sum_{l=1}^{K} v_{l m}} \\
	\phi_{n m} \mid \{ f_{k m } \}_{k=1}^{K}, \boldsymbol{z}_{n} & = & \sum_{k=1}^{K} z_{n k} f_{k m} \\
	b_{n m} \mid d_{n m}, \{ f_{k m } \}_{k=1}^{K}, \boldsymbol{z}_{n}, * & \sim & g(\cdot \mid d_{n m}, \phi_{n m}, *)
\end{eqnarray*}
Updating $v_{k m}$ was somewhat difficult for this model so we used a number of MCMC kernels which included random walk Metropolis-Hastings kernels on either individual $v_{k m}$ values or blocks.
We also used a Metropolis-Hastings kernel where the proposal was a random permutation of the values for a sample.
The final kernel was the Multiple-Try-Metropolis kernel \citep{liu2008monte}.

%% file: tables/latex/supp_tables.tex
\includeTable{tables/latex/friedman_test/pg_num_particles.tex}
\includeTable{tables/latex/nemenyi_test/pg_num_particles.tex}
\includeTable{tables/latex/friedman_test/dpf_num_particles.tex}
\includeTable{tables/latex/nemenyi_test/dpf_num_particles.tex}
\includeTable{tables/latex/friedman_test/pg_resample_threshold.tex}
\includeTable{tables/latex/nemenyi_test/pg_resample_threshold.tex}
\includeTable{tables/latex/friedman_test/pg_annealing_power.tex}
\includeTable{tables/latex/nemenyi_test/pg_annealing_power.tex}
\includeTable{tables/latex/friedman_test/dpf_annealing_power.tex}
\includeTable{tables/latex/nemenyi_test/dpf_annealing_power.tex}
\includeTable{tables/latex/friedman_test/pg_test_path.tex}
\includeTable{tables/latex/nemenyi_test/pg_test_path.tex}
\includeTable{tables/latex/friedman_test/dpf_test_path.tex}
\includeTable{tables/latex/nemenyi_test/dpf_test_path.tex}
\includeTable{tables/latex/friedman_test/lg_finite_5.tex}
\includeTable{tables/latex/nemenyi_test/lg_finite_5.tex}
\includeTable{tables/latex/friedman_test/lg_finite_20.tex}
\includeTable{tables/latex/nemenyi_test/lg_finite_20.tex}
\includeTable{tables/latex/friedman_test/lg_ibp_20.tex}
\includeTable{tables/latex/nemenyi_test/lg_ibp_20.tex}
\includeTable{tables/latex/friedman_test/lfrm_finite_5.tex}
\includeTable{tables/latex/nemenyi_test/lfrm_finite_5.tex}
\includeTable{tables/latex/friedman_test/lfrm_finite_20.tex}
\includeTable{tables/latex/nemenyi_test/lfrm_finite_20.tex}
\includeTable{tables/latex/friedman_test/lfrm_ibp_20.tex}
\includeTable{tables/latex/nemenyi_test/lfrm_ibp_20.tex}
\includeTable{tables/latex/friedman_test/pyclone_finite_4.tex}
\includeTable{tables/latex/nemenyi_test/pyclone_finite_4.tex}
\includeTable{tables/latex/friedman_test/pyclone_finite_8.tex}
\includeTable{tables/latex/nemenyi_test/pyclone_finite_8.tex}
\includeTable{tables/latex/friedman_test/pyclone_finite_12.tex}
\includeTable{tables/latex/nemenyi_test/pyclone_finite_12.tex}

%% file: pgfa.bbl
\begin{thebibliography}{21}
\providecommand{\natexlab}[1]{#1}
\providecommand{\url}[1]{\texttt{#1}}
\expandafter\ifx\csname urlstyle\endcsname\relax
  \providecommand{\doi}[1]{doi: #1}\else
  \providecommand{\doi}{doi: \begingroup \urlstyle{rm}\Url}\fi

\bibitem[Amig{\'o} et~al.(2009)Amig{\'o}, Gonzalo, Artiles, and
  Verdejo]{amigo2009comparison}
Enrique Amig{\'o}, Julio Gonzalo, Javier Artiles, and Felisa Verdejo.
\newblock A comparison of extrinsic clustering evaluation metrics based on
  formal constraints.
\newblock \emph{Information retrieval}, 12\penalty0 (4):\penalty0 461--486,
  2009.

\bibitem[Andrieu et~al.(2010)Andrieu, Doucet, and
  Holenstein]{andrieu2010particle}
Christophe Andrieu, Arnaud Doucet, and Roman Holenstein.
\newblock Particle markov chain monte carlo methods.
\newblock \emph{Journal of the Royal Statistical Society: Series B (Statistical
  Methodology)}, 72\penalty0 (3):\penalty0 269--342, 2010.

\bibitem[Barembruch et~al.(2009)Barembruch, Garivier, and
  Moulines]{barembruch2009approximate}
Steffen Barembruch, Aur{\'e}lien Garivier, and Eric Moulines.
\newblock On approximate maximum-likelihood methods for blind identification:
  How to cope with the curse of dimensionality.
\newblock \emph{IEEE Transactions on Signal Processing}, 57\penalty0
  (11):\penalty0 4247--4259, 2009.

\bibitem[Bouchard-C{\^o}t{\'e} et~al.(2017)Bouchard-C{\^o}t{\'e}, Doucet, and
  Roth]{bouchard2017particle}
Alexandre Bouchard-C{\^o}t{\'e}, Arnaud Doucet, and Andrew Roth.
\newblock Particle gibbs split-merge sampling for bayesian inference in mixture
  models.
\newblock \emph{The Journal of Machine Learning Research}, 18\penalty0
  (1):\penalty0 868--906, 2017.

\bibitem[Broderick et~al.(2013)Broderick, Jordan, Pitman,
  et~al.]{broderick2013cluster}
Tamara Broderick, Michael~I Jordan, Jim Pitman, et~al.
\newblock Cluster and feature modeling from combinatorial stochastic processes.
\newblock \emph{Statistical Science}, 28\penalty0 (3):\penalty0 289--312, 2013.

\bibitem[Chopin et~al.(2015)Chopin, Singh, et~al.]{chopin2015particle}
Nicolas Chopin, Sumeetpal~S Singh, et~al.
\newblock On particle gibbs sampling.
\newblock \emph{Bernoulli}, 21\penalty0 (3):\penalty0 1855--1883, 2015.

\bibitem[Dem{\v{s}}ar(2006)]{demvsar2006statistical}
Janez Dem{\v{s}}ar.
\newblock Statistical comparisons of classifiers over multiple data sets.
\newblock \emph{Journal of Machine learning research}, 7\penalty0
  (Jan):\penalty0 1--30, 2006.

\bibitem[Doshi-Velez and Ghahramani(2009)]{doshi2009accelerated}
Finale Doshi-Velez and Zoubin Ghahramani.
\newblock Accelerated sampling for the indian buffet process.
\newblock In \emph{Proceedings of the 26th annual international conference on
  machine learning}, pages 273--280. ACM, 2009.

\bibitem[Doucet and Johansen(2009)]{doucet2009tutorial}
Arnaud Doucet and Adam~M Johansen.
\newblock A tutorial on particle filtering and smoothing: Fifteen years later.
\newblock \emph{Handbook of nonlinear filtering}, 12\penalty0
  (656-704):\penalty0 3, 2009.

\bibitem[Fearnhead and Clifford(2003)]{fearnhead2003line}
Paul Fearnhead and Peter Clifford.
\newblock On-line inference for hidden markov models via particle filters.
\newblock \emph{Journal of the Royal Statistical Society: Series B (Statistical
  Methodology)}, 65\penalty0 (4):\penalty0 887--899, 2003.

\bibitem[Fox et~al.(2014)Fox, Hughes, Sudderth, Jordan, et~al.]{fox2014joint}
Emily~B Fox, Michael~C Hughes, Erik~B Sudderth, Michael~I Jordan, et~al.
\newblock Joint modeling of multiple time series via the beta process with
  application to motion capture segmentation.
\newblock \emph{The Annals of Applied Statistics}, 8\penalty0 (3):\penalty0
  1281--1313, 2014.

\bibitem[Ghahramani and Griffiths(2006)]{ghahramani2006infinite}
Zoubin Ghahramani and Thomas~L Griffiths.
\newblock Infinite latent feature models and the indian buffet process.
\newblock In \emph{Advances in neural information processing systems}, pages
  475--482, 2006.

\bibitem[Griffiths and Ghahramani(2011)]{griffiths2011indian}
Thomas~L Griffiths and Zoubin Ghahramani.
\newblock The indian buffet process: An introduction and review.
\newblock \emph{Journal of Machine Learning Research}, 12\penalty0
  (Apr):\penalty0 1185--1224, 2011.

\bibitem[Lindsten et~al.(2014)Lindsten, Jordan, and
  Sch{\"o}n]{lindsten2014particle}
Fredrik Lindsten, Michael~I Jordan, and Thomas~B Sch{\"o}n.
\newblock Particle gibbs with ancestor sampling.
\newblock \emph{The Journal of Machine Learning Research}, 15\penalty0
  (1):\penalty0 2145--2184, 2014.

\bibitem[Liu(2008)]{liu2008monte}
Jun~S Liu.
\newblock \emph{Monte Carlo strategies in scientific computing}.
\newblock Springer Science \& Business Media, 2008.

\bibitem[Meeds et~al.(2007)Meeds, Ghahramani, Neal, and
  Roweis]{meeds2007modeling}
Edward Meeds, Zoubin Ghahramani, Radford~M Neal, and Sam~T Roweis.
\newblock Modeling dyadic data with binary latent factors.
\newblock In \emph{Advances in neural information processing systems}, pages
  977--984, 2007.

\bibitem[Miller et~al.(2009)Miller, Jordan, and
  Griffiths]{miller2009nonparametric}
Kurt Miller, Michael~I Jordan, and Thomas~L Griffiths.
\newblock Nonparametric latent feature models for link prediction.
\newblock In \emph{Advances in neural information processing systems}, pages
  1276--1284, 2009.

\bibitem[Rosenberg and Hirschberg(2007)]{rosenberg2007v}
Andrew Rosenberg and Julia Hirschberg.
\newblock V-measure: A conditional entropy-based external cluster evaluation
  measure.
\newblock In \emph{Proceedings of the 2007 joint conference on empirical
  methods in natural language processing and computational natural language
  learning (EMNLP-CoNLL)}, pages 410--420, 2007.

\bibitem[Roth et~al.(2014)Roth, Khattra, Yap, Wan, Laks, Biele, Ha, Aparicio,
  Bouchard-C{\^o}t{\'e}, and Shah]{roth2014pyclone}
Andrew Roth, Jaswinder Khattra, Damian Yap, Adrian Wan, Emma Laks, Justina
  Biele, Gavin Ha, Samuel Aparicio, Alexandre Bouchard-C{\^o}t{\'e}, and
  Sohrab~P Shah.
\newblock Pyclone: statistical inference of clonal population structure in
  cancer.
\newblock \emph{Nature methods}, 11\penalty0 (4):\penalty0 396, 2014.

\bibitem[Whiteley et~al.(2010)Whiteley, Andrieu, and
  Doucet]{whiteley2010efficient}
Nick Whiteley, Christophe Andrieu, and Arnaud Doucet.
\newblock Efficient bayesian inference for switching state-space models using
  discrete particle markov chain monte carlo methods.
\newblock \emph{arXiv preprint arXiv:1011.2437}, 2010.

\bibitem[Wood and Griffiths(2007)]{wood2007particle}
Frank Wood and Thomas~L Griffiths.
\newblock Particle filtering for nonparametric bayesian matrix factorization.
\newblock In \emph{Advances in neural information processing systems}, pages
  1513--1520, 2007.

\end{thebibliography}
